%% file: TIT_final_arXiv.tex
\documentclass[final,twoside]{IEEEtran}

\usepackage[]{graphicx,color} 
\usepackage{cite,acronym}

\usepackage[cmex10]{amsmath}
\usepackage{amsfonts,amssymb,amsxtra,amsthm}
\usepackage{bbding,stmaryrd,dsfont}
\acrodef{CSI}{Channel State Information}

\input{rv_defs}
\input{fct_defs}
\input{th_defs}

\renewcommand{\E}[2][]{{\mathbb{E}_{#1}}{\left[#2\right]}}       
\renewcommand{\P}[2][]{{\mathbb{P}_{#1}}{\left[#2\right]}}
\renewcommand{\leq}{\leqslant}
\renewcommand{\geq}{\geqslant}

\newcommand{\metric}[1]{\mathbb{S}_{{\scriptsize#1}}}
\newcommand{\perr}{\mathbb{P}_{e}}
\newcommand{\unif}{\mathbb{U}}

\begingroup\makeatletter\ifx\SetFigFontNFSS\undefined%
\gdef\SetFigFontNFSS#1#2#3#4#5{%
  \reset@font\fontsize{#1}{#2pt}%
  \fontfamily{#3}\fontseries{#4}\fontshape{#5}%
  \selectfont}%
\fi\endgroup%

\graphicspath{{Graphics/}}

\renewcommand{\rvM}{W}

\begin{document}

\title{Strong Secrecy from Channel Resolvability \thanks{Matthieu R. Bloch is with the School of Electrical and Computer Engineering,
    Georgia Institute of Technology, Atlanta, GA, and with the GT-CNRS UMI 2958, Metz, France. J. Nicholas Laneman is
    with the Department of Electrical Engineering, University of Notre Dame, Notre Dame, IN. Parts of these results were presented at the 46th Allerton Conference on Communication, Controls, and Computing, Monticello, IL, at the 1st International ICST Workshop on Secure Wireless Networks, Cachan, France, and at the 2011 IEEE International Symposium on Information Theory, Saint-Petersburg, Russia.}  } 
\author{Matthieu~R.~Bloch,~\IEEEmembership{Member,~IEEE,} and J.~Nicholas~Laneman,~\IEEEmembership{Senior~Member,~IEEE}} 
\maketitle

\markboth{IEEE Transactions on Information Theory, To appear}{Bloch and Laneman: Strong Secrecy from Channel Resolvability}

\begin{abstract}
  We analyze physical-layer security based on the premise that the coding mechanism for secrecy over noisy channels is tied to the notion of channel resolvability. Instead of considering \emph{capacity-based} constructions, which associate to each message a sub-code that operates just below the capacity of the eavesdropper's channel, we consider \emph{channel-resolvability-based} constructions, which associate to each message a sub-code that operates just above the resolvability of the eavesdropper's channel. Building upon the work of Csisz\'ar and Hayashi, we provide further evidence that channel resolvability is a powerful and versatile coding mechanism for secrecy by developing results that hold for strong secrecy metrics and arbitrary channels. 

Specifically, we show that at least for symmetric wiretap channels, random capacity-based constructions fail to achieve the strong secrecy capacity while channel-resolvability-based constructions achieve it. We then leverage channel resolvability to establish the secrecy-capacity region of arbitrary broadcast channels with confidential messages and a cost constraint for strong secrecy metrics. Finally, we specialize our results to study the secrecy capacity of wireless channels with perfect channel state information, mixed channels and compound channels with receiver \ac{CSI}, as well as the secret-key capacity of source models for secret-key agreement. By tying secrecy to channel resolvability, we obtain achievable rates for strong secrecy metrics with simple proofs.
\end{abstract}


\section{Introduction}
\label{sec:introduction}

In virtually every communication system, the problems of reliability and secrecy are handled in fundamentally different ways. Typically, error-correcting schemes in the physical-layer guarantee reliable communications, while encryption algorithms and key-exchange protocols in the upper layers\footnote{Specific cryptographic schemes are implemented at all upper layers of the protocol stack, including MAC, transport, network, and application layers.} ensure data secrecy. Physical-layer security puts forward an alternative role for the physical layer, whereby reliability and secrecy can be handled jointly by means of appropriate coding schemes. The idea is to recognize the presence of noise in every communication channel, including the channel of a potential adversary who eavesdrops on transmitted signals, and to exploit knowledge of noise statistics to prevent eavesdroppers from retrieving information. Unlike most existing security schemes, physical-layer security can guarantee information-theoretic security, by which secrecy is measured quantitatively in terms of the statistical dependence between the messages transmitted and the observations of eavesdroppers. 

The theoretical foundation of physical-layer security is the early works of Wyner~\cite{Wyner1975} and Csisz\'ar \& K\"orner~\cite{Csiszar1978}, which prove the existence of coding schemes ensuring reliability and secrecy for the wiretap channel; however, the recent surge of information-theoretic results regarding the wiretap channel has fostered few practical engineering solutions. This state of affairs is partly due to the fact that most works extend the coding schemes of~\cite{Wyner1975,Csiszar1978}, in which the coding mechanism that guarantees secrecy is tied to channel capacity. This mechanism will be precisely defined in Section~\ref{sec:secrecy-metrics}; at this point, suffice to say that the codes in \cite{Wyner1975,Csiszar1978} are a union of sub-codes that operate just below the capacity of the eavesdropper's channel as the blocklength grows large. Although such coding schemes have been successfully used to study many multiuser information-theoretic secrecy problems~\cite{InfoTheoreticSec,Bloch2011}, deriving secrecy from channel capacity leaves open a few lingering issues:
\begin{enumerate}
\item wiretap channel models that incorporate the limitations of modern communication systems, such as the presence of memory, are difficult to analyze;
\item the results obtained by tying secrecy to channel capacity are deemed too weak for cryptographic applications.
\end{enumerate}
This paper builds upon an original observation of Csisz\'ar~\cite{Csiszar1996} and the work of Hayashi~\cite{Hayashi2006} to explore an alternative approach to physical-layer security that addresses the aforementioned issues; the premise of the approach is to relate the coding mechanism for secrecy to the notion of channel resolvability~\cite{Han1993,Han1994} and not to channel capacity.

\subsection{Motivating Examples}
\label{sec:motiv-examples}

To motivate the usefulness of channel resolvability, we start with two intuitive examples that shed light on the mechanisms one could exploit to ensure information-theoretic security. 

\begin{example}[One-time pad]
\label{ex:onte-time-pad}
  Consider a binary message $\rvM\in\{0,1\}$ that is encoded into a codeword $\rvZ$ as $\rvZ=\rvM\oplus\rvK$, where $\rvK\sim\calB\left(\tfrac{1}{2}\right)$ is a secret key and $\oplus$ denotes modulo-two addition. The crypto lemma~\cite{Forney2003} shows that the output distributions $p_{\rvZ|\rvM=0}$ and $p_{\rvZ|\rvM=1}$ are identical and equal to the uniform distribution on $\{0,1\}$; hence, messages are statistically indistinguishable for an eavesdropper only observing $\rvZ$. From an operational perspective, note that the encoder exploits the key $\rvK$ to ensure that all messages {induce} the same output distribution.
\end{example}

\begin{example}[Transmission over a noisy Gaussian channel]
\label{ex:noisy-Gaussian}
  Consider an uncoded message $\rvM$ uniformly distributed in the set $\{-1,+1\}$ and observed by an eavesdropper at the output of a real additive white Gaussian noise channel as $\rvZ=\rvM+\rvN$, where $\rvN\sim\calN(0,\sigma^2)$. As illustrated in Figure~\ref{fig:bpsk_awgn}, the output distributions $p_{\rvZ|\rvM=-1}$ and $p_{\rvZ|\rvM=+1}$ become indistinguishable from the average distribution $p_\rvZ$ as the noise variance increases.  
  \begin{figure}[htbp] 
    \centering 
    \includegraphics[width=.85\linewidth]{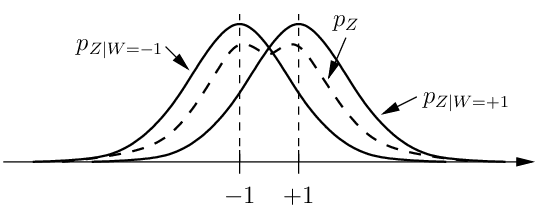}
    \caption{Distributions of channel outputs for uncoded transmission of $\{-1,+1\}$ over an AWGN channel.} 
    \label{fig:bpsk_awgn}
  \end{figure} 
  Specifically, as $\sigma$ goes to infinity, one can show that, for each $m\in\{-1,+1\}$, the variational distance between $p_{\rvZ|\rvM=m}$ and $p_\rvZ$ is at most $\calO(\sigma^{-\frac{1}{2}})$. In other words, if the noise introduces enough randomness, then the channel itself ensures that all messages approximately induce the same output distribution.
 \end{example}

In each example, statistical indistinguishability is obtained because there exists a source of randomness (key or channel noise) and a coding mechanism by which all messages induce the same distribution for the eavesdropper's observations; this mechanism is reminiscent of the codes analyzed in~\cite{Han1993,Han1994,InformationSpectrumMethods} to study the notion of \emph{channel resolvability}. At this point, the connection between secrecy and channel resolvability may seem contrived but, nevertheless, it suggests the possibility of ensuring secrecy by means that are different from those based on channel capacity and used in~\cite{Wyner1975,Csiszar1978}; this was already observed in~\cite{Csiszar1996} and more formally explored in~\cite{Hayashi2006}. In the remainder of this paper, we further expand upon this idea and we not only highlight the benefits of explicitly connecting secrecy to channel resolvability but also show the limitations of an approach based on channel capacity.

\subsection{Related Work}
\label{sec:related-work}

Most communication architectures providing information-theoretic security are based on two models of communication. The \emph{wiretap channel}, introduced by Wyner~\cite{Wyner1975} and generalized by Csisz\'ar \& K\"orner~\cite{Csiszar1978}, models an architecture in which a transmitter encodes messages $\rvM$ into codewords $\mathbf{X}$ of $n$ symbols for transmission to a receiver, in the presence of an eavesdropper that obtains noisy observations ${\mathbf{Z}}$ of ${\mathbf{X}}$. In the case of discrete memoryless channels, ~\cite{Wyner1975,Csiszar1978} have shown the existence of coding schemes simultaneously ensuring reliable transmission to the receiver and secrecy with respect to the eavesdropper. In particular, it is possible to characterize the \emph{secrecy capacity} of a wiretap channel, defined as the supremum of all reliable and secure rates. The extension of this result to Gaussian~\cite{Leung-Yan-Cheong1978} and wireless channels (see, for instance,~\cite{Liang2008a} and references therein) suggests the potential of such coding schemes to secure communication networks at the physical layer. An alternative to the wiretap channel is the \emph{source model for secret-key agreement} introduced by Maurer~\cite{Maurer1993} and Ahlswede \& Csisz\'ar~\cite{Ahlswede1993}, which considers an architecture in which two legitimate parties attempt to distill secret keys from a noisy source by communicating over a public channel. The resulting keys have to be secure with respect to an eavesdropper who obtains correlated observations from the source and observes all messages exchanged over the public channel. This architecture differs from the wiretap channel by exclusively focusing on the rate of the secret key that can be distilled from the source and by ignoring the cost of public communication. The counterpart of secrecy capacity is the \emph{secret-key capacity}, defined  as the supremum of the secret key rates that can be distilled. Although the aforementioned architectures model fundamentally different communication scenarios, they are related in that a coding scheme for the wiretap channel can be used to design a coding scheme for secret-key agreement and vice-versa.

The early information-theoretic security results obtained for the wiretap channel and source model for secret-key agreement are criticized in some circles for measuring statistical dependence in terms of the average information rate leaked to the eavesdropper $\frac{1}{n}\avgI{\rvM;{\mathbf{Z}}}$. The weakness of this metric from a cryptographic standpoint has been highlighted in multiple works~\cite{Maurer1994,Maurer2000,Bloch2011}, which have instead advocated using the average information leaked $\avgI{\rvM;{\mathbf{Z}}}$. The analysis of secure communication architectures under this more stringent secrecy metric has been performed with different methods, such as graph-coloring techniques~\cite{Csiszar1996}, privacy amplification~\cite{Maurer2000,Bennett1995}, and channel resolvability~\cite{Hayashi2006,Devetak2005}. The results presented in this paper further clarify the relation between secrecy and channel resolvability and highlight the potential of channel resolvability for solving secure communication problems.

The connection between secrecy and channel resolvability is better illustrated by studying secure communication architectures beyond the traditional memoryless setting; in particular, the distinction between the coding mechanisms for reliability and secrecy becomes apparent in the expressions of the results themselves. In this context, the information-spectrum methods pioneered by Han and Verd\'u turn out to be convenient mathematical tools, as they allow us to analyze general channels by focusing on the properties of mutual information as a random variable. We note that these tools have already been used to study some information-theoretic security problems and our results provide extensions of~\cite{Koga2000,Koga2005,Kobayashi2005,Hayashi2006}.

\subsection{Summary of Results}
\label{sec:summary-results}

In this section, we highlight the results presented in this paper, preliminary versions of which have been reported in~\cite{Bloch2008e,Bloch2011a}.
\begin{itemize}
\item We clarify the relation between information-theoretic security and statistical independence by investigating alternatives to the average mutual information rate $\frac{1}{n}\avgI{\rvM;{\mathbf{Z}}}$, which is used as the \emph{de facto} metric in most earlier works. The average mutual information rate is actually a normalized Kullback-Leibler divergence between the joint distribution $p_{\rvM{\mathbf{Z}}}$ and the product distribution $p_\rvM p_{{\mathbf{Z}}}$; the closeness of these two distributions can be measured by other means, such as the variational distance or even the cumulative distribution function (CDF) of the random variable $\I{\rvM;{\mathbf{Z}}}$. By establishing relations among different metrics (Proposition~\ref{lm:ordering_metrics}) we highlight the importance of choosing a measure of statistical dependence that is not only simple enough to be analytically tractable but also strong enough to be cryptographically relevant. This discussion also provides the basis for elegant converse proofs.
\item We provide evidence that channel resolvability is a convenient mechanism for secure communication by formalizing the ideas introduced in Example~\ref{ex:onte-time-pad} and Example~\ref{ex:noisy-Gaussian}. Specifically, we connect secrecy to channel resolvability to analyze the fundamental limits of Shannon's cipher system (Theorem~\ref{th:key_resolvability}) and of the broadcast channel with confidential messages (Theorem~\ref{th:bcc}). In the latter case, we show that at least for some specific wiretap channels, deriving secrecy from channel resolvability is more powerful than deriving secrecy from channel capacity (Proposition~\ref{prop:capac-achi-wiret}); we also derive the secrecy-capacity region for general broadcast channels with a cost constraint and for strong secrecy metrics (Theorem~\ref{th:bcc} and Theorem~\ref{th:dmc});
\item We further leverage the connection between secrecy and channel resolvability to revisit various models of secure communication. We first provide a simple proof of the strong secrecy capacity of ergodic-fading wireless channels with full channel state information (Proposition~\ref{prop:stron_secrecy_wireless}). We then show that known achievable rates for mixed channels and compound channels with receiver \ac{CSI} can be obtained with conceptually simple proofs, and that these results hold under stronger secrecy metrics than was previously established (Proposition~\ref{prop:mixed_wiretap} and Proposition~\ref{prop:compound_general}).
\item We finally exploit the general characterization of secrecy capacity to bound the secret-key capacity of a general discrete source model for secret-key agreement (Proposition~\ref{prop:bound_source_model}). The form of the result, which involves conditional entropy instead of mutual information, suggests that the mechanism behind secret-key agreement is not channel resolvability but rather channel intrinsic randomness~\cite{Csiszar1996,Bloch2010b}. 
\end{itemize}

\subsection{Outline}
\label{sec:outline}

The remainder of the paper is organized as follows. Section~\ref{sec:notation} sets the notation used throughout the paper. Section~\ref{sec:secrecy-metrics} introduces and compares several secrecy metrics that can be used to measure information-theoretic security. Section~\ref{sec:shann-ciph-syst} analyses the fundamental limits of secure communication for Shannon's cipher system. Section~\ref{sec:arbitr-wiret-chann}, which forms the core of the paper, proves the impossibility of achieving strong secrecy capacity with random codes deriving secrecy from channel capacity for some wiretap channels and establishes the secrecy-capacity region of general broadcast channels with confidential messages. Section~\ref{sec:applications} presents applications of the general results to wireless channels, mixed channels and compound channels, and secret-key agreement, which may be of independent interest. Section~\ref{sec:conclusion} offers some concluding remarks. The technical details of the proofs are organized into a series of lemmas, whose proofs are relegated to the appendices to streamline the presentation.

\section{Notation}
\label{sec:notation}

To fix notation for the sequel, consider three random variables $\rvX$, $\rvY$, and $\rvZ$ with sample values $\svx$, $\svy$, and $\svz$ taking values in alphabets $\calX$, $\calY$, and $\calZ$, respectively. The joint probability distribution is denoted $p_{\rvX\rvY\rvZ}$, and the marginal probability distributions are denoted by $p_\rvX$, $p_\rvY$, and $p_\rvZ$. Unless mentioned otherwise, alphabets are assumed to be abstract alphabets, including countably infinite or continuous alphabets. If the alphabets are finite, then the probability distributions correspond to probability mass functions; if the alphabets are uncountable, then the probability distributions correspond to probability densities, which we assume exist.\footnote{We note that more general situations can be treated with the approach of Pinsker~\cite{InformationStabilityRandomVariables}.} The \emph{mutual information} between $\rvX$ and $\rvY$ is the random variable\footnote{Unless indicated otherwise, logarithms and exponentials in the paper are taken to base two.}
\begin{align*}
  \I{\rvX;\rvY}\eqdef \log\frac{\p[\rvX\rvY]{\rvX,\rvY}}{\p[\rvX]{\rvX}\p[\rvY]{\rvY}}.
\end{align*}
The average of this random variable is the usual \emph{average mutual information}, which we denote by
$\avgI{\rvX;\rvY}$. For discrete random variables, $\avgI{\rvX;\rvY}$ has the familiar expression
\begin{align*}
  \avgI{\rvX;\rvY}&\eqdef \E[\rvX\rvY]{\I{\rvX;\rvY}}\\
  &= \sum_{\svx\in\calX}\sum_{\svy\in\calY}\p[\rvX\rvY]{\svx,\svy}
  \log\frac{\p[\rvX\rvY]{\svx,\svy}}{\p[\rvX]{\svx}\p[\rvY]{\svy}}.
\end{align*}
The conditional mutual information between $\rvX$ and $\rvY$ given $\rvZ$ and the average conditional mutual information
are accordingly defined as
\begin{align*}
  \I{\rvX;\rvY|\rvZ}&\eqdef \log\frac{\p[\rvX\rvY|\rvZ]{\rvx,\rvy|\rvz}}{\p[\rvX|\rvZ]{\rvx|\rvz}\p[\rvY|\rvZ]{\rvy|\rvz}}\\
  \text{and}\quad\avgI{\rvX;\rvY|\rvZ} &\eqdef  \E[\rvX\rvY\rvZ]{\I{\rvX;\rvY|\rvZ}},
\end{align*}
respectively. Similarly, the \emph{entropy} and \emph{average entropy} of $\rvX$ are
\begin{align*}
  \H{\rvX}&\eqdef \log\frac{1}{\p[\rvX]{\rvX}}\quad\mbox{and}\quad \avgH{\rvX}\eqdef \E[\rvX]{\H{\rvX}},
\end{align*}
and the conditional entropy and average conditional entropy of $\rvX$ given $\rvY$ are
\begin{align*}
  \H{\rvX|\rvY}&\eqdef \log\frac{1}{\p[\rvX|\rvY]{\rvX|\rvY}}\\
  \mbox{and}\quad
  \avgH{\rvX|\rvY}&\eqdef \E[\rvX\rvY]{\H{\rvX|\rvY}}.
\end{align*}
All the usual relations between average mutual information and average entropy that result from
basic properties of joint, marginal, or conditional probability distributions can be shown to hold with probability one for the
mutual information and entropy random variables. In particular, the chain rules of mutual information and entropy hold
with probability one.

The average mutual information $\avgI{\rvX;\rvX'}$ between two random variables $\rvX\in\calX$ and $\rvX'\in\calX$ is a Kullback-Leibler divergence, which measures the closeness of the distributions $p_\rvX p_{\rvX'}$ and $p_{\rvX\rvX'}$. We will often use an alternative measure in terms of the \emph{variational distance} between the distributions, defined as\footnote{This general definition of variational
  distance reduces to $\sum_{\svx\in\calX}\card{\p[\rvX]{\svx}-\p[\rvX']{\svx}}$ if $\calX$ is countable.}
\begin{align*}
  \V{p_{\rvX},p_{\rvX'}}\eqdef  2\sup_{\calA\subseteq\calX}\abs{\P[\rvX]{\calA}-\P[\rvX']{\calA}}.
\end{align*}
The variational distance is not as convenient to manipulate as the average mutual information, but we provide simple
rules for variational distance calculus in Appendix~\ref{sec:technical-lemmas}. 

Given two real numbers $a,b$ we define $\intseq{a}{b}$ as the set of integers $\{n\in\mathbb{N}:\lfloor a\rfloor\leq n\leq \lceil b\rceil\}$. To simplify notation, all vectors of length $n$ are denoted by boldface letters; for instance, $\mathbf{x}$ denotes the vector of sample values $(x_1,\dots,x_n)$ while $\mathbf{X}$ denotes the random vector $(X_1,\dots,X_n)$. Given two random vectors $\mathbf{X}$ and $\mathbf{Y}$, characterized by a joint probability distribution $p_{{\mathbf{X}}{\mathbf{Y}}}$, the probability distribution of $\frac{1}{n}\I{{\mathbf{X}};{\mathbf{Y}}}$ is referred to as the \emph{mutual information rate spectrum}. In addition, the \emph{spectral-inf mutual information rate} is defined as~\cite{InformationSpectrumMethods}
\begin{multline*}
  \pliminf[n\rightarrow\infty]{\frac{1}{n}\I{{\mathbf{X}};{\mathbf{Y}}}}
\eqdef\\ \sup\left\{\beta:\lim_{n\rightarrow\infty}\P{\frac{1}{n}\I{{\mathbf{X}};{\mathbf{Y}}}<\beta}=0\right\},
\end{multline*}
and the \emph{spectral-sup mutual information rate} is defined as
\begin{multline*}
\plimsup[n\rightarrow\infty]{\frac{1}{n}\I{{\mathbf{X}};{\mathbf{Y}}}}
\eqdef\\
\inf\left\{\alpha:\lim_{n\rightarrow\infty}\P{\frac{1}{n}\I{{\mathbf{X}};{\mathbf{Y}}}>\alpha}=0\right\}.
\end{multline*}
Operationally, the spectral-inf mutual information rate relates to channel capacity~\cite{Verdu1994} whereas the spectral-sup mutual information rate relates to the channel resolvability~\cite{Han1993}. Similarly, given an arbitrary sequence $\mathbf{X}$, the \emph{entropy rate spectrum} is the distribution of the
random variable $\frac{1}{n}\H{{\mathbf{X}}}$, and the spectral-inf entropy rate is defined as
\begin{align*}
  \pliminf{\frac{1}{n}\H{{\mathbf{X}}}}\eqdef \sup\left\{\beta:\lim_{n\rightarrow\infty}\P{\frac{1}{n}\H{{\mathbf{X}}}<\beta}=0\right\},
\end{align*}
while the spectral-sup entropy rate is
\begin{align*}
  \plimsup{\frac{1}{n}\H{{\mathbf{X}}}} \eqdef \inf\left\{\alpha:\lim_{n\rightarrow\infty}\P{\frac{1}{n}\H{{\mathbf{X}}}>\alpha}=0\right\}.
\end{align*}

The spectral-sup and spectral-inf mutual information and entropy rates play a fundamental role in the analysis of reliable communication and randomness generation~\cite{Verdu1994,Han1993,Vembu1995}. They also play a role in the analysis of secure communications, and our results combine these quantities in various ways.

\section{Preliminaries: Secrecy Metrics}
\label{sec:secrecy-metrics}

Let $n\in\mathbb{N}^*$ and $R>0$. Let $\rvM\in\intseq{1}{2^{nR}}$ be a random variable that represents a message in a communication scheme. Assume that an eavesdropper has some knowledge about $\rvM$ represented by another random variable ${\mathbf{Z}}\in\calZ^n$, characterized by the joint probability distribution $p_{\rvM{\mathbf{Z}}}$. As mentioned in the introduction, message $\rvM$ is information-theoretically secure if it is statistically independent of ${\mathbf{Z}}$; however, exact statistical independence between $\rvM$ and ${\mathbf{Z}}$ is extremely stringent and, for tractability, it is convenient to use a slightly weaker measure of secrecy, by which we only require $\rvM$ and ${\mathbf{Z}}$ to be \emph{asymptotically} independent as the parameter $n$ tends to infinity. Note that there is some leeway in the definition of asymptotic independence because one can choose how to measure the dependence between $\rvM$ and ${\mathbf{Z}}$. For instance, given any distance $d$ for the space of joint probability distributions on $\intseq{1}{2^{nR}}\times\calZ^n$, the quantity $d(p_{\rvM{\mathbf{Z}}};p_{\rvM}p_{{\mathbf{Z}}})$ could be used as a metric, and asymptotic statistical independence then amounts to the condition
\begin{align*}
  \lim_{n\rightarrow\infty}d(p_{\rvM{\mathbf{Z}}};p_{\rvM}p_{{\mathbf{Z}}}) = 0.
\end{align*}
In the following, we specify six reasonable choices for secrecy metrics. The first metric measures statistical dependence using the Kullback-Leibler divergence:
\begin{align*}
\metric{1}\left(p_{\rvM{\mathbf{Z}}},p_{\rvM}p_{{\mathbf{Z}}}\right)\eqdef \avgD{p_{\rvM{\mathbf{Z}}}}{p_{\rvM}p_{{\mathbf{Z}}}}=\avgI{\rvM;{\mathbf{Z}}}.
\end{align*}
The secrecy condition $\lim_{n\rightarrow\infty}\metric{1}\left(p_{\rvM{\mathbf{Z}}},p_{\rvM}p_{{\mathbf{Z}}}\right)=0$ corresponds to the well-known \emph{strong secrecy}~\cite{Maurer1994}. A second metric that we will find particularly useful is based on the variational distance: 
\begin{align*}
    \metric{2}\left(p_{\rvM{\mathbf{Z}}},p_{\rvM}p_{{\mathbf{Z}}}\right)\eqdef \V{p_{\rvM{\mathbf{Z}}},p_{\rvM}p_{{\mathbf{Z}}}}.
\end{align*}
For any $\epsilon>0$, the asymptotic independence of $\rvM$ and ${\mathbf{Z}}$ can also be measured in terms of the CDF of $\rvI(\rvM;{\mathbf{Z}})$:
\begin{align*}
      \metric{3}\left(p_{\rvM{\mathbf{Z}}},p_{\rvM}p_{{\mathbf{Z}}}\right)\eqdef \P{\I{{\rvM;{\mathbf{Z}}}}>\epsilon},
\end{align*}
in which case the secrecy condition
\begin{align*}
  \forall
  \epsilon>0\;\lim_{n\rightarrow\infty}\metric{3}\left(p_{\rvM{\mathbf{Z}}},p_{\rvM}p_{{\mathbf{Z}}}\right)=0
\end{align*}
means that the random variable $\I{{\rvM;{\mathbf{Z}}}}$ converges in probability to zero. Finally, one could also weaken the metrics above by introducing a normalization by a factor of $n$ as
\begin{align*}
  \metric{4}\left(p_{\rvM{\mathbf{Z}}},p_{\rvM}p_{{\mathbf{Z}}}\right)&\eqdef \tfrac{1}{n}\avgD{p_{\rvM{\mathbf{Z}}}}{p_{\rvM}p_{{\mathbf{Z}}}}\\
  \metric{5}\left(p_{\rvM{\mathbf{Z}}},p_{\rvM}p_{{\mathbf{Z}}}\right)&\eqdef \tfrac{1}{n}\V{p_{\rvM{\mathbf{Z}}},p_{\rvM}p_{{\mathbf{Z}}}},\\
  \text{for $\epsilon>0$}\quad\metric{6}\left(p_{\rvM{\mathbf{Z}}},p_{\rvM}p_{{\mathbf{Z}}}\right)&\eqdef \P{\tfrac{1}{n}\I{{\rvM;{\mathbf{Z}}}}>\epsilon}.
\end{align*}
The secrecy condition $\lim_{n\rightarrow\infty}\metric{4}\left(p_{\rvM{\mathbf{Z}}},p_{\rvM}p_{{\mathbf{Z}}}\right)=0$ corresponds to the \emph{weak secrecy} initially introduced by Wyner~\cite{Wyner1975}. 

The secrecy conditions\footnote{The limit should be understood for any $\epsilon>0$ in the case of metrics $\metric{3}$ and $\metric{6}$.} $\lim_{n\rightarrow\infty}\metric{i}(p_{\rvM{\mathbf{Z}}},p_{\rvM}p_{{\mathbf{Z}}})=0$ may not be equivalent for all $i\in\intseq{1}{6}$; by establishing an ordering among these metrics, we formalize what it means for a metric to be ``stronger'' than another. For $i,j\in\intseq{1}{6}$, we say that $\metric{i}$ is \emph{stronger} than $\metric{j}$  (or equivalently that $\metric{j}$ is \emph{weaker} than $\metric{i}$), and we write $\metric{i}\succeq \metric{j}$ if and only if
\begin{align*}
  \lim_{n\rightarrow\infty}\metric{i}(p_{\rvM{\mathbf{Z}}},p_{\rvM}p_{{\mathbf{Z}}})=0 \Rightarrow   \lim_{n\rightarrow\infty}\metric{j}(p_{\rvM{\mathbf{Z}}},p_{\rvM}p_{{\mathbf{Z}}})=0.
\end{align*}
By construction, it is clear that $\metric{1}\succeq \metric{4}$, $\metric{2}\succeq \metric{5}$ and $\metric{3}\succeq \metric{6}$; however, we establish a more precise result.
\begin{proposition}
\label{lm:ordering_metrics}
The secrecy metrics $\metric{i}$ for $i\in\intseq{1}{6}$ are ordered as follows.
  \begin{align*}
    \metric{1}\succeq \metric{2}\succeq \metric{3}\succeq \metric{4}\succeq \metric{5}\succeq \metric{6}.
  \end{align*}
\end{proposition}
\begin{IEEEproof}
 The relations $\metric{1}\succeq\metric{2}$ and $\metric{4}\succeq\metric{5}$ directly follow from Pinsker's inequality~\cite[Corollary p.16]{InformationStabilityRandomVariables}. Similarly, the relations $\metric{2}\succeq\metric{3}$ and $\metric{5}\succeq\metric{6}$ follow from~\cite[Corollary p.18]{InformationStabilityRandomVariables}; hence, we only need to prove that~$\metric{3}\succeq\metric{4}$. 

Let $\epsilon,\gamma>0$. Assume that $\lim_{n\rightarrow\infty}\metric{3}(p_{\rvM{\mathbf{Z}}},p_{\rvM}p_{{\mathbf{Z}}})=0$, so that $\lim_{n\rightarrow\infty}\P{\I{\rvM;{\mathbf{Z}}}>\epsilon}=0$. Note that metric $\metric{4}(p_{\rvM{\mathbf{Z}}},p_{\rvM}p_{{\mathbf{Z}}})$ can be written as
\begin{multline*}
\metric{4}(p_{\rvM{\mathbf{Z}}},p_{\rvM}p_{{\mathbf{Z}}})\\
\begin{split}
  &= \tfrac{1}{n}\avgI{\rvM;{\mathbf{Z}}}\\
&=\E{\tfrac{1}{n}\I{\rvM;{\mathbf{Z}}}},\\
&=\E{\tfrac{1}{n}\I{\rvM;{\mathbf{Z}}}\mathds{1}\left\{\I{\rvM;{\mathbf{Z}}}\leq-\epsilon\right\}}\\
&\phantom{--}+\E{\tfrac{1}{n}\I{\rvM;{\mathbf{Z}}}\mathds{1}\left\{-\epsilon<\I{\rvM;{\mathbf{Z}}}\leq\epsilon\right\}}\\
&\phantom{--}+ \E{\tfrac{1}{n}\I{\rvM;{\mathbf{Z}}}\mathds{1}\left\{\epsilon<\I{\rvM;{\mathbf{Z}}}\leq n(R+\gamma)\right\}}\\
&\phantom{--}+
\E{\tfrac{1}{n}\I{\rvM;{\mathbf{Z}}}\mathds{1}\left\{\I{\rvM;{\mathbf{Z}}}>n(R+\gamma)\right\}}.
\end{split}  
\end{multline*}
Clearly, we have $  \E{\tfrac{1}{n}\I{\rvM;{\mathbf{Z}}}\mathds{1}\left\{ \I{\rvM;{\mathbf{Z}}} \leq -\epsilon\right\}}<0$, $\E{\tfrac{1}{n}\I{\rvM;{\mathbf{Z}}}\mathds{1}\left\{-\epsilon<\I{\rvM;{\mathbf{Z}}}\leq\epsilon\right\}}  \leq \frac{\epsilon}{n}$, and
\begin{multline*}
\E{\tfrac{1}{n}\I{\rvM;{\mathbf{Z}}}\mathds{1}\left\{\epsilon<\I{\rvM;{\mathbf{Z}}}\leq n(R+\gamma)\right\}} \\
\leq (R+\gamma)\, \P{\I{\rvM;{\mathbf{Z}}}>\epsilon}.
\end{multline*}
Following~\cite[p. 223]{InformationSpectrumMethods}, we can also prove that 
\begin{align*}
  \lim_{n\rightarrow\infty}\E{\tfrac{1}{n}\I{\rvM;{\mathbf{Z}}}\mathds{1}\left\{\I{\rvM;{\mathbf{Z}}}>n(R+\gamma) \right\}}&=0.
\end{align*}
Therefore, $\lim_{n\rightarrow\infty}\metric{4}(p_{\rvM{\mathbf{Z}}},p_{\rvM}p_{{\mathbf{Z}}})=0$ and $\metric{3}\succeq\metric{4}$.
\end{IEEEproof}
A direct consequence of Proposition~\ref{lm:ordering_metrics} is that any secure communication scheme satisfying the secrecy condition with the strongest secrecy metric~$\metric{1}$ automatically satisfies it with the secrecy metrics~$\metric{i}$ for $i\in\intseq{2}{6}$. Conversely, any secure communication scheme that does not satisfy the secrecy condition with the weakest metric~$\metric{6}$ cannot satisfy it with any of the metrics~$\metric{i}$ for $i\in\intseq{1}{5}$. Therefore, to establish a coding theorem for a secure communication scheme, we can prove achievability with metric~$\metric{1}$ and a converse with metric~$\metric{6}$.

Although the ordering in Proposition~\ref{lm:ordering_metrics} follows strictly from mathematical properties, the idea that some metrics are stronger than others is also meaningful from a cryptographic perspective. One can construct examples of communication schemes that present obvious security loopholes while still satisfying a secrecy condition with metric~$\metric{4}$ (see, for instance, the examples in~\cite{Bloch2011,Barros2008,Hayashi2011}). It is now accepted that information-theoretic secrecy conditions\footnote{The conditions could be further strengthened by imposing an exponential convergence with $n$; however, except in the case of exponentially information stable channels~\cite{Csiszar1996}, such as memoryless channels, we were unable to prove general results with this additional constraint.} should hold at least with metrics~$\metric{1}$ or~$\metric{2}$.

\section{Secrecy from Channel Resolvability for Shannon's Cipher System}
\label{sec:shann-ciph-syst}

\begin{figure}[t]
  \centering
  \includegraphics[width=.98\linewidth]{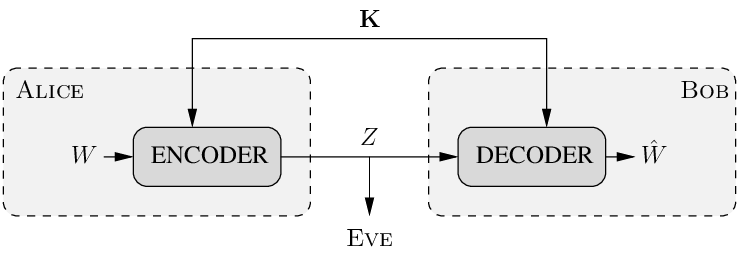}
  \caption{Shannon's cipher system for a general common source of randomness.}
  \label{fig:shannon_cipher}
\end{figure}

As a first illustration of the connection between secrecy and channel resolvability, we elaborate on Example~\ref{ex:onte-time-pad} and revisit Shannon's cipher system. We consider the model illustrated in Figure~\ref{fig:shannon_cipher}, in which a message $\rvM$ uniformly distributed in $\intseq{1}{2^{nR}}$ is to be communicated reliably from a transmitter (Alice) to a legitimate receiver (Bob) in the presence of an eavesdropper (Eve). Alice and Bob have access to a common discrete source of randomness $(\calK,\{p_{{\mathbf{K}}}\}_{n\geq 1})$, characterized by an alphabet $\calK$  and a sequence of symbol probabilities $\{p_{{\mathbf{K}}}\}_{n\geq 1}$, which is used to encode $\rvM$ into a codeword $\rvZ\in\calZ$. Bob's estimate of the message using $\rvZ$ and the source of randomness $\mathbf{K}$ is denoted by $\hat\rvM$. 

\begin{definition}
  A $(2^{nR},n)$ cipher $\calE_n$ consists of
  \begin{itemize}
  \item an encoding function $f_n:\intseq{1}{2^{nR}}\times\calK^n\rightarrow \calZ$;
  \item a decoding function $g_n:\calZ\times\calK^n\rightarrow\intseq{1}{2^{nR}}$.
  \end{itemize}
\end{definition}

The reliability performance of a cipher $\calE_n$ is measured in terms of the probability of error $\perr(\calE_n)\eqdef \P{\hat\rvM\neq\rvM|\calE_n}$ while its secrecy performance is measured in terms of the secrecy metric\footnote{We will drop the conditioning on $\calE_n$ when this is clear from the context.} $\metric{i}(\calE_n)\eqdef \metric{i}(p_{\rvM\rvZ|\calE_n},p_{\rvM}p_{\rvZ|\calE_n})$. 

\begin{definition}
  A rate $R$ is achievable for secrecy metric $\metric{i}$ for Shannon's cipher system if there exists a sequence of $\left(2^{nR},n\right)$ ciphers $\{\calE_n\}_{n\geq 1}$ such that
  \begin{align*}
    \lim_{n\rightarrow\infty}\perr(\calE_n)=0\quad\text{and}\quad\lim_{n\rightarrow\infty}\metric{i}(\calE_n)=0.
  \end{align*}
  The {secrecy capacity} $C_{\textnormal{\tiny SC}}^{(i)}$ of Shannon's cipher system is the supremum of achievable rates for secrecy metric $\metric{i}$.
\end{definition}

\begin{theorem}
\label{th:key_resolvability}
The secrecy capacity of Shannon's cipher system is the same for all metrics $\metric{i}$ with $i\in\intseq{2}{6}$ and is given by
  \begin{align}
    C_{\textnormal{\tiny SC}} = \pliminf{\frac{1}{n}\H{{\mathbf{K}}}}.
  \end{align}
If the source $(\calK,\{p_{{\mathbf{K}}}\}_{n\geq 1})$ is memoryless, then the secrecy capacity is also the same for metric $\metric{1}$.
\end{theorem}
\begin{IEEEproof}
  We first show that all rates below $\pliminf{\frac{1}{n}\H{{\mathbf{K}}}}$ are achievable for secrecy metric~$\metric{2}$. Let $\epsilon,\gamma>0$ and $R\eqdef\pliminf{\frac{1}{n}\H{{\mathbf{K}}}}-\gamma$. Let $\rvU_R$ be the random variable with uniform distribution on $\intseq{1}{2^{nR}}$. By~\cite[Lemma 3]{Vembu1995}, there exists an encoding function $f_n:\calK^n\rightarrow\intseq{1}{2^{nR}}$ such that $\V{p_{f_n({\mathbf{K}})},p_{\rvU_R}}\leq \epsilon_n$ with $\lim_{n\rightarrow\infty}\epsilon_n=0$.  A message $\rvM$ is then encoded as $\rvZ=f_n({\mathbf{K}})\oplus\rvM$, where $\oplus$ represents the addition modulo $\lceil 2^{nR}\rceil$. By construction, Bob retrieves $\rvM$ without error since $\rvM=\rvZ\oplus f_n({\mathbf{K}})$. We have
  \begin{align*}
    \metric{2}(\calE_n) &=  \V{p_{\rvM \rvZ},p_{\rvM}p_{\rvZ}}\\
    &= \E[\rvM]{\V{p_{\rvZ|\rvM},p_{\rvZ}}}\\    
    &\leq \E[\rvM]{\V{p_{\rvZ|\rvM},p_{\rvU_R}}} + \V{p_{\rvU_R},p_{\rvZ}}\\
    & \leq 2 \E[\rvM]{\V{p_{\rvZ|\rvM},p_{\rvU_R}}}\\
    & = 2 \E[\rvM]{\V{p_{f_n({\mathbf{K}})},p_{\rvU_R}}}\\
    &\leq 2\epsilon_n,
  \end{align*}
   where we have used Lemma~\ref{lm:subadditivity_basics}, the definition of $Z$, and the independence of $f_n(\mathbf{K})$ and $\rvM$. Therefore, the rate $R$ is achievable and, since $\gamma$ can be chosen arbitrarily small, we conclude that
\begin{align}
  \label{eq:lower_bound_scs}
  C_{\textnormal{\tiny SC}}^{(2)}\geq \pliminf{\frac{1}{n}\H{{\mathbf{K}}}}.
\end{align}
If the source $(\calK,p_{\rvK})$ is i.i.d., one can modify the proof of~\cite[Lemma 3]{Vembu1995} to show that, if $R\eqdef\avgH{K}-\gamma$, there exists a function $f_n:\calK^n\rightarrow\intseq{1}{2^{nR}}$ and $\alpha_\gamma>0$, such that $\V{p_{f_n({\mathbf{K}})},p_{\rvU_R}}\leq 2^{-\alpha_\gamma n}$. Following the same steps as above, we then obtain that $\metric{2}(\calE_n)\leq 2\cdot 2^{-\alpha_\gamma n}$. Finally,~\cite[Lemma 1]{Csiszar1996} shows that there exists $\beta_\gamma>0$ such that, for $n$ large enough $\metric{1}(\calE_n)\leq 2^{-\beta_\gamma n}$.

We now prove the converse part of the result. Let $R$ be an achievable rate for secrecy metric $\metric{6}$. There exists a sequence of $(2^{nR},n)$ ciphers $\{\calE_n\}_{n\geq 1}$ such that $\lim_{n\rightarrow\infty}\perr(\calE_n)=0$ and $\lim_{n\rightarrow\infty}\metric{6}(\calE_n)=0$. For every $n\in\mathbb{N}^*$, and with probability one, we have
\begin{align*}
\tfrac{1}{n}\H{\rvM} &=\tfrac{1}{n}\H{\rvM|\rvZ} + \tfrac{1}{n}\I{\rvM;\rvZ}\\
  &=\tfrac{1}{n}\I{\rvM;{\mathbf{K}}|\rvZ} +\tfrac{1}{n}\H{\rvM|\rvZ{\mathbf{K}}}+ \tfrac{1}{n}\I{\rvM;\rvZ}\\
  &=\tfrac{1}{n}\H{{\mathbf{K}}}-\tfrac{1}{n}\H{{\mathbf{K}}|\rvM\rvZ}-\tfrac{1}{n}\I{{\mathbf{K}};\rvZ}\\
  &\phantom{--------}+\tfrac{1}{n}\H{\rvM|\rvZ{\mathbf{K}}}+ \tfrac{1}{n}\I{\rvM;\rvZ}.
\end{align*}
Since $R=\pliminf{\frac{1}{n}\H{\rvM}}$, $\pliminf{\frac{1}{n}\H{{\mathbf{K}}|\rvM\rvZ}}\geq 0$, and $\pliminf{\frac{1}{n}\I{{\mathbf{K}};\rvZ}}\geq 0$, we obtain
\begin{multline*}
  R \leq \pliminf{\frac{1}{n}\H{{\mathbf{K}}}} + \plimsup{\frac{1}{n}\H{\rvM|\rvZ{\mathbf{\rvK}}}}\\
  + \plimsup{\frac{1}{n}\I{\rvM;\rvZ}}.
\end{multline*}
Note that $\plimsup{\frac{1}{n}\I{\rvM;\rvZ}}=0$ by assumption. The Verd\'u\--Han Lemma~\cite{InformationSpectrumMethods,Verdu1994} also guarantees that $\plimsup{\frac{1}{n}\H{\rvM|\rvZ{\mathbf{K}}}}=0$; hence, we have
\begin{align}
  \label{eq:upper_bound_scs}
  C_{\textnormal{\tiny SC}}^{(6)}\leq \pliminf{\frac{1}{n}\H{{\mathbf{K}}}}.
\end{align}
Combining~\eqref{eq:lower_bound_scs} and~\eqref{eq:upper_bound_scs} with Proposition~\ref{lm:ordering_metrics}, we conclude that, for every $i\in\intseq{2}{6}$, $C_{\textnormal{\tiny SC}}^{(i)}=\pliminf{\frac{1}{n}\H{{\mathbf{K}}}}$. If the source is memoryless, then for every $i\in\intseq{1}{6}$, $C_{\textnormal{\tiny SC}}^{(i)}=\avgH{\rvK}$.
\end{IEEEproof}

The coding scheme used in Theorem~\ref{th:key_resolvability} extracts the \emph{source intrinsic randomness} of $(\calK,\{p_{{\mathbf{K}}}\}_{n\geq 1})$ to protect the message with a one-time pad. Nevertheless, the message is kept secret from the eavesdropper because the encoder exploits the randomness of the source to control the distribution of the eavesdropper's observation; hence, the coding mechanism for secure communication can be interpreted as channel resolvability, which we confirm in the next section. From a cryptographic perspective, Theorem~\ref{th:key_resolvability} shows that the secure communication rate is maximized if the legitimate terminals make sure that their keys are almost perfectly uniform. This has operational significance in a practical situation if the mechanism providing secret keys is biased and does not yield perfectly uniform keys. Finally, the fact that $C_{\textnormal{\tiny SC}}^{(i)}$ remains identical for all metrics~$\metric{i}$ with $i\in\intseq{2}{6}$ suggests that asymptotic statistical independence is indeed a fundamental measure of secrecy.

\section{Secrecy from Channel Resolvability over Noisy Channels}
\label{sec:arbitr-wiret-chann}

\begin{figure}[t]
  \centering
  \includegraphics[width=.98\linewidth]{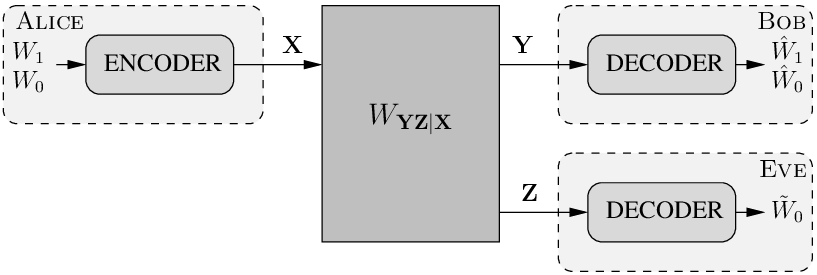}
  \caption{Broadcast channel with confidential messages.}
  \label{fig:bcc}
\end{figure}

We now turn our attention to the problem of secure communication over noisy channels. We consider a broadcast channel with confidential messages $\left(\calX,\calY,\{W_{{\mathbf{Y}}{\mathbf{Z}}|{\mathbf{X}}}\}_{n\geq 1},\calZ\right)$ characterized by an input alphabet $\calX$, two output alphabets $\calY$ and $\calZ$, and a sequence of transition probabilities $\{W_{{\mathbf{Y}}{\mathbf{Z}}|{\mathbf{X}}}\}_{n\geq 1}$. The channels $\left(\calX,\{W_{{\mathbf{Y}}|{\mathbf{X}}}\}_{n\geq 1},\calY\right)$ and $\left(\calX,\{W_{{\mathbf{Z}}|{\mathbf{X}}}\}_{n\geq 1},\calZ\right)$ obtained from the marginals are called the \emph{main channel} and the \emph{eavesdropper's channel}, respectively. The inputs to the channels are also subject to cost constraint $P\in\mathbb{R}^+$; specifically, there exists a sequence of cost functions $\{c_n\}_{n\geq 1}$ with $c_n:\calX^n\rightarrow\mathbb{R}_+$, such that any sequence $\mathbf{x}\in\calX^n$ transmitted through the channel should satisfy $\frac{1}{n}c_n(\mathbf{x})\leq P$. Following standard practice, the transmitter is named Alice, the receiver observing output $\mathbf{Y}$ is named Bob, and the receiver observing output $\mathbf{Z}$ is named Eve. As illustrated in Figure~\ref{fig:bcc}, Alice wishes to transmit a common message $\rvM_0$ to both Bob and Eve and an individual message $\rvM_1$ for Bob alone, viewing Eve as an eavesdropper for message $\rvM_1$. Bob's estimates of the messages are denoted by $\hat\rvM_0$ and $\hat\rvM_1$ while Eve's estimate is denoted by $\tilde\rvM_0$. 

\begin{definition}
  A $\left(2^{nR_0},2^{nR_1},n\right)$ wiretap code $\calC_n$ consists of
  \begin{itemize}
  \item a common message set $\calW_0=\intseq{1}{2^{nR_0}}$;
  \item an individual message set $\calW_1=\intseq{1}{2^{nR_1}}$;
  \item an auxiliary message set $\calW'=\intseq{1}{2^{nR'_n}}$, with $R'_n>0$,\footnote{Although $R_0$ and $R_1$ are fixed parameters, we allow $R'_n$ to vary with $n$.} used to randomize the encoding of messages;
  \item a source of local randomness $\left(\calR,p_\rvR\right)$, which is only known to Alice and can be used to further randomize the encoding;
  \item an encoding function $f_n:\calW_0\times\calW_1\times\calW'\times\calR\rightarrow\calX^n$, such that 
$\frac{1}{n}c_n(f_n(m_0,m_1,m',r))\leq P$;
  \item a decoding function $g_n:\calY^n\rightarrow\calW_0\times\calW_1\times\calW'$;
  \item a decoding function $h_n:\calZ^n\rightarrow\calW_0$.
  \end{itemize}
\end{definition}
The auxiliary message is denoted by $\rvM'$. All messages $\rvM_0,\rvM_1,\rvM'$ are assumed to be uniformly distributed in their respective sets. The size of the auxiliary message set and the source of local randomness $\left(\calR,p_\rvR\right)$ can be optimized as part of the code design, and the eavesdropper is assumed to know the code $\calC_n$, which includes the statistics $p_\rvR$ of the source of local randomness. In the remainder of the paper, we clearly identify the channel inputs and outputs obtained when using a code $\calC_n$ by introducing a bar in the notation of the corresponding random variables. For instance, the random variable representing a codeword chosen in $\calC_n$ is denoted $\bar{\mathbf{X}}$, those representing the corresponding channel outputs are denoted $\bar{\mathbf{Y}}$ and $\bar{\mathbf{Z}}$. The joint distribution between $\rvM_0,\rvM_1,\bar{\mathbf{X}},\bar{\mathbf{Y}},\bar{\mathbf{Z}}$ is 
\begin{multline}
  \p[\rvM_0\rvM_1\bar{\mathbf{X}} \bar{\mathbf{Y}}\bar{\mathbf{Z}}]{m_0,m_1,{\mathbf{x}},{\mathbf{y}},{\mathbf{z}}}
  =
 W_{{\mathbf{Y}}{\mathbf{Z}}|{\mathbf{X}}}({\mathbf{y}},{\mathbf{z}}|{\mathbf{x}})\\
 \p[\bar{\mathbf{X}}|\rvM_0\rvM_1]{{\mathbf{x}}|m_0,m_1}  \p[\rvM_0]{m_0}\p[\rvM_1]{m_1}.\label{eq:pdf_code_random_variables}
\end{multline} 
The reliability of a code $\calC_n$ is measured in terms of the average probability of error
\begin{multline*}
  \perr(\calC_n)\eqdef\\
  \P{(\hat\rvM_0,\hat\rvM_1,\hat\rvM')\neq (\rvM_0,\rvM_1,\rvM')\;\text{or}\;\tilde\rvM_0\neq \rvM_0\middle|\calC_n}
\end{multline*}
while its secrecy is measured in terms of the secrecy metric $\metric{i}(\calC_n)\eqdef \metric{i}(p_{\rvM_1\bar{\mathbf{Z}}|\calC_n},p_{\rvM_1}p_{\bar{\mathbf{Z}}|\calC_n})$ for $i\in\intseq{1}{6}$. 
\begin{definition}
  A rate pair $(R_0,R_1)$ is achievable for secrecy metric $\metric{i}$ over a broadcast channel if there exists a sequence of
  $(2^{nR_0},2^{nR_1},n)$ codes $\{\calC_n\}_{n\geq 1}$ such that
  \begin{align*}
    \lim_{n\rightarrow\infty}\perr(\calC_n)=0\quad\text{and}\quad \lim_{n\rightarrow\infty}\metric{i}(\calC_n)=0.
  \end{align*}
  The {secrecy-capacity region} $\calR^{\text{\tiny($i$)}}_{\textnormal{\tiny BCC}}$ is the closure of the set of rate pairs achievable for secrecy metric $\metric{i}$, and
the {secrecy capacity} for secrecy metric $\metric{i}$ is 
  \begin{align*}
    C_{\textnormal{\tiny WT}}^{\text{\tiny($i$)}}\eqdef \sup\{R_1:(0,R_1)\,\mbox{is achievable for secrecy metric $\metric{i}$}\}.
  \end{align*}
\end{definition}
In the absence of a common message $(R_0=0)$, a broadcast channel with confidential messages is concisely called a wiretap channel, and a $(1,2^{nR_1},n)$ code is simply denoted as a $(2^{nR_1},n)$ code. Note that our definition of a wiretap code explicitly introduces the randomness used in the encoding process. The randomness is split between a source of local randomness and an auxiliary message with uniform distribution that we require the legitimate receiver to decode. This allows us to distinguish the part of the randomness that merely acts as artificial random noise from the part that helps secrecy without reducing the reliable communication rate. Since the source of local randomness can be arbitrarily chosen, our definition incurs no loss of generality and allows us to explicitly define the class of \emph{capacity-based wiretap codes} in Section~\ref{sec:resolv-vs.-capac}.

\begin{remark}
  Csisz\'ar and K\"orner~\cite{Csiszar1978} analyze the fundamental limits of secure communication more precisely by studying the rate-equivocation region $(R_0,R_1,R_e)$, where $R_e\leq R_1$ represents the equivocation-rate $\frac{1}{n}\avgH{\rvM_1|\bar{\mathbf{Z}}}$ of the eavesdropper about the individual message. Unlike the rates $R_0$ and $R_1$, the notion of equivocation depends on the secrecy metric considered; therefore, we restrict ourselves to the special case of full secrecy rates $R_1=R_e$, for which we can leverage the result of Proposition~\ref{lm:ordering_metrics}.
\end{remark}

\subsection{Capacity-Based Wiretap Codes and Strong Secrecy}
\label{sec:resolv-vs.-capac}
We now define the subclass of \emph{capacity-based} wiretap codes.
\begin{definition}
  \label{def:capacity-based-wiretap-code}
    A $\left(2^{nR_0},2^{nR_1},n\right)$ capacity-based wiretap code $\calC_n$ is a $\left(2^{nR_0},2^{nR_1},n\right)$ wiretap code such that :
    \begin{itemize}
    \item the auxiliary message rate is $R'_n=C_e-\epsilon_n$, where $C_e$ is the eavesdropper's channel capacity and $\{\epsilon_n\}_{n\geq 1}$ satisfies $\lim_{n\rightarrow\infty}\epsilon_n=0$;
    \item there exists a decoding function $h'_n:\calZ^n\times\calW_1\rightarrow\calW'$, which allows the eavesdropper to estimate the auxiliary message $\rvM'$ from the observation of $\bar{\mathbf{Z}}$ and $\rvM_1$.
    \end{itemize}
\end{definition}
We let $\tilde\rvM'$ denote Eve's estimate of $\rvM'$. The reliability of a capacity-based wiretap code $\calC_n$ is then measured in terms of the modified average probability of error
\begin{multline*}
  \perr^*(\calC_n)\eqdef \mathbb{P}\left[(\hat\rvM_0,\hat\rvM_1,\hat\rvM')\neq (\rvM_0,\rvM_1,\rvM')\right.\\
  \left.\;\text{or}\;(\tilde\rvM_0,\tilde\rvM')\neq (\rvM_0,\rvM')\middle|\calC_n\right].
\end{multline*}
\begin{definition}
  A rate pair $(R_0,R_1)$ is achievable for secrecy metric $\metric{i}$ with capacity-based wiretap codes if there exists a sequence of 
  $(2^{nR_0},2^{nR_1},n)$ capacity-based wiretap codes $\{\calC_n\}_{n\geq 1}$ such that
  \begin{align*}
    \lim_{n\rightarrow\infty}\perr^*(\calC_n)=0\quad\text{and}\quad \lim_{n\rightarrow\infty}\metric{i}(\calC_n)=0.
  \end{align*}
\end{definition}
The constraint $\lim_{n\rightarrow\infty}\perr^*(\calC_n)=0$ ensures that, given knowledge of $\bar{\mathbf{Z}}$ and $\rvM_1$, the eavesdropper could reliably decode the auxiliary message $\rvM'$. Nevertheless, since the eavesdropper does not have access to the message $\rvM_1$, this property is solely used to impose structure on the code. However, note that this also imposes $\lim_{n\rightarrow\infty} \epsilon_n \sqrt{n}=\infty$~\cite[Theorem 49]{Polyanskiy2010}. The denomination ``capacity-based code'' is used because the set of codewords associated to a known pair of messages $(\rvM_0,\rvM_1)$ forms a sub-code of rate $R'_n=C_e-\epsilon_n$, which stems from a sequence of capacity-achieving codes for Eve's channel. 

As formalized in~\cite[Theorem 1]{Thangaraj2007}, capacity-based wiretap codes are implicitly used in most works that show the existence of wiretap codes achieving secrecy rates for metric $\metric{4}$. In this section, we show that this may be an intrinsic limitation, by proving that sequences of random capacity-based wiretap codes that achieve the weak secrecy capacity \emph{cannot} achieve the strong secrecy capacity. 

Specifically, we consider a discrete memoryless wiretap channel $\left(\calX,\calY,W_{{Y}{Z}|{X}},\calZ\right)$ without cost constraint ($\forall {\mathbf{x}}\in\calX^n\;c_n({\mathbf{x}})=n$ and $P=1$) in which the eavesdropper's channel and the main channel are both symmetric.\footnote{More specifically, we use Gallager's notion of symmetry~\cite[p. 94]{InformationTheoryReliableCommunication}.} We further assume that the main channel is more capable than the eavesdropper's channel and has capacity $C_m<\tfrac{1}{2}\log\card{\calX}$ bits. The former assumption ensures that, without loss of optimality, we can assume no source of local randomness $(\calR,p_R)$ is available~\cite{Csiszar1978} and that the secrecy capacity is $C_s=C_m-C_e$; the latter one is a technical assumption required to simplify the analysis.
\begin{proposition}
  \label{prop:capac-achi-wiret}
  Let $\{\rvC_n\}_{n\geq 1}$ be a sequence of $(2^{nR},n)$ random capacity-based wiretap codes, obtained by generating codeword symbols independently and uniformly at random. Let the rate $R'_n$ of the auxiliary message be such that $R'_n=C_e-\epsilon_n$ and $R+R'_n=C_m-\epsilon_n$. Then, there exists $\eta,\alpha>0$, such that, for $n$ sufficiently large,
  \begin{multline*}
    \P{\metric{2}(\rvC_n)>\eta,\;\;\perr^*(\rvC_n)\leq \epsilon'_n \;\;\text{and}\;\;\metric{4}(\rvC_n)\leq 3\epsilon'_n}\\
    \geq 1-2^{-\alpha n \epsilon_n^2},
  \end{multline*}
  with $\epsilon_n' \eqdef \max(\epsilon_n,\log\card{\calX} 2^{-\alpha n \epsilon_n^2},n^{-1})$, i.e., with high probability over the random code ensemble, a sequence of capacity-based random codes achieves the weak secrecy capacity but does not achieve the strong secrecy capacity.
\end{proposition}
\begin{IEEEproof}
  See Appendix~\ref{sec:proof-proposition-weak-strong}
\end{IEEEproof}
We conjecture that the inability to achieve strong secrecy holds for any capacity-based wiretap codes, and not just random codes, as well as for any discrete memoryless channel, and not just symmetric channels. Despite its lack of generality, Proposition~\ref{prop:capac-achi-wiret} shows that a random coding argument with capacity-based wiretap codes is not powerful enough to prove strong secrecy results, which suggests exploiting a more powerful mechanism to ensure secrecy. In the remainder of the paper, we derive secrecy from channel resolvability and show that the resulting codes do not suffer from the limitations of capacity-based wiretap codes.

\begin{remark}
  If the main channel is noiseless and the eavesdropper's channel is symmetric, a slight modification of the proof of Proposition~\ref{prop:capac-achi-wiret} shows that no capacity-based wiretap code (including non-random codes) achieves secrecy capacity for metrics $\metric{2}$ and $\metric{1}$. This fact was independently noted in~\cite{Mahdavifar2011} for metric $\metric{1}$ using results for finite blocklength channel coding~\cite{Polyanskiy2010}. Our approach builds on a similar result established for secret-key agreement in~\cite{Watanabe2010a}.
\end{remark}

\subsection{General Broadcast Channels with Confidential Messages and Cost Constraint}
\label{sec:arbitr-wiret-chann-1}

\begin{figure}[htb]
  \centering
    \includegraphics[width=.98\linewidth]{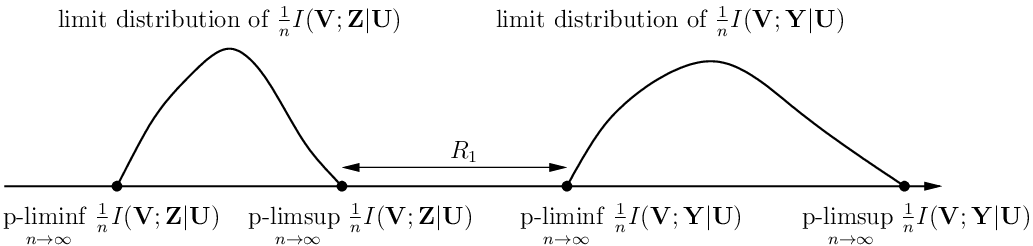}
  \caption{Illustration of secure rates in Theorem~\ref{th:bcc}.}
  \label{fig:illustration_secure_rates}
\end{figure}

In this section, we establish the secrecy-capacity region of a general broadcast channel with confidential messages for secrecy metrics $\metric{i}$ with $i\in\intseq{2}{6}$; the alphabets and transition probabilities of the channel $\{W_{{\mathbf{Y}}{\mathbf{Z}}|{\mathbf{X}}}\}_{n\geq 1}$ are arbitrary, so that the model includes continuous channels and channels with memory. Following the conclusions drawn from Proposition~\ref{prop:capac-achi-wiret}, we analyze codes that are more powerful than capacity-based wiretap codes and whose secrecy is tied to the notion of channel resolvability.
\begin{theorem}
  \label{th:bcc}
  The secrecy-capacity region of a broadcast channel $\left(\calX,\calY,\{W_{{\mathbf{Y}}{\mathbf{Z}}|{\mathbf{X}}}\}_{n\geq 1},\calZ\right)$ with confidential messages and cost constraint $P$ is the same for all secrecy metrics~$\metric{i}$ with $i\in\intseq{2}{6}$ and is given by
  \begin{multline}
    \calR_{\textnormal{\tiny BCC}}=\\
    \bigcup_{\left\{{\mathbf{U}}{\mathbf{V}}{\mathbf{X}}\right\}_{n\geq 1}\in\calP}\left\{\begin{array}{l} (R_0,R_1)\in\mathbb{R}_+^2:\\\vspace{5pt}
        \displaystyle R_0\leq \min\left(\pliminf{\frac{1}{n}\I{{\mathbf{U}};{\mathbf{Y}}}},\right.\\
        \phantom{-----}\left.\pliminf{\frac{1}{n}\I{{\mathbf{U}};{\mathbf{Z}}}}\right),\\
        \displaystyle R_1\leq
        \pliminf{\frac{1}{n}\I{{\mathbf{V}};{\mathbf{Y}}|{\mathbf{U}}}}\\
        \phantom{-----}-\plimsup{\frac{1}{n}\I{{\mathbf{V}};{\mathbf{Z}}|{\mathbf{U}}}}
    \end{array}\right\}
  \label{eq:general_secrecy_capacity_region}
\end{multline}
where
  \begin{multline*}
    \calP\eqdef\left\{\smash{\left\{{\mathbf{U}}{\mathbf{V}}{\mathbf{X}}\right\}_{n\geq 1}}:\forall n\in\mathbb{N}^* \, {\mathbf{U}}\rightarrow{\mathbf{V}}\rightarrow{\mathbf{X}}\rightarrow{\mathbf{Y}}{\mathbf{Z}}\text{ forms}\right.\\
      \left.\text{ a Markov chain and } \P{\tfrac{1}{n}c_n({\mathbf{X}})\leq P}=1 \right\}.
  \end{multline*}
\end{theorem}
Notice that the form of the secrecy-capacity region is the natural generalization of that obtained for memoryless channels in~\cite[Corollary 1]{Csiszar1978}; however, the main channel statistics affect the secure rate $R_1$ through their ``worst realization'' $\pliminf{\frac{1}{n}\I{{\mathbf{V}};{\mathbf{Y}}|{\mathbf{U}}}}$ while the eavesdropper's channel statistics affect it through their ``best realization'' $\plimsup{\frac{1}{n}\I{{\mathbf{V}};{\mathbf{Z}}|{\mathbf{U}}}}$. Intuitively, as illustrated in Figure~\ref{fig:illustration_secure_rates}, this occurs because the worst case for secure communication is when the main channel conveys the smallest information rate to the legitimate receiver while the eavesdropper's channel leaks the largest information rate to the eavesdropper. It will be apparent in the proof that this asymmetry, which disappears in the case of memoryless channels, arises because the coding mechanisms used to ensure reliability and secrecy are different.

\begin{IEEEproof}[Proof of Theorem~\ref{th:bcc}]
We start with the achievability part of the proof, for which we create a codebook by combining superposition coding and binning schemes. Let $n\in\mathbb{N}^*$ and $\epsilon,\gamma,R_0,R_1,R'>0$. Define $M_0\eqdef \lceil 2^{nR_0}\rceil$, $M_1\eqdef \lceil 2^{nR_1}\rceil$ and $M'\eqdef \lceil 2^{nR'}\rceil$. Let $\calU$ be an arbitrary alphabet and fix a distribution $p_{{\mathbf{U}}}$ on $\calU^n$. Fix a conditional distribution $p_{{\mathbf{X}}|{\mathbf{U}}}$ on $\calX^n\times\calU^n$ such that $\P{\frac{1}{n}c_n({\mathbf{X}})\leq P}=1$. Let ${\mathbf{U}},{\mathbf{X}},{\mathbf{Y}},{\mathbf{Z}}$ be the random variables with joint distribution
\begin{align}
    \p[{\mathbf{U}}{\mathbf{X}}{\mathbf{Y}}{\mathbf{Z}}]{{\mathbf{u}},{\mathbf{x}},{\mathbf{y}},{\mathbf{z}}} \eqdef W_{{\mathbf{Y}} {\mathbf{Z}}|{\mathbf{X}}}({\mathbf{y}},{\mathbf{z}}|{\mathbf{x}}) p_{{\mathbf{X}}|{\mathbf{U}}}({\mathbf{x}}|{\mathbf{u}})p_{{\mathbf{U}}}({\mathbf{u}}). \label{eq:def_zn}
\end{align}

\begin{itemize}
\item \textbf{Code generation:} Randomly generate $M_0$ sequences ${\mathbf{u}}_k\in\calU^n$ with $k\in\intseq{1}{M_0}$ according to $p_{{\mathbf{U}}}$. For each $k\in\intseq{1}{M_0}$, generate $M_1M'$ sequences ${\mathbf{x}}_{klm}\in\calX^n$ with $(l,m)\in\intseq{1}{M_1}\times\intseq{1}{M'}$ according to $p_{{\mathbf{X}}|{\mathbf{U}}={\mathbf{u}}_k}$. We denote by $\rvC_n$ the random variable representing the generated code and by $\calC_n$ one of its realizations.
\item \textbf{Encoding:} To transmit a message pair $(k,l)\in\intseq{1}{M_0}\times\intseq{1}{M_1}$, Alice generates an auxiliary message $m$ uniformly at random in $\intseq{1}{M'}$ and sends the codeword $\mathbf{x}_{klm}$ through the channel.
\item \textbf{Bob's decoding:} Define the sets
\begin{align*}
  \calT_1^n&\eqdef\left\{({\mathbf{u}},{\mathbf{y}})\in\calU^n\times\calY^n:\right.\\
  &\phantom{------}\left.\frac{1}{n}\log\frac{p_{{\mathbf{Y}}|{\mathbf{U}}}({\mathbf{y}}|{\mathbf{u}})}{\p[{\mathbf{Y}}]{{\mathbf{y}}}}\geq \frac{1}{n}\log M_0+\gamma\right\},\\
  \calT_2^n&\eqdef\left\{({\mathbf{u}},{\mathbf{x}},{\mathbf{y}})\in\calU^n\times\calX^n\times\calY^n:\right.\\
&\phantom{---} \left. \frac{1}{n}\log\frac{p_{{\mathbf{Y}}|{\mathbf{X}}{\mathbf{U}}}({\mathbf{y}}|{\mathbf{x}},{\mathbf{u}})}{p_{{\mathbf{Y}}|{\mathbf{U}}}({\mathbf{y}}|{\mathbf{u}})}\geq \frac{1}{n}\log M_1M'+\gamma\right\}.
\end{align*}
Upon observing ${\mathbf{y}}$, Bob decodes $k$ as the received common message if ${\mathbf{u}}_k$ is the unique sequence in $\calC_n$ such that $\left({\mathbf{u}}_k,{\mathbf{y}}\right)\in\calT_1^n$; otherwise, a random message is chosen. Similarly, he decodes $l$ as the received individual message and $m$ as the received auxiliary message if there exists a unique codeword ${\mathbf{x}}_{klm}$ such that $\left({\mathbf{u}}_k,{\mathbf{x}}_{klm},{\mathbf{y}}\right)\in\calT_2^n$; otherwise, random messages are chosen.
\item \textbf{Eve's decoding:} Define the set
  \begin{align*}
    \calT_3^n&\eqdef\left\{({\mathbf{u}},{\mathbf{z}})\in\calU^n\times\calZ^n:\right.\\
    &\phantom{------}\left.\frac{1}{n}\log\frac{p_{{\mathbf{Z}}|{\mathbf{U}}}({\mathbf{z}}|{\mathbf{u}})}{p_{{\mathbf{Z}}}({\mathbf{z}})}\geq \frac{1}{n}\log M_0+\gamma\right\}.
  \end{align*}
Upon observing ${\mathbf{z}}$, Eve decodes $k$ as the received common message if ${\mathbf{u}}_k$ is the unique sequence such that $\left({\mathbf{u}}_k,{\mathbf{z}}\right)\in\calT_3^n$; otherwise, a random message is chosen.
\end{itemize}
The following lemmas, whose proofs are relegated to Appendix~\ref{sec:proof-theorem-bcc-achievability}, provide sufficient conditions to guarantee reliability and secrecy.
\begin{lemma}[Reliability conditions]
  \label{lm:reliability_bcc}
   \begin{align*}
    &\text{If }R_0 \leq \min\left(\pliminf{\frac{1}{n}\I{{\mathbf{U}};{\mathbf{Y}}}}-2\gamma,\right.\\
          &\left.\phantom{-----------}\pliminf{\frac{1}{n}\I{{\mathbf{U}};{\mathbf{Z}}}}-2\gamma\right)\\
          &\text{and }R_1+R'\leq \pliminf{\frac{1}{n}\I{{\mathbf{X}};{\mathbf{Y}}|{\mathbf{U}}}}-2\gamma,
  \end{align*}
then $\lim_{n\rightarrow\infty}\E{\perr(\rvC_n)}\leq \epsilon$.
\end{lemma}
\begin{lemma}[Secrecy from channel resolvability condition]
  \label{lm:secrecy_bcc}
  \begin{multline*}
        \text{If }R'\geq \plimsup{\frac{1}{n}\I{{\mathbf{X}};{\mathbf{Z}}|{\mathbf{U}}}}+2\gamma\\
        \text{ then } \lim_{n\rightarrow\infty}\E{\metric{2}(\rvC_n)}\leq \epsilon.
      \end{multline*}
    \end{lemma}
Combining Lemma~\ref{lm:reliability_bcc} and Lemma~\ref{lm:secrecy_bcc}, we obtain that if
\begin{multline*}
  R_0 \leq \min\left(\pliminf{\frac{1}{n}\I{{\mathbf{U}};{\mathbf{Y}}}}-2\gamma\right.\\
  \left.,\pliminf{\frac{1}{n}\I{{\mathbf{U}};{\mathbf{Z}}}}-2\gamma\right)
\end{multline*}
\begin{multline*}
  \text{and }      R_1\leq \pliminf{\frac{1}{n}\I{{\mathbf{X}};{\mathbf{Y}}|{\mathbf{U}}}}\\
-\plimsup{\frac{1}{n}\I{{\mathbf{X}};{\mathbf{Z}}|{\mathbf{U}}}}-4\gamma,
\end{multline*}
then, $\lim_{n\rightarrow\infty}\E{\perr(\rvC_n)}\leq \epsilon$ and $\lim_{n\rightarrow\infty}\E{\metric{2}(\rvC_n)}\leq \epsilon$. By Markov's inequality and the union bound, there exists at least one sequence of $(2^{nR_0},2^{nR_1},n)$ codes $\{\calC_n\}_{n\geq 1}$ such that $\lim_{n\rightarrow\infty}{\perr(\calC_n)}\leq 3\epsilon$ and $\lim_{n\rightarrow\infty}{\metric{2}(\calC_n)}\leq 3\epsilon$. Since $\epsilon$ and $\gamma$ can be chosen arbitrarily small, we conclude that
  \begin{align}
    \bigcup_{\left\{{\mathbf{U}}{\mathbf{X}}\right\}_{n\geq 1}\in\calP}\left\{\begin{array}{l} (R_0,R_1)\in\mathbb{R}_+^2:\\\vspace{5pt}
        \displaystyle R_0\leq \min\left(\pliminf{\frac{1}{n}\I{{\mathbf{U}};{\mathbf{Y}}}},\right.\\
                  \phantom{-----}\left. \pliminf{\frac{1}{n}\I{{\mathbf{U}};{\mathbf{Z}}}}\right),\\
        \displaystyle R_1\leq
        \pliminf{\frac{1}{n}\I{{\mathbf{X}};{\mathbf{Y}}|{\mathbf{U}}}}\\
                \phantom{----}-\plimsup{\frac{1}{n}\I{{\mathbf{X}};{\mathbf{Z}}|{\mathbf{U}}}}
    \end{array}\right\}\subseteq \calR^{\text{\tiny($2$)}}_{\text{\tiny BCC}}
  \label{eq:general_secrecy_capacity_region_2}
  \end{align}
  where
  \begin{multline*}
    \calP\eqdef\left\{\smash{\left\{{\mathbf{U}}{\mathbf{X}}\right\}_{n\geq 1}}:\forall n\in\mathbb{N}^*\, {\mathbf{U}}\rightarrow{\mathbf{X}}\rightarrow{\mathbf{Y}}{\mathbf{Z}}\text{ forms}\right.\\
\left.        \text{a Markov chain and } \P{\tfrac{1}{n}c_n({\mathbf{X}})\leq P}=1  \right\}.
\end{multline*}
Finally, note that the source of local randomness $(\calR,p_{\rvR})$ can be used to prefix an arbitrary channel $\left(\calV,\{p_{{\mathbf{X}}|{\mathbf{V}}}\}_{n\geq 1},\calX\right)$ to the broadcast channel $\left(\calX,\calY,\{W_{{\mathbf{Y}}{\mathbf{Z}}|{\mathbf{X}}}\}_{n\geq 1},\calZ\right)$. That this prefix is useful for secrecy applications is well established~\cite{Csiszar1978}. By applying the proof above to the concatenated channel $\left(\calV,\calY,\{p_{{\mathbf{Y}}{\mathbf{Z}}|{\mathbf{V}}}\}_{n\geq 1},\calZ\right)$, we conclude that the region given in Theorem~\ref{th:bcc} is included in the capacity region $\calR^{\text{\tiny($2$)}}_{\text{\tiny BCC}}$.

We now turn to the converse part of the proof. Consider a sequence of codes $\{\calC_n\}_{n\geq 1}$ achieving the rate pair $\left(R_0,R_1\right)$ for secrecy metric $\metric{6}$. For $n\in\mathbb{N}^*$, let $\bar{\mathbf{U}}$ denote the choice of a common message uniformly at random in $\intseq{1}{2^{nR_0}}$ and let $\bar{\mathbf{\rvW}}$ denote the choice of an individual message uniformly at random in $\intseq{1}{2^{nR_1}}$. Let $\bar{\mathbf{Y}}$ and $\bar{\mathbf{Z}}$ denote the channel outputs corresponding to the transmission of the message pair $(\bar{\mathbf{U}},\bar{\mathbf{W}})$. As shown in Appendix~\ref{sec:proof-theorem-bcc-converse}, the following lemma holds.

\begin{lemma}
  \label{lm:bcc_converse}
  If $\lim_{n\rightarrow\infty}\perr(\calC_n)=0$ and $\lim_{n\rightarrow\infty}\metric{6}(\calC_n)=0$, then
  \begin{align*}
    R_0 &\leq \min\left(\pliminf{\frac{1}{n}\I{\bar{\mathbf{U}};\bar{\mathbf{Y}}}},\pliminf{\frac{1}{n}\I{\bar{\mathbf{U}};\bar{\mathbf{Z}}}}\right)\\
    R_1&\leq \pliminf{\frac{1}{n}\I{\bar{\mathbf{W}};\bar{\mathbf{Y}}|\bar{\mathbf{U}}}}-\plimsup{\frac{1}{n}\I{\bar{\mathbf{W}};\bar{\mathbf{Z}}|\bar{\mathbf{U}}}}.
  \end{align*}
 \end{lemma}
Note that, by assumption, $\bar{\mathbf{U}}\bar{\mathbf{W}}\rightarrow\bar{\mathbf{X}}\rightarrow\bar{\mathbf{Y}}\bar{\mathbf{Z}}$ forms a Markov chain. Define $\bar{\mathbf{V}}\eqdef (\bar{\mathbf{U}},\bar{\mathbf{W}})$, which is such that
$\bar{\mathbf{U}}\rightarrow\bar{\mathbf{V}}\rightarrow\bar{\mathbf{X}}\rightarrow\bar{\mathbf{Y}}\bar{\mathbf{Z}}$ forms a Markov chain. With probability one, we have
\begin{align*}
  \I{\bar{\mathbf{W}};\bar{\mathbf{Y}}|\bar{\mathbf{U}}} = \I{\bar{\mathbf{V}};\bar{\mathbf{Y}}|\bar{\mathbf{U}}}\quad\mbox{and}\quad\I{\bar{\mathbf{W}};\bar{\mathbf{Z}}|\bar{\mathbf{U}}} = \I{\bar{\mathbf{V}};\bar{\mathbf{Z}}|\bar{\mathbf{U}}};
\end{align*}
therefore, an achievable pair $(R_0,R_1)$ must satisfy
\begin{align*}
  R_0 &\leq \min\left(\pliminf{\frac{1}{n}\I{\bar{\mathbf{U}};\bar{\mathbf{Y}}}},\pliminf{\frac{1}{n}\I{\bar{\mathbf{U}};\bar{\mathbf{Z}}}}\right),\\
  \mbox{and}\quad R_1 &\leq \pliminf{\frac{1}{n}\I{\bar{\mathbf{V}};\bar{\mathbf{Y}}|\bar{\mathbf{U}}}}-\plimsup{\frac{1}{n}\I{\bar{\mathbf{V}};\bar{\mathbf{Z}}|\bar{\mathbf{U}}}},
\end{align*}
where $\bar{\mathbf{U}}\rightarrow\bar{\mathbf{V}}\rightarrow\bar{\mathbf{X}}\rightarrow\bar{\mathbf{Y}}\bar{\mathbf{Z}}$ forms a Markov chain, $p_{\bar{\mathbf{Y}}\bar{\mathbf{Z}}|\bar{\mathbf{X}}}=W_{{\mathbf{Y}}{\mathbf{Z}}|{\mathbf{X}}}$, and $\P{\frac{1}{n}c_n(\bar{\mathbf{X}})\leq P}=1$. Taking the union over all possible processes $\{\bar{\mathbf{U}}\bar{\mathbf{V}}\bar{\mathbf{X}}\}_{n\geq 1}$ gives the desired outer bound for the secrecy-capacity region $\calR^{\text{\tiny($6$)}}_{\text{\tiny BCC}}$.

Since the outer bound for $\calR^{\text{\tiny($6$)}}_{\text{\tiny BCC}}$ and the inner bound for $\calR^{\text{\tiny($2$)}}_{\text{\tiny BCC}}$ match, we conclude using Proposition~\ref{lm:ordering_metrics} that the secrecy-capacity region is the same for all metrics $i\in\intseq{2}{6}$.
\end{IEEEproof}

A few comments regarding Theorem~\ref{th:bcc} are now in order. First, the achievability part of the proof is based on an explicit operational interpretation of secrecy in terms of channel resolvability; in Lemma~\ref{lm:secrecy_bcc}, codes are constructed so that, for a given message $\rvM_0$ and taking the average over the random codebook selection, the probability distribution induced at the eavesdropper's channel output by all messages $\rvM_1$ is asymptotically the same in the sense of variational distance. Second, the existence of a sequence of codes simultaneously satisfying the reliability and secrecy conditions is obtained by handling the constraints separately, as illustrated by the separate results of Lemma~\ref{lm:reliability_bcc} and Lemma~\ref{lm:secrecy_bcc}. This contrasts with the approach of~\cite{Wyner1975,Csiszar1978}, in which the two constraints are handled somewhat simultaneously by using capacity-based wiretap codes. As should be clear from the condition $R'>\plimsup{\frac{1}{n}\I{{\mathbf{X}};{\mathbf{Z}}|{\mathbf{U}}}}$ obtained in Lemma~\ref{lm:secrecy_bcc}, the codes constructed are not capacity-based wiretap codes, for which the condition would read $R'<\pliminf{\frac{1}{n}\I{{\mathbf{X}};{\mathbf{Z}}|{\mathbf{U}}}}$; essentially, channel resolvability enables the analysis of codes operating at rates beyond the capacity of the eavesdropper's channel. Finally, we note that, as in Section~\ref{sec:shann-ciph-syst}, the secrecy-capacity region is invariant with respect to the metrics $\metric{i}$ for $i\in\intseq{2}{6}$; nevertheless, practical coding schemes should be designed to provide secrecy with respect to the strongest metric.

\begin{remark}
\label{rmk:exponential_information_stable}
  If the eavesdropper's channel is exponentially information stable, so that 
  \begin{align*}
     \P[{\mathbf{U}}{\mathbf{X}}{\mathbf{Z}}]{\frac{1}{n}\I{{\mathbf{X}};{\mathbf{Z}}|{\mathbf{U}}}>\frac{1}{n}\log M'+\epsilon}
  \end{align*}
  decays exponentially fast with $n$ for any $\epsilon>0$, then a closer look at the proof of Theorem~\ref{th:bcc} shows that $\metric{2}(\calC_n)$, and consequently $\metric{1}(\calC_n)$, would also decay exponentially fast with $n$. We do not explore this issue further for arbitrary channels but we analyze it more precisely in the next section for memoryless channels.
\end{remark}

Without a common message ($R_0=0$), we obtain in a similar way the secrecy capacity of a general wiretap
channel established by Hayashi~\cite[Theorem 5]{Hayashi2006}.
\begin{corollary}
  \label{cor:wtc}
  The secrecy capacity of a wiretap channel $\left(\calX,\calY,\{W_{{\mathbf{Y}}{\mathbf{Z}}|{\mathbf{X}}}\}_{n\geq 1},\calZ\right)$ with cost constraint $P$ is identical for secrecy
  metrics~$\metric{i}$ with $i\in\intseq{2}{6}$ and is given by
  \begin{multline}
    C_s = \sup_{\{{\mathbf{V}}{\mathbf{X}}\}_{n\geq 1}\in\calP}\left(\pliminf{\frac{1}{n}\I{{\mathbf{V}};{\mathbf{Y}}}}\right.\\
      \left.-\plimsup{\frac{1}{n}\I{{\mathbf{V}};{\mathbf{Z}}}}\right),
    \end{multline}
    where
\begin{multline*}
     \calP\eqdef\left\{\smash{\left\{{\mathbf{V}}{\mathbf{X}}\right\}_{n\geq 1}}:\forall n\in\mathbb{N}^*, {\mathbf{V}}\rightarrow{\mathbf{X}}\rightarrow{\mathbf{Y}}{\mathbf{Z}}\text{ forms}\right.\\
\left.        \text{a Markov chain and } \P{\tfrac{1}{n}c_n({\mathbf{X}})\leq P}=1   \right\}.
\end{multline*}
\end{corollary}

\subsection{Memoryless Broadcast Channels with Additive Cost Constraint}
\label{sec:discr-memoryl-chann}
We now consider memoryless channels (not necessarily discrete) with an additive cost constraint. This is a special case of the general model, in which the transition probabilities factor as
\begin{align*}
    W_{{\mathbf{Y}}{\mathbf{Z}}|{\mathbf{X}}}({{\mathbf{y}},{\mathbf{z}}|{\mathbf{x}}})=\prod_{i=1}^nW_{\rvY\rvZ|\rvX}({y_i,z_i|x_i})
\end{align*}
and the cost constraint satisfies $c_n({\mathbf{x}}) = \sum_{i=1}^nc(x_i)$ for some cost function $c:\calX\rightarrow\mathbb{R}^+$. For this special class of channels and constraints, and under mild conditions, the result of Section~\ref{sec:arbitr-wiret-chann-1} extends to metric $\metric{1}$. For discrete memoryless channels without cost constraint, this result was obtained independently in~\cite{Matsumoto2010arXiv,CodingTheoremsDMC2} using secure multiplex coding and privacy amplification.
\begin{theorem}
  \label{th:dmc}
  The secrecy-capacity region of a memoryless broadcast channel $\left(\calX,\calY,W_{\rvY\rvZ|\rvX},\calZ\right)$ with confidential messages and additive cost constraint $P$ is the same for all secrecy
  metrics~$\metric{i}$ with $i\in\intseq{2}{4}$ and is given by
  \begin{align}
\calR_{\textnormal{\tiny BCC}}=    \bigcup_{(\rvU\rvV\rvX)\in\calP}\left\{\begin{array}{l}
        (R_0,R_1)\in\mathbb{R}_+^2:\\
        \displaystyle R_0\leq \min\left(\avgI{\rvU;\rvY},\avgI{\rvU;\rvZ}\right)\\
        \displaystyle R_1\leq \avgI{\rvV;\rvY|\rvU}-\avgI{\rvV;\rvZ|\rvU}
    \end{array}\right\},
  \end{align}
  where
  \begin{multline*}
     \calP\eqdef\left\{(\rvU\rvV\rvX):
       \rvU\rightarrow\rvV\rightarrow\rvX\rightarrow\rvY\rvZ\text{ forms a }\right.\\\left.\text{Markov chain and }\E{c(\rvX)}\leq P\right\}.
   \end{multline*}
   If the rates on the boundary of $\calR_{\textnormal{\tiny BCC}}$ are obtained for some random variables $\rvU\rvV\rvX\rvY\rvZ$ such that the integrals defining the moment generating functions of $\I{\rvV;\rvZ|\rvU}$ and $c(\rvX)$ converge uniformly in a neighborhood of $0$ and are differentiable at $0$, then $\calR_{\textnormal{\tiny BCC}}$ is also the secrecy-capacity region for metric $\metric{1}$.
\end{theorem}
\begin{IEEEproof}
  See Appendix~\ref{sec:proof-theorem-dmc-bcc}.  
\end{IEEEproof}
  The conditions that yield $\calR_{\textnormal{\tiny BCC}}$ for metric $\metric{1}$ are sufficient conditions required to obtain exponential upper bounds when applying Chernov bounds. These conditions are not too restrictive and are automatically satisfied for discrete memoryless channels and for Gaussian channels with additive power constraint. Improved exponents can be obtained in such cases using techniques as in~\cite{Hayashi2012}. 

In the absence of a common message $(R_0=0)$, we obtain in a similar way the following result, which was already obtained for discrete memoryless channels by Csisz\'ar~\cite{Csiszar1996} and Maurer and Wolf~\cite{Maurer2000} with different techniques. 
\begin{corollary}
  \label{cor:strong_secrecy_memoryless}
  The secrecy capacity of a memoryless wiretap channel $\left(\calX,\calY,W_{\rvY\rvZ|\rvX},\calZ\right)$ with additive cost constraint $P$ is the same for all secrecy metrics~$\metric{i}$ with $i\in\intseq{2}{4}$ and is given by
  \begin{align*}
    C_s=\sup_{(\rvV\rvX)\in\calP}\left(\avgI{\rvV;\rvY}-\avgI{\rvV;\rvZ}\right),
  \end{align*}
  where 
  \begin{multline*}
    \calP\eqdef\left\{(\rvV\rvX):
      \rvV\rightarrow\rvX\rightarrow\rvY\rvZ\text{ forms a Markov
        chain}\right.\\\left.\text{ and }\E{c(\rvX)}\leq P\right\}.
  \end{multline*}
  If the random variables $\rvV\rvX\rvY\rvZ$ maximizing $C_s$ are such that the integrals defining the moment generating functions of $\I{\rvV;\rvZ}$ and $c(\rvX)$ converge uniformly in a neighborhood of $0$ and are differentiable at $0$, then $C_s$ is also the secrecy capacity for metric $\metric{1}$.
\end{corollary}
For general memoryless channels, the converse part of Theorem~\ref{th:dmc} and Corollary~\ref{cor:strong_secrecy_memoryless} follows from standard arguments with metric $\metric{4}$~\cite{Csiszar1978}; however, for discrete memoryless channels, the converse is obtained by specializing Theorem~\ref{th:bcc} and holds for metric $\metric{6}$.

\begin{remark}
  In the proof of Theorem~\ref{th:dmc}, we can actually establish a stronger result than the one stated. If the conditions for the moment generating functions of $\I{\rvV;\rvZ|\rvU}$ and $c(\rvX)$ are satisfied, we can show that $\metric{1}(\calC_n)$ vanishes exponentially fast with $n$.
\end{remark}

\section{Applications}
\label{sec:applications}

In this section, we illustrate the usefulness of deriving secrecy from channel resolvability by considering several problems in which the derivation of achievable secrecy rates is tremendously simplified. In particular, results for wireless channels, mixed wiretap channels and compound wiretap channels come almost ``for free''. For simplicity, we only consider cases in which the common message rate is zero ($R_0=0$).

\subsection{Ergodic Wireless Channels with Full \ac{CSI}}
\label{sec:wireless-channels}

We consider the situation in which Alice and Bob communicate over an ergodic-fading wiretap channel and have access to the instantaneous fading gains for both the main channel and the eavesdropper's channel. Specifically, at each time $k\geq 1$, the relationships between input and outputs are given by
\begin{align*}
  \rvY_k & = \rvH_{m,k}\rvX_k+\rvN_{m,k},\\
   \rvZ_k & = \rvH_{e,k}\rvX_k+\rvN_{e,k},
\end{align*}
where $\{\rvH_{m,k}\}_{k\geq 1}$, $\{\rvH_{e,k}\}_{k\geq 1}$ are fading gains known to all parties and $\{\rvN_{m,k}\}_{k\geq 1}$, $\{\rvN_{e,k}\}_{k\geq 1}$ are i.i.d. complex Gaussian zero-mean noise processes with respective variance $\sigma_m^2$ and $\sigma_e^2$. In addition, the channel inputs are subject to the long-term power constraint $\frac{1}{n}\sum_{k=1}^n\E{\rvX_k^2}\leq P$.
\begin{proposition}
  \label{prop:stron_secrecy_wireless}
  The secrecy capacity of the ergodic wireless channel with full \ac{CSI} for secrecy metric $\metric{1}$ is
  \begin{multline}
    C_s = \max_{\gamma}\mathbb{E}\left[\log\left(1+\frac{\abs{\rvH_m}^2\gamma(\rvH_m,\rvH_e)}{\sigma_m^2}\right)\right.\\
    \left.-\log\left(1+\frac{\abs{\rvH_e}^2\gamma(\rvH_m,\rvH_e)}{\sigma_e^2}\right)\right],
  \end{multline}
  where the maximization is over all power allocation functions $\gamma:\mathbb{C}^2\rightarrow\mathbb{R}^+$ such that $\E{\gamma(\rvH_m,\rvH_e)}\leq P$.
\end{proposition}
\begin{IEEEproof}[Sketch of proof]
  We only sketch the achievability part of the proof; the converse for secrecy metric $\metric{4}$ is established in~\cite{Liang2008a}. Because the channel gains are instantaneously known to all parties, the ergodic wireless channel can be demultiplexed into a set of independent Gaussian wiretap channels, each characterized by a specific realization $(\svh_m,\svh_e)$ of the channel gains and subject to a power constraint $\gamma(\svh_m,\svh_e)$.  Upon substituting $\rvV=0$ and $\rvX\sim\calN(0,\gamma(\svh_m,\svh_e))$ in Corollary~\ref{cor:strong_secrecy_memoryless}, we obtain the following achievable rate for metric $\metric{1}$ and for each channel:
  \begin{align*}
    \log\left(1+\frac{\abs{\svh_m}^2\gamma(\svh_m,\svh_e)}{\sigma_m^2}\right)-\log\left(1+\frac{\abs{\svh_e}^2\gamma(\svh_m,\svh_e)}{\sigma_e^2}\right).
  \end{align*}
Hence, using the ergodicity of the channel, we conclude that all the rates $R\geq 0$ such that
\begin{multline*}
R<\max_{\gamma}\mathbb{E}\left[\log\left(1+\frac{\abs{\rvH_m}^2\gamma(\rvH_m,\rvH_e)}{\sigma_m^2}\right)\right.\\
  \left.-\log\left(1+\frac{\abs{\rvH_e}^2\gamma(\rvH_m,\rvH_e)}{\sigma_e^2}\right)\right]
\end{multline*}
are achievable for metric $\metric{1}$,  where $\gamma:\mathbb{C}^2\rightarrow\mathbb{R}^+$ satisfies $\E{\gamma(\rvH_m,\rvH_e)\leq P}$.
\end{IEEEproof}
The result of Proposition~\ref{prop:stron_secrecy_wireless} has already been established in~\cite{Barros2008} with a completely different approach; deriving secrecy from channel resolvability and leveraging Corollary~\ref{cor:strong_secrecy_memoryless} provides a much simpler and direct proof, which can be generalized to include the effect of imperfect \ac{CSI}~\cite{Bloch2009,Bloch2012b}. 

\subsection{Mixed and Compound Channels with receiver \ac{CSI}}
\label{sec:mixed-wiret-chann}

As another application, we study mixed and compound wiretap channels with receiver \ac{CSI}. These models have practical relevance since they allow one to analyze situations in which the channel is imperfectly known to the transmitter, either because the channel estimation mechanism is imperfect or because the channel is partially controlled by the eavesdropper.

Let $K\in\mathbb{N}^*$ and let $\{\alpha_{k}\}_{k\in\intseq{1}{K}}$ be such that $\forall k\in\intseq{1}{K}\;\alpha_k>0$ and $\sum_{k=1}^K\alpha_k=1$.  Consider the wiretap channels $\left(\smash{\calX,\calY,\left\{W_{{\mathbf{Y}}_k{\mathbf{Z}}_k|{\mathbf{X}}}\right\}_{n\geq1},\calZ}\right)$ for
  $k\in\intseq{1}{K}$. The \emph{mixed wiretap channel} is the channel $\left(\calX,\calY,W_{{\mathbf{Y}}{\mathbf{Z}}|{\mathbf{X}}},\calZ\right)$ whose transition probabilities satisfy
\begin{align*} 
W_{{\mathbf{Y}}{\mathbf{Z}}|{\mathbf{X}}}({{\mathbf{y}},{\mathbf{z}}|{\mathbf{x}}})=\sum_{k=1}^K \alpha_kW_{{\mathbf{Y}}_k{\mathbf{Z}}_k|{\mathbf{X}}}({\mathbf{y}},{\mathbf{z}}|{\mathbf{x}}). 
\end{align*}

\begin{proposition}
  \label{prop:mixed_wiretap}
  The secrecy capacity of the mixed wiretap channel with power constraint $P$ is the same for all secrecy metrics $\metric{i}$ with $i\in\intseq{2}{6}$ and is given by
\begin{multline}
  \sup_{{\left\{{\mathbf{V}},{\mathbf{X}}\right\}_{n\geq 1}}\in\calP}\left(\min_{k\in\intseq{1}{K}}\pliminf{\frac{1}{n}\I{{\mathbf{V}};{\mathbf{Y}}_k}}\right.\\
\left.  -\max_{k\in\intseq{1}{K}}\plimsup{\frac{1}{n}\I{{\mathbf{V}};{\mathbf{Z}}_k}}\right),
\end{multline}
where
\begin{multline*}
  \calP\eqdef\left\{\smash{\left\{{\mathbf{V}}{\mathbf{X}}\right\}_{n\geq 1}}:\forall n\in\mathbb{N}^*, \;\forall k\in\intseq{1}{K},\,{\mathbf{V}}\rightarrow{\mathbf{X}}\rightarrow{\mathbf{Y}}_k{\mathbf{Z}}_k\right.\\
  \left.        \text{ forms a Markov chain }
        \text{and } \P{\tfrac{1}{n}c_n({\mathbf{X}})\leq P}=1  \right\}.
\end{multline*}
\end{proposition}
\begin{IEEEproof}
 Using~\cite[Lemma 1.4.2]{InformationSpectrumMethods}, we obtain
  \begin{align*}
    \pliminf{\frac{1}{n}\I{{\mathbf{V}};{\mathbf{Y}}}}&= \min_{k\in\intseq{1}{K}}\left(\pliminf{\frac{1}{n}\I{{\mathbf{V}};{\mathbf{Y}}_k}}\right)\\
    \plimsup{\frac{1}{n}\I{{\mathbf{V}};{\mathbf{Z}}}}&= \max_{k\in\intseq{1}{K}}\left(\plimsup{\frac{1}{n}\I{{\mathbf{V}};{\mathbf{Z}}_k}}\right).
  \end{align*}
  The result follows by substituting these equalities in Corollary~\ref{cor:wtc}.
\end{IEEEproof}
Note that for $i\in\intseq{1}{2}$, we have $\metric{i}(p_{\rvM\bar{\mathbf{Z}}},p_{\rvM}p_{\bar{\mathbf{Z}}})\leq \sum_{k=1}^K\alpha_k\metric{i}(p_{\rvM\bar{\mathbf{Z}}_k},p_{\rvM}p_{\bar{\mathbf{Z}}_k})$. Therefore, a code ensuring secrecy for the mixed wiretap channel may not guarantee secrecy over each individual wiretap channel. If one wants to ensure secrecy over all possible $K$ channels, one must consider a \emph{compound wiretap channel}, in which the transmitter has no knowledge (even statistical knowledge) of which channel in the set is used for transmission; however, to avoid unnecessary mathematical complications, we assume that receivers can estimate channel statistics perfectly and always know from which channel they obtain observations; hence, we refer to this model as a compound channel with receiver \ac{CSI}. For every channel $k\in\intseq{1}{K}$, the performance of a code $\calC_n$ is measured in terms of the average probability of error $\perr^{\text{\tiny($k$)}}(\calC_n)$ and in terms of the secrecy metric $\metric{i}^{\text{\tiny($k$)}}(\calC_n)\eqdef\metric{i}(p_{\rvM\bar{\mathbf{Z}}_k},p_{\rvM}p_{\bar{\mathbf{Z}}_k})$; the notion of achievable rate is accordingly modified as follows.
\begin{definition}
  A rate $R$ is achievable over a compound wiretap channel with receiver \ac{CSI} for secrecy metric $\metric{i}$ if there exists a sequence of $(2^{nR},n)$ wiretap codes $\{\calC_n\}_{n\geq 1}$ such that
  \begin{align*}
    \forall k\in\intseq{1}{K}\quad\lim_{n\rightarrow\infty}\perr^{\text{\tiny($k$)}} (\calC_n)=0\quad\text{and}\quad\lim_{n\rightarrow\infty}\metric{i}^{\text{\tiny($k$)}} (\calC_n)=0.
  \end{align*}
\end{definition}
Unlike the mixed wiretap channel, there is no distribution associated to the choice of the channel in the set, and secrecy and reliability must be guaranteed for any realized channel.

\begin{proposition}
\label{prop:compound_general}
  The secrecy capacity of a compound wiretap channel with receiver \ac{CSI} and with cost constraint $P$ is the same for all secrecy metrics $\metric{i}$ with $i\in\intseq{2}{6}$ and is given by
\begin{multline}
  \label{eq:compound_general_secrecy_capacity}
  \sup_{{\left\{{\mathbf{V}},{\mathbf{X}}\right\}_{n\geq 1}}\in\calP}\left(\min_{k\in\intseq{1}{K}}\pliminf{\frac{1}{n}\I{{\mathbf{V}};{\mathbf{Y}}_k}}\right.\\
  \left.  -\max_{k\in\intseq{1}{K}}\plimsup{\frac{1}{n}\I{{\mathbf{V}};{\mathbf{Z}}_k}}\right),
\end{multline}
where
\begin{multline*}
  \calP\eqdef\left\{\smash{\left\{{\mathbf{V}}{\mathbf{X}}\right\}_{n\geq 1}}:\forall n\in\mathbb{N}^*, \;\forall k\in\intseq{1}{K},\,    {\mathbf{V}}\rightarrow{\mathbf{X}}\rightarrow{\mathbf{Y}}_k{\mathbf{Z}}_k\right.\\
\left.  \text{ forms a Markov chain and } \P{\tfrac{1}{n}c_n({\mathbf{X}})\leq P}=1  \right\}.
\end{multline*}
\end{proposition}
\begin{IEEEproof}
We start with the achievability part of the proof, which is similar to that of Theorem~\ref{th:bcc}. Let $n\in\mathbb{N}^*$ and $\epsilon,\gamma,R_1,R'>0$. Define $M_1\eqdef \lceil 2^{nR_1}\rceil$ and $M'\eqdef \lceil 2^{nR'}\rceil$. Fix a distribution $p_{{\mathbf{X}}}$ on $\calX^n$ such that $\P{\frac{1}{n}c_n({\mathbf{X}})\leq P}=1$. Let ${\mathbf{X}}$, $\{{\mathbf{Y}}_k\}_{k\in\intseq{1}{K}}$, $\{{\mathbf{Z}}_k\}_{k\in\intseq{1}{K}}$ be the random variables with joint distribution
\begin{align*}
  \forall k\in\intseq{1}{K}\;\;
  \p[{\mathbf{X}}{\mathbf{Y}}_k{\mathbf{Z}}_k]{{\mathbf{x}},{\mathbf{y}},{\mathbf{z}}} \eqdef W_{{\mathbf{Y}}_k{\mathbf{Z}}_k|{\mathbf{X}}}({\mathbf{y}},{\mathbf{z}}|{\mathbf{x}})\p[{\mathbf{X}}]{{\mathbf{x}}}.
\end{align*}

\begin{itemize}
\item \textbf{Code generation:} Randomly generate $M_1M'$ sequences ${\mathbf{x}}_{lm}\in\calX^n$ with $(l,m)\in\intseq{1}{M_1}\times\intseq{1}{M'}$ according to $p_{{\mathbf{X}}}$. We denote by $\rvC_n$ the random random variable representing the generated code and by $\calC_n$ one of its realizations.
\item \textbf{Encoding:} To transmit a message $l\in\intseq{1}{M_1}$, Alice generates an auxiliary message $m$ uniformly at random in $\intseq{1}{M'}$ and transmits the codeword $\mathbf{x}_{lm}$ through the channel.
\item \textbf{Bob's decoding for channel $k\in\intseq{1}{K}$:} Define the set
\begin{align*}
  \calT_k^n&\eqdef\left\{({\mathbf{x}},{\mathbf{y}})\in\calX^n\times\calY_k^n:\right. \\
  &\left.\phantom{----}\frac{1}{n}\log\frac{W_{{\mathbf{Y}}_k|{\mathbf{X}}}({\mathbf{y}}|{\mathbf{x}})}{\p[{\mathbf{Y}}_k]{{\mathbf{y}}}}\geq \frac{1}{n}\log M_1M'+\gamma\right\}.
\end{align*}
Note that the decoding rule depends on the channel index $k$ since we have assumed that Bob knows which channel is being observed. Upon observing $\mathbf{y}_k$, Bob decodes $l$ as the received individual message and $m$ as the received auxiliary message if there exists a unique codeword ${\mathbf{x}}_{lm}$ such that $\left({\mathbf{x}}_{lm},\mathbf{y}_k\right)\in\calT_k^n$; otherwise, a random message is chosen.
\end{itemize}
The following lemmas provide sufficient conditions to guarantee reliability and secrecy. Their proofs are similar to those provided in Appendix~\ref{sec:proof-theorem-bcc-achievability} and are omitted.
\begin{lemma}[Reliability conditions]
  \label{lm:reliability_compound}
  For each $k\in\intseq{1}{K}$,
  \begin{multline*}
    \text{If }R_1+R'\leq \pliminf{\frac{1}{n}\I{{\mathbf{X}};{\mathbf{Y}}_k}}-2\gamma\\
    \text{then} \lim_{n\rightarrow\infty}\E{\perr^{\text{\tiny($k$)}}(\rvC_n)}\leq \epsilon.
  \end{multline*}
\end{lemma}
\begin{lemma}[Secrecy from channel resolvability condition] For each $k\in\intseq{1}{K}$,
  \label{lm:secrecy_compound}
  \begin{align*}
        \text{If }R'\geq \plimsup{\frac{1}{n}\I{{\mathbf{X}};{\mathbf{Z}}_k}}+2\gamma\text{ then } \lim_{n\rightarrow\infty}\E{\metric{2}^{\text{\tiny($k$)}}(\rvC_n)}\leq \epsilon.
  \end{align*}
\end{lemma}
Using Lemmas~\ref{lm:reliability_compound} and~\ref{lm:secrecy_compound}, we obtain that if
\begin{multline*}
      R_1\leq \min_{k\in\intseq{1}{K}}\pliminf{\frac{1}{n}\I{{\mathbf{X}};{\mathbf{Y}}_k}}\\
      -\max_{k\in\intseq{1}{K}}\pliminf{\frac{1}{n}\I{{\mathbf{X}};{\mathbf{Z}}_k}} -4\gamma\\
      \text{then } \forall k\in\intseq{1}{K}\;
      \left\{
    \begin{array}{l}
      \lim_{n\rightarrow\infty}\E{\perr^{\text{\tiny($k$)}} (\rvC_n)}\leq \epsilon\\
      \lim_{n\rightarrow\infty}\E{\metric{2}^{\text{\tiny($k$)}} (\rvC_n)}\leq \epsilon
    \end{array}\right. .
\end{multline*}

Using Markov's inequality and the union bound, we can show there exists at least one sequence of $(2^{nR_1},n)$ codes $\{\calC_n\}_{n\geq 1}$ such that 
\begin{multline*}
  \forall k\in\intseq{1}{K}\quad \lim_{n\rightarrow\infty}{\perr^{\text{\tiny($k$)}} (\calC_n)}\leq (K+1)\epsilon\\
  \text{ and } \lim_{n\rightarrow\infty}{\metric{2}^{\text{\tiny($k$)}} (\calC_n)}\leq (K+1)\epsilon.
\end{multline*}
Since $K$ is fixed and $\epsilon,\gamma$ can be chosen arbitrarily small, we conclude that all rates $R_1$ such that
\begin{multline}
0\leq   R_1<\sup_{{\left\{{\mathbf{X}}\right\}_{n\geq 1}}\in\calP}\left(\min_{k\in\intseq{1}{K}}\pliminf{\frac{1}{n}\I{{\mathbf{X}};{\mathbf{Y}}_k}}\right.\\
\left.  -\max_{k\in\intseq{1}{K}}\plimsup{\frac{1}{n}\I{{\mathbf{X}};{\mathbf{Z}}_k}}\right)
\end{multline}
are achievable, where 
\begin{align*}
  \calP\eqdef\left\{\{\rvX_n\}_{n\geq
      1}:\P{\tfrac{1}{n}c_n({\mathbf{X}})\leq P}=1 \right\}.
\end{align*}
The achievability of the rates below the secrecy capacity $C_s^{\text{\tiny($2$)}}$ in~\eqref{eq:compound_general_secrecy_capacity} is then obtained by introducing a prefix channel as in the proof of Theorem~\ref{th:bcc}.

We now turn to the converse part of the proof. Consider a sequence of wiretap codes $\{\calC_n\}_{n\geq 1}$ achieving rate $R_1$ for secrecy metric $\metric{6}$. For $n\in\mathbb{N}^*$, let $\bar{\mathbf{V}}$ denote the choice of a message uniformly at random in $\intseq{1}{2^{nR_1}}$. By definition, for every $n\in\mathbb{N}^*$ and $k\in\intseq{1}{K}$, $\bar{\mathbf{V}}\rightarrow\bar{\mathbf{X}}\rightarrow\bar{\mathbf{Y}}_k\bar{\mathbf{Z}}_k$ forms a Markov chain and $\P{\frac{1}{n}c(\bar{\mathbf{X}})\leq P}=1$. By the Verd\'u-Han Lemma~\cite[Theorem 4]{Verdu1994}, we obtain
\begin{align}
  \label{eq:compound_converse_1}
  R_1\leq\min_{k\in\intseq{1}{K}}\pliminf{\frac{1}{n}\I{\bar{\mathbf{V}};\bar{\mathbf{Y}}_k}}.
\end{align}
By definition of the metric $\metric{6}$, we also have
\begin{align}
  \label{eq:compound_converse_2}
  \max_{k\in\intseq{1}{K}}\plimsup{\frac{1}{n}\I{\bar{\mathbf{V}};\bar{\mathbf{Z}}_k}}=0.
\end{align}
Subtracting~\eqref{eq:compound_converse_2}~to~\eqref{eq:compound_converse_1}, and maximizing over all processes $\{\bar{\mathbf{V}}\bar{\mathbf{X}}\}$, we obtain the desired result. 
\end{IEEEproof} 
Although the secrecy capacity of a compound wiretap channel with receiver \ac{CSI} is identical to that of a mixed wiretap channel, the coding schemes achieving it may be fundamentally different. 

\begin{proposition}
\label{prop:compound_memoryless}
  Given a memoryless compound wiretap channel with receiver \ac{CSI} and additive cost constraint $P$, all rates $R_1$ such that 
\begin{align}
  \label{eq:compound_memoryless_rates}
0\leq  R_1<\sup_{{(\rvV\rvX)}\in\calP}\left(\min_{k\in\intseq{1}{K}}\avgI{\rvV;\rvY_k}
  -\max_{k\in\intseq{1}{K}}\avgI{\rvV;\rvZ_k}\right)
\end{align}
are achievable for secrecy metrics $\metric{i}$ with $i\in\intseq{2}{6}$, where
\begin{multline*}
  \calP\eqdef\left\{\rvV\rvX: \;\forall k\in\intseq{1}{K},\, \rvV\rightarrow\rvX\rightarrow\rvY_k\rvZ_k\text{ forms}\right.\\
  \left.\text{ a Markov chain and } \E{c(\rvX)}\leq P \right\}.
\end{multline*}
If the random variables maximizing~\eqref{eq:compound_memoryless_rates} are such that, for all $k\in\intseq{1}{K}$, the integrals defining the moment generating functions of $\I{\rvV;\rvY_k}$ and $c(\rvX)$ converge uniformly in a neighborhood of $0$ and are differentiable at $0$, then the rates are also achievable for metric $\metric{1}$.
\end{proposition}
\begin{IEEEproof}
  The proof of Proposition~\ref{prop:compound_memoryless} follows from steps similar to those used in the proof of Proposition~\ref{prop:compound_general} and Theorem~\ref{th:dmc} and is omitted.
\end{IEEEproof}
If the receivers do not know which channel they observe, the counterpart of Proposition~6 was independently derived in~\cite{Bjelakovic2011}. Note that deriving secrecy from channel resolvability circumvents the enhancement argument used in~\cite[Theorem 1]{Liang2009}, which is required to show achievability using capacity-based wiretap codes. Similarly, when applied to Gaussian compound wiretap channels with power constraint, Proposition~\ref{prop:compound_memoryless} strengthens~\cite[Theorem 1]{Liu2008} with receiver \ac{CSI}.

\begin{remark}
  The general result of Proposition~\ref{prop:compound_general} holds provided the number of channels $K$ is fixed and independent of the number $n$ of channel uses; nevertheless, in the special case of Proposition~\ref{prop:compound_memoryless}, for which we establish secrecy for metric $\metric{1}$, we can show that, for each $k\in\intseq{1}{K}$, $\metric{1}^{\text{\tiny($k$)}}\leq (K+1)2^{-\epsilon_k n}$ for some $\epsilon_k>0$. Therefore, Proposition~\ref{prop:compound_memoryless} also holds if the number of compound channels grows exponentially with $n$ as $K=2^{\beta n}$ with $\beta<\min_{k\in\intseq{1}{K}}\epsilon_k$.
\end{remark}

\subsection{Secret-Key Agreement from General Sources.}
\label{sec:key-generation}

\begin{figure}[t]
  \centering
    \includegraphics[width=.98\linewidth]{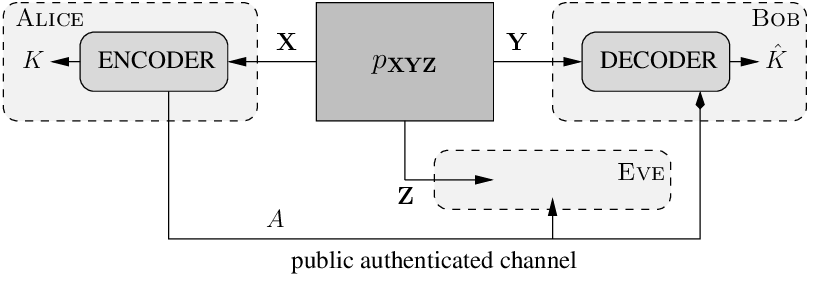}
  \caption{Secret-key agreement from general source.}
  \label{fig:ska}
\end{figure}

As a last application, we exploit a connection between secret-key agreement and wiretap coding to analyze the fundamental limits of secret-key agreement for a general source model. Specifically, we consider a \emph{discrete} source $\left(\smash{\calX,\calY,\calZ,\left\{p_{{\mathbf{X}}{\mathbf{Y}}{\mathbf{Z}}}\right\}_{n\geq 1}}\right)$ with three components taking values in {discrete} alphabets. As illustrated in Figure~\ref{fig:ska}, Alice and Bob attempt to distill a secret key from their correlated observations
${\mathbf{X}}$ and ${\mathbf{Y}}$, respectively, and a message transmitted by Alice over public authenticated channel with unlimited capacity. The key should remain secret from an eavesdropper who observes $\mathbf{Z}$ and the public message.

\begin{definition}
    A $(2^{nR},n)$ key-distillation strategy $\calS_n$ consists of:
  \begin{itemize}
  \item a key alphabet $\calK=\intseq{1}{2^{nR}}$;
  \item an alphabet $\calA$ used by Alice to communicate over the public channel;
    \item a source of local randomness\index{local~randomness} for Alice $(\calR_\calX,p_{\rvR_\rvX})$;
  \item a source of local randomness\index{local~randomness} for Bob $(\calR_\calY,p_{\rvR_\rvY})$;
    \item an encoding function $f:\calX^n\times\calR_\calX\rightarrow \calA$;
    \item a key-distillation function $\kappa_a:\calX^n\times\calR_\calX\rightarrow \calK$;
  \item a key-distillation function $\kappa_b:\calY^n\times\calA\times\calR_\calY\rightarrow \calK$;
  \end{itemize}
 \end{definition}
The random variables corresponding to the public message, Alice's key, and Bob's key are denoted by $\rvA$, $\rvK$, and $\hat\rvK$, respectively. The performance of a secret-key distillation strategy $\calS_n$ is measured in terms of the average probability of error $\perr(\calS_n)\eqdef\P{\rvK\neq\hat\rvK\middle\vert\calS_n}$, the secrecy of the key $\metric{i}(\calS_n)\eqdef \metric{i}(p_{\rvK{\mathbf{Z}}\rvA|\calS_n},p_{\rvK|\calS_n}p_{{\mathbf{Z}}\rvA|\calS_n})$ for $i\in\intseq{1}{6}$, and the uniformity of the key $\unif(\calS_n)\eqdef \log\lceil{2^{nR}}\rceil-\avgH{\rvK}$.
\begin{definition}
  A key rate $R$ is achievable for secrecy metric $\metric{i}$ for a source if there exists a sequence $\{\calS_n\}_{n\geq 1}$ of $\left(2^{nR},n\right)$ key-distillation strategies such that
  \begin{align*}
    \lim_{n\rightarrow\infty}\perr(\calS_n)=0,\quad     \lim_{n\rightarrow\infty}\metric{i}(\calS_n)=0,\quad    \lim_{n\rightarrow\infty}\unif(\calS_n)=0.
  \end{align*}
  The forward secret-key capacity $C^{\text{\tiny($i$)}}_{\textnormal{\tiny SK}}$ is the supremum of achievable key rates for metric $\metric{i}$.
\end{definition}

\begin{proposition}
  \label{prop:bound_source_model}
  The forward secret-key capacity of a discrete source $\left(\smash{\calX,\calY,\calZ,\left\{p_{{\mathbf{X}}{\mathbf{Y}}{\mathbf{Z}}}\right\}_{n\geq 1}}\right)$ for secrecy metrics $\metric{i}$ with $i\in\intseq{2}{6}$ satisfies
  \begin{multline}
    \pliminf{\frac{1}{n}\H{{\mathbf{X}}|{\mathbf{Z}}}}-\plimsup{\frac{1}{n}\H{{\mathbf{X}}|{\mathbf{Y}}}}
    \leq     C^{\text{\tiny($i$)}}_{\textnormal{\tiny SK}}\\
    \leq \min\left(\pliminf{\frac{1}{n}\I{{\mathbf{X}};{\mathbf{Y}}}},\pliminf{\frac{1}{n}\I{{\mathbf{X}};{\mathbf{Y}}|{\mathbf{Z}}}}\right). \label{eq:ska_bounds}
  \end{multline}
\end{proposition}
If the discrete source is i.i.d., Proposition~\ref{prop:bound_source_model} holds for secrecy metric $\metric{1}$, as already known from~\cite{Maurer2000,Csiszar1996}.
\begin{corollary}
  \label{cor:secret-key-agreement-memoryless}
  The secret-key capacity of an i.i.d discrete source $\left(\calX,\calY,\calZ,p_{\rvX\rvY\rvZ}\right)$ for secrecy metric $\metric{1}$ satisfies
  \begin{multline*}
    \max\left(\avgI{\rvX;\rvY}-\avgI{\rvX;\rvZ},\avgI{\rvX;\rvY}-\avgI{\rvY;\rvZ}\right) \\
    \leq C^{\text{\tiny(1)}}_{\textnormal{\tiny SK}}\leq \min\left(\avgI{\rvX;\rvY},\avgI{\rvX,\rvY|\rvZ}\right).
  \end{multline*}
\end{corollary}
\begin{IEEEproof}[Proof of Theorem~\ref{prop:bound_source_model} and Corollary~\ref{cor:secret-key-agreement-memoryless}]
  The achievability part of Theorem~\ref{prop:bound_source_model} is based on the construction of a conceptual wiretap channel as in~\cite{Maurer1993}. Assume that Alice, Bob and Eve observe $n$ realizations ${\mathbf{X}}$, ${\mathbf{Y}}$ and ${\mathbf{Z}}$ of the source, respectively. Consider an arbitrary process $\{\mathbf{U}\}_{n\geq1}$ such that $\mathbf{U}\in\calX^n$. Assume that Alice forms the signal ${\mathbf{U}}\oplus{\mathbf{X}}$ on the public channel, in which $\oplus$ denotes the symbol-wise modulo-$\calX$ addition. This operation creates a conceptual wiretap channel with input ${\mathbf{U}}$, in which Bob observes the outputs ${\mathbf{Y}}$ and ${\mathbf{U}}\oplus{\mathbf{X}}$ while Eve observes the outputs ${\mathbf{Z}}$ and ${\mathbf{U}}\oplus{\mathbf{X}}$. From Corollary~\ref{cor:wtc}, the secrecy
  capacity of this conceptual channel for secrecy metrics $\metric{i}$ with $i\in\intseq{2}{6}$ is at least
  \begin{multline*}
    \sup_{{\mathbf{U}}}\left(\pliminf{\frac{1}{n}\I{{\mathbf{U}};{\mathbf{Y}},{\mathbf{U}}\oplus{\mathbf{X}}}}\right.\\
\left.        -\plimsup{\frac{1}{n}\I{{\mathbf{U}};{\mathbf{Z}},{\mathbf{U}}\oplus{\mathbf{X}}}}\right).
    \end{multline*}
    In particular, we can choose for $\mathbf{U}$ an i.i.d. process such that, for all $j\in\mathbb{N}^*$, $\rvU_j$ is independent of ${\mathbf{X}}{\mathbf{Y}}{\mathbf{Z}}$  and uniformly distributed in $\calX$. Then, with probability one,
\begin{align*}
  \I{{\mathbf{U}};{\mathbf{Y}},{\mathbf{U}}\oplus{\mathbf{X}}} &= \log
  \frac{\p[{\mathbf{U}}\oplus{\mathbf{X}},{\mathbf{Y}}|{\mathbf{U}}]{{\mathbf{U}}\oplus{\mathbf{X}},{\mathbf{Y}}|{\mathbf{U}}}}{\p[{\mathbf{U}}\oplus{\mathbf{X}},{\mathbf{Y}}]{{\mathbf{U}}\oplus{\mathbf{X}},{\mathbf{Y}}}}\\
    &= \log \frac{\p[{\mathbf{X}}|{\mathbf{Y}}{\mathbf{U}}]{{\mathbf{X}}|{\mathbf{Y}}{\mathbf{U}}}\p[{\mathbf{Y}}|{\mathbf{U}}]{{\mathbf{Y}}|{\mathbf{U}}}}{\p[{\mathbf{U}}\oplus{\mathbf{X}}|{\mathbf{Y}}]{{\mathbf{U}}\oplus{\mathbf{X}}|{\mathbf{Y}}}\p[{\mathbf{Y}}]{{\mathbf{Y}}}}\\
    &= \log \p[{\mathbf{X}}|{\mathbf{Y}}]{{\mathbf{X}}|{\mathbf{Y}}} - \log\frac{1}{\card{\calX}^n},
\end{align*}
where the last inequality follows from $\p[{\mathbf{Y}}|{\mathbf{U}}]{{\mathbf{Y}}|{\mathbf{U}}}=\p[{\mathbf{Y}}]{{\mathbf{Y}}}$,
$\p[{\mathbf{X}}|{\mathbf{Y}}{\mathbf{U}}]{{\mathbf{X}}|{\mathbf{Y}}{\mathbf{U}}}=\p[{\mathbf{X}}|{\mathbf{Y}}]{{\mathbf{X}}|{\mathbf{Y}}}$ since ${\mathbf{U}}$ is independent of
${\mathbf{X}}{\mathbf{Y}}$ and $\p[{\mathbf{U}}\oplus{\mathbf{X}}|{\mathbf{Y}}]{{\mathbf{U}}\oplus{\mathbf{X}}|{\mathbf{Y}}}=\frac{1}{\card{\calX}^n}$ by the crypto lemma~\cite{Forney2003}. Therefore,
\begin{multline}
  \pliminf{\frac{1}{n}\I{{\mathbf{U}};{\mathbf{Y}},{\mathbf{U}}\oplus{\mathbf{X}}}}\\
  =\log\card{\calX}-\plimsup{\frac{1}{n}\H{{\mathbf{X}}|{\mathbf{Y}}}}. \label{eq:ska_lb_1}
\end{multline}
Similarly, one obtains
\begin{multline}
  \plimsup{\frac{1}{n}\I{{\mathbf{U}};{\mathbf{Z}},{\mathbf{U}}\oplus{\mathbf{X}}}}\\
  =\log\card{\calX}-\pliminf{\frac{1}{n}\H{{\mathbf{X}}|{\mathbf{Z}}}}.
  \label{eq:ska_lb_2}
\end{multline}
Combining~\eqref{eq:ska_lb_1}
and~\eqref{eq:ska_lb_2}, we conclude that any rate $R$ such that 
\begin{align*}
  R<\pliminf{\frac{1}{n}\H{{\mathbf{X}}|{\mathbf{Z}}}}-\plimsup{\frac{1}{n}\H{{\mathbf{X}}|{\mathbf{Y}}}}
\end{align*} 
is an achievable rate for the conceptual wiretap channel. Since this channel allows one to transmit uniformly distributed messages, $R$ is also an achievable secret-key rate for the source model. For i.i.d. discrete sources, a similar proof based on Corollary~\ref{cor:strong_secrecy_memoryless} in place of Corollary~\ref{cor:wtc} shows that the result holds for metric $\metric{1}$ as well. The proof of the converse is an information-spectrum version of the converse in~\cite{Ahlswede1993} and is omitted for brevity.
\end{IEEEproof}

In Proposition~\ref{prop:bound_source_model},  achievable key rates are expressed in terms of conditional entropy; except in some special cases, such as i.i.d. sources, this is rather different from the achievable secrecy rates for wiretap channels in Corollary~\ref{cor:wtc}, which are expressed in terms of mutual information. In particular, if $\pliminf{\frac{1}{n}\H{{\mathbf{X}}}}=\plimsup{\frac{1}{n}\H{{\mathbf{X}}}}$, then, 
  \begin{multline*}
    \pliminf{\frac{1}{n}\H{{\mathbf{X}}|{\mathbf{Z}}}}-\plimsup{\frac{1}{n}\H{{\mathbf{X}}|{\mathbf{Y}}}}\\
    \geq \pliminf{\frac{1}{n}\I{{\mathbf{X}};{\mathbf{Y}}}}-\plimsup{\frac{1}{n}\I{{\mathbf{X}};{\mathbf{Z}}}}.
  \end{multline*}
This distinction suggests that the coding mechanism for secret-key distillation, which one would have to exploit to design secret-key distillation strategies without relying on the existence of wiretap codes, is not linked to channel resolvability; indeed, the first author has argued in a previous work that secret-key distillation is more easily understood in terms of \emph{channel intrinsic randomness}~\cite{Csiszar1996,Bloch2010b} and privacy amplification~\cite{Bennett1995,Hayashi2012}. In that respect, the proof of Proposition~\ref{prop:bound_source_model} provides limited insight into the design of practical secret-key distillation strategies.

\section{Conclusion}
\label{sec:conclusion}
We have analyzed several models of secure communication by building upon the work of Csisz\'ar~\cite{Csiszar1996} and Hayashi~\cite{Hayashi2006} and by exploiting the idea that the coding mechanism to ensure secrecy can be tied to channel resolvability. This approach has allowed us to establish several results for generic channels and for stronger secrecy metrics than the usual average mutual information rate between messages and eavesdropper's observations.

From a technical point of view, deriving secrecy from channel resolvability provides a conceptually simple approach to analyze the secure achievable rates of many models. Although we have limited applications to mixed, compound, and wireless channels, the connection between secrecy and channel resolvability is useful in many other settings. Examples of secure communication models for which deriving secrecy from channel resolvability simplifies the analysis include queuing channels~\cite{Dunn2009}, wireless channels with imperfect state information~\cite{Bloch2009,Bloch2012b}, runlength-limited channels ~\cite{Sankarasubramaniam2009}, and two-way wiretap channels~\cite{Pierrot2011a}. 

From a practical perspective, we believe that the connection between strong secrecy and channel resolvability opens intriguing perspectives for code design. In particular, we have provided evidence that this connection circumvents a weakness of capacity-based wiretap codes, which cannot always achieve the strong secrecy capacity. This observation is consistent with practical code constructions achieving strong secrecy rates~\cite{Mahdavifar2011,Subramanian2011} and other approaches based on privacy amplification~\cite{Maurer2000,Matsumoto2010arXiv,Hayashi2011}.

Our results can be extended in several directions. For instance, the coding mechanisms for secrecy presented in Section~\ref{sec:shann-ciph-syst} for Shannon's cipher system and in Section~\ref{sec:arbitr-wiret-chann} for wiretap channels can be combined without much difficulty using a coding scheme similar to that proposed in~\cite{Yamamoto1997}. One could also further investigate the nature of the coding mechanisms for secrecy in secret-key agreement models. Some results along these lines are already available, for instance in~\cite{Csiszar1996,Bloch2010b,Watanabe2010a}.

\appendices

\section{Supporting Lemmas}
\label{sec:technical-lemmas}

\begin{lemma}[Chernov bound]
  \label{lm:chernov}
  Let $\rvX$ be a real-valued random variable with moment generating function $\phi_\rvX:\mathbb{R}\rightarrow\mathbb{R}:s\mapsto \E{e^{s\rvX}}$. Let $\{\rvX_i\}_{i=1}^n$ be i.i.d. with distribution $p_\rvX$. If the integral defining $\phi_\rvX$ converges uniformly in a neighborhood of $0$ and is differentiable at $0$ then, $      \forall \epsilon>0\;\exists \alpha_\epsilon>0$ such that
    \begin{align*}
\P{\frac{1}{n}\sum_{i=1}^n\rvX_i>\E{\rvX}+\epsilon}\leq 2^{-\alpha_\epsilon n}.
    \end{align*}
\end{lemma}

\begin{lemma}[Basic properties of variational distance]
  \label{lm:subadditivity_basics}
  Let $\rvX_1$, $\rvX_2$, and $\rvX_3$ be random variables defined on the same alphabet $\calX$. Then,
  \begin{align*}
    \V{p_{\rvX_1};p_{\rvX_3}}&\leq \V{p_{\rvX_1};p_{\rvX_2}}+\V{p_{\rvX_2};p_{\rvX_3}},\\
    \mbox{and}\quad\V{p_{\rvX_1};p_{\rvX_2}} &\leq \V{p_{\rvX_1}p_{\rvX_3};p_{\rvX_2\rvX_3}}\\&=\E[\rvX_3]{\V{p_{\rvX_1},p_{\rvX_2|\rvX_3}}}.
  \end{align*}
\end{lemma}

\begin{lemma}[Data-processing inequality for variational distance]
  \label{lm:dpi_variational}
  Let $\rvX_1$ and $\rvX_2$ be random variables defined on the same alphabet $\calX$. Let $W_{\rvZ|\rvX}$ be the transition probability from $\calX$ to $\calZ$ and define the random variables $\rvZ_1$ and $\rvZ_2$ such that
  \begin{multline*}
    \forall (\svz,\svx)\in\calZ\times\calX\quad p_{\rvZ_1\rvX_1}(\svz,\svx) = W_{\rvZ|\rvX}(\svz|\svx)p_{\rvX_1}(\svx)\\
    \text{ and }p_{\rvZ_2\rvX_2}(\svz,\svx) = W_{\rvZ|\rvX}(\svz|\svx)p_{\rvX_2}(\svx).
  \end{multline*}
  Then, $\V{p_{\rvZ_1},p_{\rvZ_2}} \leq     \V{p_{\rvX_1},p_{\rvX_2}}$.
\end{lemma}

\section{Proof of Proposition~\ref{prop:capac-achi-wiret}}
\label{sec:proof-proposition-weak-strong}
Let $\rvC_n$ be the random variable that denotes a randomly generated capacity-based wiretap code, whose codeword symbols are generated i.i.d. according to the uniform distribution $q_{X}$.  Let $p_{Z}$ be the output distribution of the eavesdropper's channel corresponding to the input on $\calX$, i.e. 
\begin{align*}
\forall  z\in\calZ, \;p_Z(z) = \sum_{x\in\calX}W_{\rvZ|\rvX}(z|x)\frac{1}{\card{\calX}}.
\end{align*}
Let $p_{\mathbf{Z}}$ be the distribution of $n$ i.i.d. random variables distributed according to $p_Z$. The proof of the proposition relies on the following lemmas.

\begin{lemma}
  \label{lm:no_redundancy}
  Consider $M_n\eqdef 2^{nR}$ codewords of length $n$, obtained by generating codeword symbols independently and uniformly at random in $\calX$. If $R<\tfrac{1}{2}\log{\card{\calX}}$,  there exists $\alpha_0>0$ such that the probability that all $M_n$ codewords are distinct satisfies
  \begin{align*}
    \P{\text{all $M_n$ codewords are distinct}}\geq 1- 2^{-\alpha_0 n}.
  \end{align*}
\end{lemma}
\begin{IEEEproof}
  The proof follows from the same technique as in~\cite[Lemma 6]{Ozarow1984}, which we recall for convenience. Note that,
    \begin{multline*}
      \mathbb{P}\left(\text{all $M_n$ codewords are distinct}\right)\\
      = \prod_{i=0}^{M_n-1}\frac{\card{\calX}^n-i}{\card{\calX}^n} = \prod_{i=0}^{M_n-1}\left(1-\frac{i}{\card{\calX}^n}\right)
    \end{multline*}
    Since $\ln(1-x)\geq \frac{-x}{1-x}$ for $x\in[0,1)$, we have
    \begin{multline*}
      \mathbb{P}\left(\text{all $M_n$ codewords are distinct}\right)\\
      \begin{split}
        &\geq \exp\left(-\sum_{i=0}^{M_n-1}\frac{i}{\card{\calX}^n-i}\right)\\
        &\geq
        \exp\left(-\frac{(M_n-1)(M_n-1)}{\card{\calX}^n-(M_n-1)}\right).
      \end{split}
    \end{multline*}
    Since $e^{-x}\geq 1-x$, we obtain
    \begin{multline*}
      \mathbb{P}\left(\text{all $M_n$ codewords are distinct}\right) \\
      \geq 1-\frac{(M_n-1)^2}{\card{\calX}^n-(M_n-1)}\geq 1-\frac{M_n^2}{\card{\calX}^{n}-M_n}
    \end{multline*}
    Substituting $M_n=2^{nR}$, we obtain
    \begin{align*}
      \mathbb{P}\left(\text{all $M_n$ codewords are distinct}\right) \geq 1-\frac{2^{2nR}}{\card{\calX}^n-2^{nR}},
    \end{align*}
    which goes to $1$ as $n$ goes to infinity provided $R<\tfrac{1}{2}\log{\card{\calX}}$.
\end{IEEEproof}
\begin{lemma}
  \label{lm:hp_weak_secrecy}
  There exists $\alpha_1>0$, such that, for $n$ sufficiently large,
  \begin{align*}
    \P{\perr^*(\rvC_n)\leq \epsilon'_n\;\;\text{and}\;\;\metric{4}(\rvC_n)\leq 3\epsilon'_n} \geq 1-2^{-\alpha_1 n\epsilon_n^2},
  \end{align*}
  with $\epsilon'_n\eqdef\max(\epsilon_n, \log\card{\calX}2^{-\alpha_1 n\epsilon_n^2}, n^{-1})$.
\end{lemma}
\begin{IEEEproof}
  The existence of $\alpha_1>0$  such that $\P{\perr^*(\rvC_n)\leq 2^{-\alpha_1 n\epsilon_n^2}} \geq 1-2^{-\alpha_1 n\epsilon_n^2}$ follows from a standard random coding argument. Consider a code $\calC_n$ such that ${\perr^*(\calC_n)\leq 2^{-\alpha_1 n\epsilon_n^2}}$. Then, for $n$ large enough,
    \begin{align*}
      \metric{4}(\calC_n) &= \tfrac{1}{n}\avgI{\rvM_1;\bar{\mathbf{Z}}}\\
      &= \tfrac{1}{n}\avgI{\rvM_1\rvM';\bar{\mathbf{Z}}}-\tfrac{1}{n}\avgI{\rvM';\bar{\mathbf{Z}}|\rvM_1}\\
      &\stackrel{(a)}{\leq}\tfrac{1}{n}\avgI{\bar{\mathbf{X}};\bar{\mathbf{Z}}} -\tfrac{1}{n}\avgH{\rvM'|\rvM_1}+\tfrac{1}{n}\avgH{\rvM'|\rvM_1\bar{\mathbf{Z}}}\\
      &\stackrel{(b)}{\leq} C_e - (C_e-\epsilon_n) + R'\perr^*(\calC_n)+\tfrac{1}{n}\\
      &\leq 3\epsilon'_n
    \end{align*}
    where $(a)$ follows because $\rvM_1\rvM'\rightarrow \bar{\mathbf{X}}\rightarrow\bar{\mathbf{Z}}$ forms a Markov chain, and $(b)$ follows from Fano's inequality.
\end{IEEEproof}
\begin{lemma}
  \label{lm:hp_resolvability}
There exists $\beta,\alpha_2>0$, such that, for $n$ large enough
  \begin{align*}
    \P{\V{p_{\bar{\mathbf{Z}}},p_{{\mathbf{Z}}}}\leq 2^{-\beta n}} \geq 1-2^{-\alpha_2 n}.
  \end{align*}
\end{lemma}
\begin{IEEEproof}
  This result follows from~\cite[Theorem 6.3.1]{InformationSpectrumMethods} by remarking that memoryless channels are exponentially information stable or, alternatively, from~\cite[Lemma 19]{Cuff2009}.
\end{IEEEproof}
For $n\in\mathbb{N}^*$, let $\calC_n$ denote a randomly generated code such that all codewords are distinct and
\begin{align}
  \perr^*(\calC_n)\leq \epsilon'_n,\;\metric{4}(\calC_n)\leq 3\epsilon'_n, \;\text{and}\;\V{p_{\bar{\mathbf{Z}}},p_{{\mathbf{Z}}}}\leq 2^{-\beta n}.\label{eq:conditions_cn}
\end{align}
For $n$ large enough, Lemma~\ref{lm:no_redundancy}, Lemma~\ref{lm:hp_weak_secrecy}, and Lemma~\ref{lm:hp_resolvability} guarantee that this occurs with probability at least $1-2^{-\alpha n\epsilon_n^2}$ for some  $\alpha<\alpha_1$ and $n$ large enough. With a slight abuse of notation, we also let $\calC_n\subset\calX^n$ denote the codebook and let $f_n^{-1}:\calC_n\rightarrow\calM_1$ be the restriction to $\calM_1$ of the inverse mapping of $f_n$, which is well defined since codewords are distinct. Let us introduce the functions $\phi_n$ and $\psi_n$ as
  \begin{align*}
    \phi_n:&
      \calC_n\rightarrow\calM_1: {\mathbf{x}}\mapsto f_n^{-1}({\mathbf{x}})\\
    \text{and}\quad
    \psi_n:&
      \calZ^n\times \calM_1\rightarrow\calC_n:      ({\mathbf{z}},m)\mapsto f_n(m,h_n({\mathbf{z}},m)).
  \end{align*}
  The functions $\phi_n$ and $\psi_n$ define the encoder and decoder of a source code for the compression of the source $\bar{\mathbf{X}}\in\calC_n$ (the choice of codewords uniformly at random in the code) with $\bar{\mathbf{Z}}$ as correlated side information at the receiver, whose probability of decoding error is $\perr^*(\calC_n)$. We now leverage the results obtained by Hayashi~\cite{Hayashi2008} and generalized by Watanabe~\emph{et al.}~\cite{Watanabe2010a} that establish a tradeoff between probability and error and resolvability for source coding of arbitrary sources. Combining~\cite[Theorem 10]{Watanabe2010a} and the proof of~\cite[Theorem 11]{Watanabe2010a}, we obtain, $\forall b>0$, $\forall n\in\mathbb{N}^*$
    \begin{align}
\perr^*(\calC_n) + \metric{2}(\calC_n)
      \geq 1-\left(2^{-b\sqrt{n}+1}+\P[\bar{\mathbf{X}}\bar{\mathbf{Z}}]{\calA_n}\right),\label{eq:bound_tradeoff}
    \end{align}
    with
    \begin{align*}
      \calA_n\eqdef\left\{({\mathbf{x}},{\mathbf{z}})\in\calC_n\times\calZ^n:\frac{2^{-b\sqrt{n}}}{\card{\calM_1}}<\p[\bar{\mathbf{X}}|\bar{\mathbf{Z}}]{{\mathbf{x}}|{\mathbf{z}}}\leq \frac{2^{b\sqrt{n}}}{\card{\calM_1}}\right\}.
\end{align*}
Note that $\card{\calM'}=2^{n(C_e-\epsilon_n)}$ and $\p[\bar{\mathbf{X}}]{\bar{\mathbf{X}}}=\frac{1}{\card{\calM_1}\card{\calM'}}$. Therefore, by Bayes's rule
  \begin{align*}
    \P[\bar{\mathbf{X}}\bar{\mathbf{Z}}]{\calA_n} 
      &= \P[\bar{\mathbf{X}}\bar{\mathbf{Z}}]{\frac{2^{-b\sqrt{n}}}{\card{\calM_1}}<\p[\bar{\mathbf{X}}|\bar{\mathbf{Z}}]{\bar{\mathbf{X}}|\bar{\mathbf{Z}}}\leq \frac{2^{b\sqrt{n}}}{\card{\calM_1}}}\nonumber{}\\
      &= \P[\bar{\mathbf{X}}\bar{\mathbf{Z}}]{\frac{2^{-b\sqrt{n}}}{\card{\calM_1}}<W_{{\mathbf{Z}}|{\mathbf{X}}}(\bar{\mathbf{Z}}|\bar{\mathbf{X}})\frac{\p[\bar{\mathbf{X}}]{\bar{\mathbf{X}}}}{\p[\bar{\mathbf{Z}}]{\bar{\mathbf{Z}}}}\leq
        \frac{2^{b\sqrt{n}}}{\card{\calM_1}}} \nonumber{}\\
      &=\P[\bar{\mathbf{X}}\bar{\mathbf{Z}}]{\calQ_n^+}-\P[\bar{\mathbf{X}}\bar{\mathbf{Z}}]{\calQ_n^-},
    \end{align*}
    where we have defined
    \begin{multline*}
      \calQ_n^\pm\eqdef\left\{({\mathbf{x}},{\mathbf{z}})\in\calC_n\times\calZ^n:\log\frac{W_{{\mathbf{Z}}|{\mathbf{X}}}({\mathbf{z}}|{\mathbf{x}})}{\p[\bar{\mathbf{Z}}]{{\mathbf{z}}}}\right.\\
      \left.\leq
        \pm b\sqrt{n}+n(C_e-\epsilon_n)\right\}.
    \end{multline*}
    
    We analyze $\P[\bar{\mathbf{X}}\bar{\mathbf{Z}}]{\calQ_n^+}$ and $\P[\bar{\mathbf{X}}\bar{\mathbf{Z}}]{\calQ_n^-}$ by introducing the sets
    \begin{align*}
      \calA_n^{\pm}&\eqdef\left\{({\mathbf{x}},{\mathbf{z}})\in\calC_n\times\calZ^n:\log\frac{W_{{\mathbf{Z}}|{\mathbf{X}}}({\mathbf{z}}|{\mathbf{x}})}{\p[{\mathbf{Z}}]{{\mathbf{z}}}}\right.\\
      & \left. \phantom{----------}\leq \pm 2b\sqrt{n}+n(C_e-\epsilon_n)\right\}\\
      \calB_n&\eqdef \left\{({\mathbf{x}},{\mathbf{z}})\in\calC_n\times\calZ^n:\log\frac{\p[\bar{\mathbf{Z}}]{{\mathbf{z}}}}{\p[{\mathbf{Z}}]{{\mathbf{z}}}}< b\sqrt{n}\right\}\\
      \text{and}\quad   \calD_n&\eqdef \left\{({\mathbf{x}},{\mathbf{z}})\in\calC_n\times\calZ^n:\log\frac{\p[\bar{\mathbf{Z}}]{{\mathbf{z}}}}{\p[{\mathbf{Z}}]{{\mathbf{z}}}}> -b\sqrt{n}\right\}.
    \end{align*}

    Using the law of total probability and the fact that $\calQ_n^+\cap\calB_n\subset\calA_n^+$, we now upper bound $\P[\bar{\mathbf{X}}\bar{\mathbf{Z}}]{\calQ_n^+}$ as follows.
    \begin{align}
      \P[\bar{\mathbf{X}}\bar{\mathbf{Z}}]{\calQ_n^+} &=  \P[\bar{\mathbf{X}}\bar{\mathbf{Z}}]{\calQ_n^+\cap \calB_n} +       \P[\bar{\mathbf{X}}\bar{\mathbf{Z}}]{\calQ_n^+\cap\calB_n^c}\nonumber{}\\
      & \leq \P[\bar{\mathbf{X}}\bar{\mathbf{Z}}]{\calA_n^+} + \P[\bar{\mathbf{X}}\bar{\mathbf{Z}}]{\calB_n^c}\label{eq:bound_qn_+}.
    \end{align}
    We first establish a bound on $\P[\bar{\mathbf{X}}\bar{\mathbf{Z}}]{\calB_n^c}$.
    \begin{align}
     \P[\bar{\mathbf{X}}\bar{\mathbf{Z}}]{\calB_n^c} &=\frac{1}{b\sqrt{n}}\sum_{{\mathbf{z}}\in\calZ^n}b\sqrt{n}\,\p[\bar{\mathbf{Z}}]{{\mathbf{z}}}\mathds{1}\left\{\log\frac{\p[\bar{\mathbf{Z}}]{{\mathbf{z}}}}{\p[{\mathbf{Z}}]{{\mathbf{z}}}}\geq b\sqrt{n}\right\}\nonumber{}\\
      &\leq \frac{1}{b\sqrt{n}}\sum_{{\mathbf{z}}\in\calZ^n}\p[\bar{\mathbf{Z}}]{{\mathbf{z}}}\log\frac{\p[\bar{\mathbf{Z}}]{{\mathbf{z}}}}{\p[{\mathbf{Z}}]{{\mathbf{z}}}}\nonumber{}\\
      &=\frac{1}{b\sqrt{n}}\avgD{p_{\bar{\mathbf{Z}}}}{p_{{\mathbf{Z}}}}.
    \end{align}
    We define $\mu_Z\eqdef \min_{z\in\calZ:p_{Z}(z)>0}p_{Z}(z)$ and we upper bound the divergence as follows.
    \begin{align}
      \avgD{p_{\bar{\mathbf{Z}}}}{p_{{\mathbf{Z}}}} &= -\avgH{\bar{\mathbf{Z}}} + \sum_{\mathbf{z}\in\calZ^n}p_{\bar{\mathbf{Z}}}(\mathbf{z})\log \frac{1}{p_{{\mathbf{Z}}}(\mathbf{z})}\nonumber{}\\
      &=\avgH{{\mathbf{Z}}}-\avgH{\bar{\mathbf{Z}}} + \sum_{\mathbf{z}\in\calZ^n}\left(p_{\bar{\mathbf{Z}}}(\mathbf{z})-p_{\mathbf{Z}}(\mathbf{z})\right)\log \frac{1}{p_{{\mathbf{Z}}}(\mathbf{z})}\nonumber{}\\
      &\leq \abs{\avgH{{\mathbf{Z}}}-\avgH{\bar{\mathbf{Z}}}} + n \V{p_{\bar{\mathbf{Z}}},p_{{\mathbf{Z}}}}\log\frac{1}{\mu_Z}\nonumber{}\\
      &\stackrel{(a)}{\leq} \V{p_{\bar{\mathbf{Z}}},p_{{\mathbf{Z}}}}\log\frac{\card{\calZ}^n}{\V{p_{\bar{\mathbf{Z}}},p_{{\mathbf{Z}}}}}+ n \V{p_{\bar{\mathbf{Z}}},p_{{\mathbf{Z}}}}\log\frac{1}{\mu_Z}\nonumber{}\\
      &\stackrel{(b)}{\leq} \left(\log \card{\calZ} + \beta+\log\frac{1}{\mu_Z}\right)n2^{-\beta n}\label{eq:bound_bnc}
    \end{align}
    where $(a)$ follows from~\cite[Lemma 2.7]{CodingTheoremsDMC2} and $(b)$ follows from the fact that $x\mapsto x\log \frac{\card{\calZ}^n}{x}$ is monotonously increasing for $x$ small enough. 

    To upper bound $\P[\bar{\mathbf{X}}\bar{\mathbf{Z}}]{\calA_n^+}$, recall that the eavesdropper's channel is symmetric; hence, there exists a partition $\{\calZ_i\}_{i\in\intseq{1}{k}}$ of $\calZ$ such that:
    \begin{enumerate}
    \item $\forall x,\tilde{x}\in\calX$, there exists a permutation $\pi_{x\tilde{x}}:\calZ\rightarrow\calZ$ that satisfies
      \begin{multline*}
        \forall i\in\intseq{1}{k}\;\pi_{x\tilde{x}}(\calZ_i) = \calZ_i\\
        \forall z\in\calZ\;W_{\rvZ|\rvX}(z|x) = W_{\rvZ|\rvX}(\pi_{x\tilde{x}}(z)|\tilde{x})
      \end{multline*}
    \item The output distribution $p_Z$ correponding to a uniform input distribution is locally uniform, i.e.
      \begin{align*}
        \forall i\in\intseq{1}{k}\;,\forall z,z'\in\calZ_i\; p_Z(z)=p_Z(z').
      \end{align*}
    \end{enumerate}
    Consequently, upon defining $b_n\eqdef 2b\sqrt{n}+n(C_e-\epsilon_n)$ and for any $\tilde{x}\in\calX$, we can rewrite $\P[\bar{\mathbf{X}}\bar{\mathbf{Z}}]{\calA_n^+}$ as shown on top of the next page,
\begin{figure*}[!t]
\normalsize
\begin{align*} \P[\bar{\mathbf{X}}\bar{\mathbf{Z}}]{\calA_n^+}&=\sum_{\mathbf{x}\in\calC_n}\frac{1}{\card{\calC_n}}\sum_{\mathbf{z}\in\calZ^n}W_{\mathbf{Z}|\mathbf{X}}(\mathbf{z}|\mathbf{x})\mathds{1}\left\{\log\frac{W_{{\mathbf{Z}}|{\mathbf{X}}}({\mathbf{z}}|{\mathbf{x}})}{\p[{\mathbf{Z}}]{{\mathbf{z}}}}\leq b_n\right\}\\
  &=\sum_{\mathbf{x}\in\calC_n}\frac{1}{\card{\calC_n}}\sum_{\mathbf{z}\in\calZ^n}\left(\prod_{i=1}^n W_{\rvZ|\rvX}(z_i|x_i)\right)\mathds{1}\left\{\sum_{i=1}^n \log\frac{W_{\rvZ|\rvX}(z_i|x_i)}{\p[\rvZ]{z_i}}\leq b_n\right\}\\
  &\stackrel{(a)}{=}\sum_{\mathbf{x}\in\calC_n}\frac{1}{\card{\calC_n}}\sum_{\mathbf{z}\in\calZ^n}\left(\prod_{i=1}^n W_{\rvZ|\rvX}(\pi_{x_i\tilde{x}}(z_i)|\tilde{x})\right)\mathds{1}\left\{\sum_{i=1}^n \log\frac{W_{\rvZ|\rvX}(\pi_{x_i\tilde{x}}(z_i)|\tilde{x})}{\p[\rvZ]{\pi_{x_i\tilde{x}}(z_i)}}\leq b_n\right\}\\
  &\stackrel{(b)}{=}\sum_{\mathbf{x}\in\calC_n}\frac{1}{\card{\calC_n}}\sum_{\mathbf{z}\in\calZ^n}\left(\prod_{i=1}^n W_{\rvZ|\rvX}(z_i|\tilde{x})\right)\mathds{1}\left\{\sum_{i=1}^n \log\frac{W_{\rvZ|\rvX}(z_i|\tilde{x})}{\p[\rvZ]{z_i}}\leq b_n\right\}\\
  &\stackrel{(c)}{=}\P{\sum_{i=1}^n\log\frac{W_{\rvZ|\rvX}(\tilde{\rvZ}_i|\tilde{x})}{p_Z(\tilde{\rvZ}_i)}\leq b_n}
\end{align*}
\hrulefill
\vspace*{4pt}
\end{figure*}
where $(a)$ follows because the eavesdropper's channel is symmetric, $(b)$ follows because the functions $\pi_{x_i\tilde{x}}$ are permutations, and $(c)$ follows by defining the i.i.d. random variables $\tilde{Z}_i$ as the eavesdropper's channel output when the channel input is the symbol $\tilde{x}$. Note that the random variables $\log\frac{W_{\rvZ|\rvX}(\tilde{\rvZ}_i|\tilde{x})}{p_Z(\tilde{\rvZ}_i)}$ are also i.i.d., with mean $C_e$ since the channel is symmetric, variance $\sigma>0$, and third moment $\rho<\infty$; therefore,
    \begin{multline*}
      \P[\bar{\mathbf{X}}\bar{\mathbf{Z}}]{\calA_n^+}=\mathbb{P}\left[\frac{1}{\sqrt{n}\sigma}\sum_{i=1}^n\left(\log\frac{W_{\rvZ|\rvX}(\tilde{\rvZ}_i|\tilde{x})}{p_Z(\tilde{\rvZ}_i)}-C_e\right)\right.\\
      \left.\leq \frac{2b}{\sigma}-\frac{\sqrt{n}\epsilon_n}{\sigma}\right].
    \end{multline*}
    From the Berry-Esseen Theorem~\cite{Shiryaev1995}, there exists a universal constant $c>0$ such that
      \begin{align}
            \P[\bar{\mathbf{X}}\bar{\mathbf{Z}}]{\calA_n^+}&\leq \frac{1}{\sqrt{2\pi}}\int_{-\infty}^{\frac{2b}{\sigma}-\frac{\sqrt{n}\epsilon_n}{\sigma}}e^{-\frac{x^2}{2}}\dd x +\frac{c}{\sqrt{n}}\frac{\rho}{\sigma^3}.\label{eq:berry_esseen_1}
      \end{align}

    Similarly, using the law of total probability, the fact that $\calA_n^-\cap\calD_n\subset \calQ_n^-\cap\calD_n$, and the inclusion-exclusion principle, we lower bound $\P[\bar{\mathbf{X}}\bar{\mathbf{Z}}]{\calQ_n^-}$ as follows.
    \begin{align}
      \P[\bar{\mathbf{X}}\bar{\mathbf{Z}}]{\calQ_n^-} &=  \P[\bar{\mathbf{X}}\bar{\mathbf{Z}}]{\calQ_n^-\cap \calD_n} + \P[\bar{\mathbf{X}}\bar{\mathbf{Z}}]{\calQ_n^-\cap\calD_n^c}\nonumber{}\\
      & \geq \P[\bar{\mathbf{X}}\bar{\mathbf{Z}}]{\calA_n^-\cap\calD_n}\nonumber{}\\
      & = \P[\bar{\mathbf{X}}\bar{\mathbf{Z}}]{\calA_n^-}+\P[\bar{\mathbf{X}}\bar{\mathbf{Z}}]{\calD_n}-\P[\bar{\mathbf{X}}\bar{\mathbf{Z}}]{\calA_n^-\cup\calD_n}\nonumber{}\\
      & \geq \P[\bar{\mathbf{X}}\bar{\mathbf{Z}}]{\calA_n^-}+\P[\bar{\mathbf{X}}\bar{\mathbf{Z}}]{\calD_n}-1\nonumber{}\\
      & \geq \P[\bar{\mathbf{X}}\bar{\mathbf{Z}}]{\calA_n^-}-\P[\bar{\mathbf{X}}\bar{\mathbf{Z}}]{\calD_n^c}\label{eq:bound_qn_-}
    \end{align}
    Note that,
    \begin{align}
            \P[\bar{\mathbf{X}}\bar{\mathbf{Z}}]{\calD_n^c}  &= \sum_{{\mathbf{z}}\in\calZ^n} \p[\bar{\mathbf{Z}}]{{\mathbf{z}}}\mathds{1}\left\{\log\frac{\p[\bar{\mathbf{Z}}]{{\mathbf{z}}}}{\p[{\mathbf{Z}}]{{\mathbf{z}}}}\leq -b\sqrt{n}\right\}\nonumber{}\\
            & \leq 2^{-b\sqrt{n}} \sum_{{\mathbf{z}}\in\calZ^n} \p[{\mathbf{Z}}]{{\mathbf{z}}}\nonumber{}\\
            & \leq 2^{-b\sqrt{n}}.\label{eq:bound_dnc}
    \end{align}
    and, following the reasoning leading to~\eqref{eq:berry_esseen_1}, 
      \begin{align}
        \P[\bar{\mathbf{X}}\bar{\mathbf{Z}}]{\calA_n^-}&\geq \frac{1}{\sqrt{2\pi}}\int_{-\infty}^{-\frac{2b}{\sigma}-\frac{\sqrt{n}\epsilon_n}{\sigma}}e^{-\frac{x^2}{2}}\dd x -\frac{c}{\sqrt{n}}\frac{\rho}{\sigma^3}.\label{eq:berry_esseen_2}
      \end{align}
    
      Combining equations \eqref{eq:bound_qn_+}-\eqref{eq:berry_esseen_2}, we obtain
      \begin{align}
        \P[\bar{\mathbf{X}}\bar{\mathbf{Z}}]{\calA_n}&= \P[\bar{\mathbf{X}}\bar{\mathbf{Z}}]{\calQ_n^+}-\P[\bar{\mathbf{X}}\bar{\mathbf{Z}}]{\calQ_n^-}\nonumber{}\\
        &\leq \frac{1}{\sqrt{2\pi}}\int_{-\frac{2b}{\sigma}-\frac{\sqrt{n}\epsilon_n}{\sigma}}^{\frac{2b}{\sigma}-\frac{\sqrt{n}\epsilon_n}{\sigma}}e^{-\frac{x^2}{2}}\dd x +\frac{2c}{\sqrt{n}}\frac{\rho}{\sigma^3}\nonumber\\
        &\phantom{--}+\frac{\sqrt{n}}{b}\left(\log \card{\calZ} + \beta+\log\frac{1}{\mu_Z}\right)2^{-\beta n} + 2^{-b\sqrt{n}}\nonumber{}\\
        &\leq \frac{4b}{\sigma\sqrt{2\pi}}         +\frac{2c}{\sqrt{n}}\frac{\rho}{\sigma^3} \nonumber\\
        &\phantom{-}+\frac{\sqrt{n}}{b}\left(\log \card{\calZ} + \beta+\log\frac{1}{\mu_Z}\right)2^{-\beta n} + 2^{-b\sqrt{n}}.\label{eq:final_bound_a0}
      \end{align}
      Combining~\eqref{eq:final_bound_a0} with~\eqref{eq:bound_tradeoff}, and using the assumption $\lim_{n\rightarrow\infty}\perr^*(\calC_n)=0$ from~\eqref{eq:conditions_cn}, we have
      \begin{align*}
\forall b>0\quad        \lim_{n\rightarrow\infty}\metric{2}(\calC_n)\geq 1-\frac{4b}{\sigma\sqrt{2\pi}}.
      \end{align*}
     Therefore, there exists $\eta>0$ such that, for $n$ large enough, $\metric{2}(\calC_n)\geq\eta$. Notice that Proposition~\ref{lm:ordering_metrics} immediately implies that there exists $\eta^*>0$ such that $\lim_{n\rightarrow\infty} \metric{1}(\calC_n)\geq\eta^*$.

\section{Lemmas used in the Achievability Proof of Theorem~\ref{th:bcc}}
\label{sec:proof-theorem-bcc-achievability}

The following notation is used throughout this appendix. We recall that ${\mathbf{U}},{\mathbf{X}},{\mathbf{Y}},{\mathbf{Z}}$ are the random variables defined by the random code generation with distribution given in~\eqref{eq:def_zn}. For any $(k,l,m)\in\intseq{1}{M_0}\times\intseq{1}{M_1}\times\intseq{1}{M'}$, the random variables representing the codewords ${\mathbf{u}}_k$ and ${\mathbf{x}}_{klm}$ obtained with the random code generation are denoted by ${\mathbf{U}}_k$ and ${\mathbf{X}}_{klm}$. 

The random variables that correspond to the use of a specific code $\calC_n$ are denoted by $\bar{\mathbf{U}},\bar{\mathbf{X}},\bar{\mathbf{Y}},\bar{\mathbf{Z}}$ with distribution given by~\eqref{eq:pdf_code_random_variables}. The channel outputs that correspond to the transmission of ${\mathbf{u}}_k$ and ${\mathbf{x}}_{klm}$ are denoted by $\bar{\mathbf{Y}}_{klm}$, and $\bar{\mathbf{Z}}_{klm}$, respectively. 

\subsection{Proof of Lemma~\ref{lm:reliability_bcc}}
By symmetry of the random code construction, we have
\begin{align*}
  &\E{\perr(\rvC_n)}\\
  &= \sum_{k=1}^{M_0}\sum_{l=1}^{M_1}\sum_{m=1}^{M'}\frac{\E{\perr(\rvC_n|\rvM_0=k,\rvM_1=l,\rvM'=m)}}{M_0M_1M'} \\
  &= \E{\perr(\rvC_n|\rvM_0=1,\rvM_1=1,\rvM'=1)},
\end{align*}
which can be analyzed in terms of the events
\begin{align*}
  E_1(k)\eqdef& \left\{(\bar{\mathbf{U}}_k,\bar{\mathbf{Y}}_{111})\in\calT_1^n|\rvM_0=\rvM_1=\rvM'=1\right\}\\
  E_2(k)\eqdef& \left\{(\bar{\mathbf{U}}_k,\bar{\mathbf{Z}}_{111})\in\calT_3^n|\rvM_0=\rvM_1=\rvM'=1\right\}\\
  E_3(k,l,m)&\\ 
 \eqdef &\phantom{}\left\{(\bar{\mathbf{U}}_k,\bar{\mathbf{X}}_{klm},\bar{\mathbf{Y}}_{111})\in\calT_2^n|\rvM_0=\rvM_1=\rvM'=1\right\}.
\end{align*}
It follows from standard arguments (see, for instance,~\cite[Chapter 3]{InformationSpectrumMethods}) that $\E{\perr(\rvC_n)}<\epsilon$ for $n$ large enough provided
\begin{align}
  \begin{array}{l}
  \frac{1}{n}\log M_0 \leq \pliminf{\frac{1}{n}\I{{\mathbf{U}};{\mathbf{Y}}}}-2\gamma\\
  \frac{1}{n}\log M_0 \leq \pliminf{\frac{1}{n}\I{{\mathbf{U}};{\mathbf{Z}}}}-2\gamma\\
  \frac{1}{n}\log M_1M' \leq \pliminf{\frac{1}{n}\I{{\mathbf{X}};{\mathbf{Y}}|{\mathbf{U}}}}-2\gamma.
  \end{array}
  \label{eq:reliability_conditions}
\end{align}
\subsection{Proof of Lemma~\ref{lm:secrecy_bcc}}
We start by developing an upper bound for $\metric{2}(\calC_n)$ that will be simpler to analyze. First, we have
\begin{align*}
  \metric{2}(\calC_n)\eqdef\V{p_{\rvM_1\bar{\mathbf{Z}}},p_{\rvM_1}p_{\bar{\mathbf{Z}}}}&\leq \V{p_{\bar{\mathbf{U}}\rvM_1\bar{\mathbf{Z}}},p_{\rvM_1}p_{\bar{\mathbf{U}}\bar{\mathbf{Z}}}}\\
  &=\E[\bar{\mathbf{U}}\rvM_1]{\V{p_{\bar{\mathbf{Z}}|\bar{\mathbf{U}}\rvM_1},p_{\bar{\mathbf{Z}}|\bar{\mathbf{U}}}}}.
\end{align*}
Next, we use Lemma~\ref{lm:subadditivity_basics} to further bound $\metric{2}(\calC_n)$ as follows.
\begin{align}
  \metric{2}(\calC_n)&\leq \E[\bar{\mathbf{U}}\rvM_1]{\V{p_{\bar{\mathbf{Z}}|\bar{\mathbf{U}}\rvM_1},p_{{\mathbf{Z}}|{\mathbf{U}}}}+\V{p_{{\mathbf{Z}}|{\mathbf{U}}},p_{\bar{\mathbf{Z}}|\bar{\mathbf{U}}}}}\nonumber\\
  &= \E[\bar{\mathbf{U}}\rvM_1]{\V{p_{\bar{\mathbf{Z}}|\bar{\mathbf{U}}\rvM_1},p_{{\mathbf{Z}}|{\mathbf{U}}}}}+\E[\bar{\mathbf{U}}]{\V{p_{{\mathbf{Z}}|{\mathbf{U}}},p_{\bar{\mathbf{Z}}|\bar{\mathbf{U}}}}}\nonumber\\
  &\leq \E[\bar{\mathbf{U}}\rvM_1]{\V{p_{\bar{\mathbf{Z}}|\bar{\mathbf{U}}\rvM_1},p_{{\mathbf{Z}}|{\mathbf{U}}}}}\nonumber\\
  &\phantom{---------}+\E[\bar{\mathbf{U}}]{\V{p_{\rvM_1}p_{{\mathbf{Z}}|{\mathbf{U}}},p_{\bar{\mathbf{Z}}\rvM_1|\bar{\mathbf{U}}}}}\nonumber\\
  &= 2 \E[\bar{\mathbf{U}}\rvM_1]{\V{p_{\bar{\mathbf{Z}}|\bar{\mathbf{U}}\rvM_1},p_{{\mathbf{Z}}|{\mathbf{U}}}}}.
\label{eq:final_bound_s2}
\end{align}
Notice that the term in brackets on the right hand side is a variational distance between the following two distributions:
\begin{itemize}
\item $\p[\bar{\mathbf{Z}}|\bar{\mathbf{U}}={\mathbf{u}}_k,\rvM_1=l]{{\mathbf{z}}} = \sum_{m=1}^{M'}\frac{1}{M'}W_{{\mathbf{Z}}|{\mathbf{X}}}({\mathbf{z}}|{\mathbf{x}}_{klm})$, which represents the distribution induced at the eavesdropper's channel output by the $M'$ codewords $\{{\mathbf{x}}_{kli}\}_{i\in\intseq{1}{M'}}$ selected with a uniform distribution;
\item $p_{{\mathbf{Z}}|{\mathbf{U}}={\mathbf{u}}_k}({\mathbf{z}})=\sum_{{\mathbf{x}}}W_{{\mathbf{Z}}|{\mathbf{X}}}({\mathbf{z}}|{\mathbf{x}})p_{{\mathbf{X}}|{\mathbf{U}}={\mathbf{u}}_k}({\mathbf{x}})$, which represents the distribution induced at the eavesdropper's channel output by an input process with distribution $p_{{\mathbf{X}}|{\mathbf{U}}={\mathbf{u}}_k}({\mathbf{x}})$.
\end{itemize}
Therefore, a \emph{sufficient condition} for $\metric{2}(\calC_n)$ to vanish is that, for
every pair $(k,l)\in\intseq{1}{M_0}\times\intseq{1}{M_1}$, the variational distance between the two distributions
vanishes as well. This is possible if each set of codewords $\{{\mathbf{x}}_{kli}\}_{i\in\intseq{1}{M'}}$ approximates the same process
with distribution $p_{{\mathbf{Z}}|{\mathbf{U}}={\mathbf{u}}_k}({\mathbf{z}})$ at the eavesdropper's output, which is exactly what the concept
of channel resolvability reviewed in Section~\ref{sec:notation} is about. In other words, \emph{a sufficient
  condition to guarantee secrecy is for each sub-codebook $\{{\mathbf{x}}_{kli}\}_{i\in\intseq{1}{M'}}$ to be a ``channel resolvability
  code''}.

We establish the existence of such codebooks with a random coding argument following that used in~\cite{Han1993,InformationSpectrumMethods}. The presence of a common message makes the proof slightly more involved but the steps remain essentially the same. On taking the average over $\rvC_n$ for both sides of~\eqref{eq:final_bound_s2}, we obtain
\begin{align}
  \E[\rvC_n]{\metric{2}(\rvC_n)} \leq 2 \E[\bar{\mathbf{U}}\rvM_1]{\E[\rvC_n]{\V{p_{\bar{\mathbf{Z}}|\bar{\mathbf{U}}\rvM_1},p_{{\mathbf{Z}}|{\mathbf{U}}}}}}
  \label{eq:average_s2}
\end{align}
By symmetry of the random code construction, the inner expectation in~\eqref{eq:average_s2} is the same for all values of $\bar{\mathbf{U}}={\mathbf{u}}_k$ and $\rvM_1=l$; hence, we have
\begin{align}
  \E[\rvC_n]{\metric{2}(\rvC_n)} \leq   2\E[\rvC_n]{\V{p_{\bar{\mathbf{Z}}|\bar{\mathbf{U}}={\mathbf{U}}_1 \rvM_1=1},p_{{\mathbf{Z}}|{\mathbf{U}}={\mathbf{U}}_1}}}.
  \label{eq:average_simplified_s2}
\end{align}
Let $\tau>0$. On using~\cite[Lemma
6.3.1]{InformationSpectrumMethods} we finally upper bound~\eqref{eq:average_simplified_s2} as
\begin{align}
  \E[\rvC_n]{\metric{2}(\rvC_n)}\leq 4\frac{\tau}{\log e} + 4A_n \label{eq:bound_delta_n}
\end{align}
with
\begin{align*}
  A_n\eqdef
  \E[\rvC_n]{\P[\bar{\mathbf{Z}}|\bar{\mathbf{U}}={\mathbf{U}}_1
    \rvM_1=1]{\log\frac{\p[\bar{\mathbf{Z}}|\bar{\mathbf{U}}={\mathbf{U}}_1
        \rvM_1=1]{\bar{\mathbf{Z}}}}{p_{{\mathbf{Z}}|{\mathbf{U}}={\mathbf{U}}_1}(\bar{\mathbf{Z}})}>\tau}}.
\end{align*}

Note that the expectation over $\rvC_n$ reduces to the expectation over ${\mathbf{U}}_1$ and $\{{\mathbf{X}}_{11j}\}_{j\in\intseq{1}{M'}}$. Writing $A_n$ explicitly, we obtain Equation~\eqref{eq:factorization_random_code_gen} shown on top of the next page,
\begin{figure*}[!t]
\normalsize
\setcounter{equation}{34}
\begin{align}
  A_n&=\sum_{{\mathbf{u}}_1\in\calU^n}p_{{\mathbf{U}}}({\mathbf{u}}_1)\sum_{{\mathbf{x}}_{111}\in\calX^n}p_{{\mathbf{X}}|{\mathbf{U}}}({\mathbf{x}}_{111}|{\mathbf{u}}_1)\dots \sum_{\svx_{11M'}\in\calX^n}p_{{\mathbf{X}}|{\mathbf{U}}}({\mathbf{x}}_{11M'}|{\mathbf{u}}_1)\nonumber\\
  &\phantom{---------------}  \sum_{{\mathbf{z}}\in\calZ^n}\p[\bar{\mathbf{Z}}|\bar{\mathbf{U}}={\mathbf{u}}_1 \rvM_1=1]{{\mathbf{z}}}
  \mathds{1}\left\{\log\frac{\p[\bar{\mathbf{Z}}|\bar{\mathbf{U}}={\mathbf{u}}_1 \rvM_1=1]{{\mathbf{z}}}}{p_{{\mathbf{Z}}|{\mathbf{U}}}({\mathbf{z}}|{\mathbf{u}}_1)}>\tau\right\}\nonumber\\
  &\stackrel{(a)}{=}\frac{1}{M'}\sum_{m=1}^{M'}\sum_{{\mathbf{u}}_1\in\calU^n}p_{{\mathbf{U}}}({\mathbf{u}}_1)\sum_{{\mathbf{x}}_{111}\in\calX^n}p_{{\mathbf{X}}|{\mathbf{U}}}({\mathbf{x}}_{111}|{\mathbf{u}}_1)\dots \sum_{\svx_{11M'}\in\calX^n}p_{{\mathbf{X}}|{\mathbf{U}}}({\mathbf{x}}_{11M'}|{\mathbf{u}}_1) \nonumber\\
  &\phantom{---------------} 
  \sum_{{\mathbf{z}}\in\calZ^n}W_{{\mathbf{Z}}|{\mathbf{X}}}({{\mathbf{z}}|{\mathbf{x}}_{11m}})\mathds{1}\left\{\log\frac{\p[\bar{\mathbf{Z}}|\bar{\mathbf{U}}={\mathbf{u}}_1 \rvM_1=1]{{\mathbf{z}}}}{p_{{\mathbf{Z}}|{\mathbf{U}}}({\mathbf{z}}|{\mathbf{u}}_1)}>\tau\right\}\nonumber\displaybreak[0]\\
  &\stackrel{(b)}{=}\sum_{{\mathbf{u}}_1\in\calU^n}p_{{\mathbf{U}}}({\mathbf{u}}_1)\sum_{{\mathbf{x}}_{112}\in\calX^n}p_{{\mathbf{X}}|{\mathbf{U}}}({\mathbf{x}}_{112}|{\mathbf{u}}_1)\dots \sum_{\svx_{11M'}\in\calX^n}p_{{\mathbf{X}}|{\mathbf{U}}}({\mathbf{x}}_{11M'}|{\mathbf{u}}_1) \nonumber\\
  &\phantom{-----------} 
  \sum_{{\mathbf{x}}_{111}\in\calX^n}\sum_{{\mathbf{z}}\in\calZ^n}W_{{\mathbf{Z}}|{\mathbf{X}}}({{\mathbf{z}}|{\mathbf{x}}_{111}}) p_{{\mathbf{X}}|{\mathbf{U}}}({\mathbf{x}}_{111}|{\mathbf{u}}_1)\mathds{1}\left\{\log\frac{\p[\bar{\mathbf{Z}}|\bar{\mathbf{U}}={\mathbf{u}}_1 \rvM_1=1]{{\mathbf{z}}}}{p_{{\mathbf{Z}}|{\mathbf{U}}}({\mathbf{z}}|{\mathbf{u}}_1)}>\tau\right\},\nonumber\\
  &\stackrel{(c)}{=}\sum_{{\mathbf{u}}_1\in\calU^n}p_{{\mathbf{U}}}({\mathbf{u}}_1)\sum_{{\mathbf{x}}_{112}\in\calX^n}p_{{\mathbf{X}}|{\mathbf{U}}}({\mathbf{x}}_{112}|{\mathbf{u}}_1)\dots \sum_{\svx_{11M'}\in\calX^n}p_{{\mathbf{X}}|{\mathbf{U}}}({\mathbf{x}}_{11M'}|{\mathbf{u}}_1) \nonumber\\
  &\phantom{--------}  \sum_{{\mathbf{x}}_{111}\in\calX^n}\sum_{{\mathbf{z}}\in\calZ^n}p_{{\mathbf{Z}}{\mathbf{X}}|{\mathbf{U}}}({{\mathbf{z}},{\mathbf{x}}_{111}|{\mathbf{u}}_1})\mathds{1}\left\{\log\left(\frac{1}{M'}\sum_{m=1}^{M'}\frac{p_{{\mathbf{Z}}|{\mathbf{X}}{\mathbf{U}}}({{\mathbf{z}}|{\mathbf{x}}_{11m}{\mathbf{u}}_1})}{p_{{\mathbf{Z}}|{\mathbf{U}}}({{\mathbf{z}}|{\mathbf{u}}_1})}\right)>\tau\right\}\label{eq:factorization_random_code_gen}
\end{align}
\hrulefill
\vspace*{4pt}
\end{figure*}
\setcounter{equation}{35}
where equality $(a)$ follows from the definition of $\p[\bar{\mathbf{Z}}|\bar{\mathbf{U}}={\mathbf{u}}_1 \rvM_1=1,\calC_n]{{\mathbf{z}}}$, equality $(b)$ follows by remarking that all codewords are generated according to the same density $p_{{\mathbf{X}}|{\mathbf{U}}}$  and equality $(c)$ follows by noting that
\begin{itemize}
\item $W_{{\mathbf{Z}}|{\mathbf{X}}}({{\mathbf{z}}|{\mathbf{x}}_{111}})p_{{\mathbf{X}}|{\mathbf{U}}}({\mathbf{x}}_{111}|{\mathbf{u}}_1)=p_{{\mathbf{Z}}{\mathbf{X}}|{\mathbf{U}}}({\mathbf{z}},{\mathbf{x}}_{111}|{\mathbf{u}}_1)$ according to~\eqref{eq:def_zn};
\item for any ${\mathbf{u}}_1$ such that $\p[{\mathbf{X}}|{\mathbf{U}}]{{\mathbf{x}}_{11m}|{\mathbf{u}}_1}>0$,
  \begin{align*}
p_{\bar{\mathbf{Z}}|\bar{\mathbf{U}}={\mathbf{u}}_1 \rvM_1=1}({\mathbf{z}})
    &=\frac{1}{M'}\sum_{m=1}^{M'}W_{{\mathbf{Z}}|{\mathbf{X}}}({\mathbf{z}}|{\mathbf{x}}_{11m})\\
    &=\frac{1}{M'}\sum_{m=1}^{M'}p_{{\mathbf{Z}}|{\mathbf{X}}{\mathbf{U}}}({\mathbf{z}}|{\mathbf{x}}_{11m}{\mathbf{u}}_1).
  \end{align*}
\end{itemize}
By adapting the proof technique developed in~\cite[Chapter 6]{InformationSpectrumMethods} and after some calculations, one can further bound $A_n$ to obtain
\begin{multline}
  \E[\rvC_n]{\metric{2}(\rvC_n)}\leq 4\frac{\tau}{\log e}\\
  + 4\P[{\mathbf{U}}{\mathbf{X}}{\mathbf{Z}}]{\frac{1}{n}\I{{\mathbf{X}};{\mathbf{Z}}|{\mathbf{U}}} \geq\frac{\log
      M'}{n}+\frac{\log\rho}{n}}\\
  + 4\P[{\mathbf{U}}{\mathbf{X}}{\mathbf{Z}}]{\frac{1}{n}\I{{\mathbf{X}};{\mathbf{Z}}|{\mathbf{U}}}\geq\frac{\log M'}{n}}+ \frac{4\cdot 2^{-n\gamma}}{\rho^2}\\
  +\frac{4}{\rho^2}\P[{\mathbf{U}}{\mathbf{X}}{\mathbf{Z}}]{\frac{1}{n}\I{{\mathbf{X}};{\mathbf{Z}}|{\mathbf{U}}}\geq\frac{\log
      M'}{n}-\gamma}\label{eq:bound_delta_total}.
\end{multline}
where $\rho\eqdef \frac{2^\tau-1}{2}$. Therefore, $  \E[\rvC_n]{\metric{2}(\rvC_n)}<\epsilon$ for $n$ large enough provided
\begin{align}
  \frac{1}{n}\log M'\geq \plimsup{\frac{1}{n}\I{{\mathbf{X}};{\mathbf{Z}}|{\mathbf{U}}}}+2\gamma.\label{eq:resolvability_conditions}
\end{align}

\section{Lemma used in the Converse Proof of Theorem~\ref{th:bcc}}
\label{sec:proof-theorem-bcc-converse}

To prove Lemma~\ref{lm:bcc_converse}, note that, with probability one, 
\begin{align*}
  \tfrac{1}{n}\I{\bar{\mathbf{W}};\bar{\mathbf{Z}}} &= \tfrac{1}{n}\I{\bar{\mathbf{W}};\bar{\mathbf{Z}}\bar{\mathbf{U}}}-\tfrac{1}{n}\I{\bar{\mathbf{W}};\bar{\mathbf{U}}|\bar{\mathbf{Z}}}\\
  &=\tfrac{1}{n}\I{\bar{\mathbf{W}};\bar{\mathbf{Z}}|\bar{\mathbf{U}}}-\tfrac{1}{n}\I{\bar{\mathbf{W}};\bar{\mathbf{U}}|\bar{\mathbf{Z}}}\\
  & = \tfrac{1}{n}\I{\bar{\mathbf{W}};\bar{\mathbf{Z}}|\bar{\mathbf{U}}}-\tfrac{1}{n}\H{\bar{\mathbf{U}}|\bar{\mathbf{Z}}}+\tfrac{1}{n}\H{\bar{\mathbf{U}}|\bar{\mathbf{W}}\bar{\mathbf{Z}}},
  \end{align*}
where the second equality follows from the independence of $\bar{\mathbf{W}}$ and $\bar{\mathbf{U}}$. Consequently,
\begin{align*}
  \lim_{n\rightarrow\infty}\metric{6}(\calC_n)&=\plimsup{\frac{1}{n}\I{\bar{\mathbf{W}};\bar{\mathbf{Z}}}}\\
  &\geq\plimsup{\frac{1}{n}\I{\bar{\mathbf{W}};\bar{\mathbf{Z}}|\bar{\mathbf{U}}}}-\plimsup{\frac{1}{n}\H{\bar{\mathbf{U}}|\bar{\mathbf{Z}}}}\\
  &\phantom{-------}+\pliminf{\frac{1}{n}\H{\bar{\mathbf{U}}|\bar{\mathbf{W}}\bar{\mathbf{Z}}}}.
\end{align*}
Since $\lim_{n\rightarrow\infty}\perr(\calC_n)=0$, note that $\pliminf{\frac{1}{n}\H{\bar{\mathbf{U}}|\bar{\mathbf{W}}\bar{\mathbf{Z}}}}= 0$ and $\pliminf{\frac{1}{n}\H{\bar{\mathbf{U}}|\bar{\mathbf{Z}}}}= 0$ by the Verd\'u-Han Lemma. As $\lim_{n\rightarrow\infty}\metric{6}(\calC_n)=0$, we finally obtain
\begin{align*}
  \plimsup{\frac{1}{n}\I{\bar{\mathbf{W}};\bar{\mathbf{Z}}|\bar{\mathbf{U}}}}=0.
\end{align*}

Note that, with probability one,
\begin{align*}
  \H{\bar{\mathbf{W}}} &= \H{\bar{\mathbf{W}}} - \H{\bar{\mathbf{W}}|\bar{\mathbf{Y}}\mathbf{\bar{\mathbf{U}}}}+\H{\bar{\mathbf{W}}|\bar{\mathbf{Y}}\bar{\mathbf{U}}}\\
  &=\I{\bar{\mathbf{W}};\bar{\mathbf{Y}}|\bar{\mathbf{U}}} - \I{\bar{\mathbf{W}};\bar{\mathbf{Z}}|\bar{\mathbf{U}}} + \I{\bar{\mathbf{W}};\bar{\mathbf{Z}}|\bar{\mathbf{U}}} \\
  &\phantom{---------------}+ \H{\bar{\mathbf{W}}|\bar{\mathbf{Y}}\bar{\mathbf{U}}}
\end{align*}
Hence, 
\begin{align*}
  R_1 &\leq \pliminf{\frac{1}{n}\H{\bar{\mathbf{W}}}} \\
  &\leq \pliminf{\frac{1}{n}\I{\bar{\mathbf{W}};\bar{\mathbf{Y}}|\bar{\mathbf{U}}}}-\pliminf{\frac{1}{n}\I{\bar{\mathbf{W}};\bar{\mathbf{Z}}|\bar{\mathbf{U}}}}\\
  &\phantom{--}+\plimsup{\frac{1}{n}\I{\bar{\mathbf{W}};\bar{\mathbf{Z}}|\bar{\mathbf{U}}}}+\plimsup{\frac{1}{n}\H{\bar{\mathbf{W}}|\mathbf{Y}\bar{\mathbf{U}}}}
\end{align*}
As seen above, $\plimsup{\frac{1}{n}\I{\bar{\mathbf{W}};\bar{\mathbf{Z}}|\bar{\mathbf{U}}}}=0$ and, since $\lim_{n\rightarrow\infty}\perr(\calC_n)=0$, we have $\pliminf{\frac{1}{n}\H{\bar{\mathbf{W}}|\bar{\mathbf{Y}}\bar{\mathbf{U}}}}=0$. Therefore,
\begin{align*}
  R_1 &\leq  \pliminf{\frac{1}{n}\I{\mathbf{W};\bar{\mathbf{Y}}|\bar{\mathbf{U}}}}-\plimsup{\frac{1}{n}\I{\mathbf{W};\bar{\mathbf{Z}}|\bar{\mathbf{U}}}}.
\end{align*}

Finally, with probability one,
\begin{align*}
    \H{\bar{\mathbf{U}}} = \I{\bar{\mathbf{U}};\bar{\mathbf{Y}}} +\H{\bar{\mathbf{Y}}|\bar{\mathbf{U}}}=\I{\bar{\mathbf{U}};\bar{\mathbf{Z}}} +\H{\bar{\mathbf{Z}}|\bar{\mathbf{U}}}
\end{align*}
from which conclude after a similar reasoning that
\begin{align*}
  R_0\leq \min\left(\pliminf{\I{\bar{\mathbf{U}};\bar{\mathbf{Y}}}},\pliminf{\I{\bar{\mathbf{U}};\bar{\mathbf{Z}}}}\right).
\end{align*}

\section{Proof of Theorem~\ref{th:dmc}}
\label{sec:proof-theorem-dmc-bcc}

We prove Theorem~\ref{th:dmc} with small modifications of the proof of Theorem~\ref{th:bcc}. Specifically, we establish secrecy for $\metric{1}$ by showing that there exist sequences of codes $\{\calC_n\}_{n\geq 1}$ for which $\metric{2}(\calC_n)$ decreases exponentially fast with $n$ and by using~\cite[Lemma 1]{Csiszar1996} to obtain an upper bound for $\metric{1}(\calC_n)$. We handle the power constraint by using an appropriate distribution during the random code generation process as in~\cite[Section 3.2]{InformationSpectrumMethods}. We note that a similar technique has been used by He and Yener in~\cite{He2010a}.

Let $\gamma,\delta,\epsilon>0$. Let $\calU$ be an arbitrary discrete alphabet and fix a distribution $p_{\tilde\rvU}$ on $\calU$. Fix a conditional distribution $p_{\tilde\rvX|\tilde\rvU}$ on $\calX\times\calU$ such that $\E{c(\tilde\rvX)}\leq P-\delta$. Let $\tilde{\mathbf{U}},\tilde{\mathbf{X}},\tilde{\mathbf{Z}}$ be the random variables with joint distribution
\begin{align*}
    p_{ \tilde{\mathbf{Z}} \tilde{\mathbf{X}} \tilde{\mathbf{U}}}({{\mathbf{z}},{\mathbf{x}},{\mathbf{u}}}) = \prod_{i=1}^nW_{\rvZ|\rvX}(\svz_i|\svx_i)p_{\tilde\rvX|\tilde\rvU}(\svx_i|\svu_i)p_{\tilde\rvU}(\svu_i).
\end{align*}
We assume that $\tilde{\mathbf{U}},\tilde{\mathbf{X}},\tilde{\mathbf{Z}}$ are such that the integrals defining the moment generating functions of $c(\tilde\rvX)$ and $\I{\tilde\rvX;\tilde\rvZ|\tilde\rvU}$ converge uniformly in a neighborhood of $0$ and are differentiable at $0$. 

Define the set $\calP_n$ as
\begin{align*}
  \calP_n\eqdef\left\{{\mathbf{x}}\in\calX^n:\frac{1}{n}\sum_{i=1}^nc(\svx_i) \leq P\right\}.
\end{align*}
Lemma~\ref{lm:chernov} shows that there exists $\alpha_\delta>0$ such that $\P{\tilde{\mathbf{X}}\in\calP_n}\geq 1- 2^{-\alpha_\delta n}$. In the sequel, we define $\gamma_n \eqdef 1-2^{-n\frac{\alpha_\delta}{2}}$. Define the set $\calG_n\subset\calU^n$ as follows:
\begin{align*}
  \calG_n \eqdef \left\{{\mathbf{u}}: \P[\tilde{\mathbf{X}}|\tilde{\mathbf{U}}={\mathbf{u}}]{\tilde{\mathbf{X}}\notin\calP_n|\tilde{\mathbf{U}}={\mathbf{u}}}< 2^{-n\tfrac{\alpha_\delta}{2}}\right\}.
\end{align*}
Upon using Markov's inequality, we obtain
\begin{align}
  \P[\tilde{\mathbf{U}}]{\tilde{\mathbf{U}}\notin \calG_n} & =   \P[\tilde{\mathbf{U}}]{\P[\tilde{\mathbf{X}}|\tilde{\mathbf{U}}]{\tilde{\mathbf{X}}\notin\calP_n|\tilde{\mathbf{U}}}\geq  2^{-n\tfrac{\alpha_\delta}{2}}}\nonumber{}\\
  &\leq \E[\tilde{\mathbf{U}}]{\P[\tilde{\mathbf{X}}|\tilde{\mathbf{U}}]{\tilde{\mathbf{X}}\notin\calP_n|\tilde{\mathbf{U}}}}2^{n\tfrac{\alpha_\delta}{2}}\nonumber{}\\
  &= \P[\tilde{\mathbf{X}}]{\tilde{\mathbf{X}}\notin\calP_n}2^{n\tfrac{\alpha_\delta}{2}}\nonumber{}\displaybreak[0]\\
  &\leq 2^{-n(\alpha_\delta-\tfrac{\alpha_\delta}{2})}\nonumber{}\\
  &= 1-\gamma_n.\label{eq:bound_not_Gn}
\end{align}

Now, we define the random variables ${\mathbf{U}},{\mathbf{X}},{\mathbf{Z}}$ as follows. First,
\begin{align*}
  \forall {\mathbf{u}}\in\calU^n\quad\p[{\mathbf{U}}]{{\mathbf{u}}}=  \left\{  
\begin{array}{l}
      \frac{1}{\P[\tilde{\mathbf{U}}]{\tilde{\mathbf{U}}\in\calG_n}} \p[\tilde{\mathbf{U}}]{{\mathbf{u}}}\text{ if ${\mathbf{u}}\in\calG_n$}\\
      0\;\text{else}.
    \end{array}
\right.
\end{align*}
From Eq.~\eqref{eq:bound_not_Gn}, we have
\begin{align}
  \forall {\mathbf{u}}\in\calU^n\quad \p[{\mathbf{U}}]{{\mathbf{u}}} \leq \frac{\p[\tilde{\mathbf{U}}]{{\mathbf{u}}}}{\gamma_n}.\label{eq:bound_pun}
\end{align}
Next, $\forall ({\mathbf{x}},{\mathbf{u}})\in\calX^n\times\calG_n$
\begin{align*}
  \p[{\mathbf{X}}|{\mathbf{U}}]{{\mathbf{x}}|{\mathbf{u}}}=  \left\{  
\begin{array}{l}
      \frac{1}{\P[\tilde{\mathbf{X}}|\tilde{\mathbf{U}}={\mathbf{u}}]{\tilde{\mathbf{X}}\in\calP_n|\tilde{\mathbf{U}}={\mathbf{u}}}} \p[\tilde{\mathbf{X}}|\tilde{\mathbf{U}}]{{\mathbf{x}}|{\mathbf{u}}}\text{ if ${\mathbf{x}}\in\calP_n$}\\
      0\;\text{else}.
    \end{array}
\right.
\end{align*}
By construction, we have
\begin{align}
   \forall ({\mathbf{x}},{\mathbf{u}})\in\calX^n\times\calG_n\quad \p[{\mathbf{X}}|{\mathbf{U}}]{{\mathbf{x}}|{\mathbf{u}}} \leq \frac{\p[\tilde{\mathbf{X}}|\tilde{\mathbf{U}}]{{\mathbf{x}}|{\mathbf{u}}}}{\gamma_n}.\label{eq:bound_pxnun}
\end{align}
Finally, $\forall ({\mathbf{z}},{\mathbf{x}},{\mathbf{u}})\in\calZ^n\times\calX^n\times\calG_n$
\begin{align}
  p_{{\mathbf{Z}}{\mathbf{X}}{\mathbf{U}}}({\mathbf{z}},{\mathbf{x}},{\mathbf{u}}) = W_{{\mathbf{Z}}|{\mathbf{X}}}({\mathbf{z}}|{\mathbf{x}})\p[{\mathbf{X}}|{\mathbf{U}}]{{\mathbf{x}}|{\mathbf{u}}}\p[{\mathbf{U}}]{{\mathbf{u}}}. \label{eq:def_trunc_proba}
\end{align}
We repeat the random coding argument in the proof of Theorem~\ref{th:bcc} using the distribution $p_{{\mathbf{X}}{\mathbf{U}}}$ defined by~\eqref{eq:def_trunc_proba} and with the following lemmas. 

\begin{lemma}[Reliability conditions]
\label{lm:reliabilty-dmc}
    \begin{align*}
        \text{If } R_0 &\leq \min\left(\avgI{\tilde\rvU;\tilde\rvY}-2\gamma,\avgI{\tilde\rvU;\tilde\rvZ}-2\gamma\right)\\
        \text{and }R_1+R_1'&\leq \avgI{\tilde\rvX;\tilde\rvY|\tilde\rvU}-2\gamma,
\end{align*}
then $ \lim_{n\rightarrow\infty}\E{\perr(\rvC_n)}\leq \epsilon$.
\end{lemma}
\begin{IEEEproof}
Following~\cite[Proof of Theorem 3.6.2]{InformationSpectrumMethods}, one can show that
\begin{align*}
  \pliminf{\frac{1}{n}\I{{\mathbf{U}};{\mathbf{Y}}}} &\geq \avgI{\tilde\rvU;\tilde\rvY},\\
  \quad \pliminf{\frac{1}{n}\I{{\mathbf{U}};{\mathbf{Z}}}}&\geq \avgI{\tilde\rvU;\tilde\rvZ}\\
  \mbox{and}\quad\pliminf{\frac{1}{n}\I{{\mathbf{X}};{\mathbf{Y}}|{\mathbf{U}}}} &\geq
  \avgI{\tilde\rvX;\tilde\rvY|\tilde\rvU}.
\end{align*}
Hence, the result follows directly from Lemma~\ref{lm:reliability_bcc}.
\end{IEEEproof}

\begin{lemma}[Secrecy from channel resolvability conditions]
  \label{lm:secrecy-dmc}
  There exists $\alpha_{\delta,\gamma}>0$, such that
      \begin{align*}
\text{If }R_1'\geq \avgI{\tilde\rvX;\tilde\rvZ|\tilde\rvU}+2\gamma
\text{ then } \lim_{n\rightarrow\infty}\E{\metric{2}(\rvC_n)}\leq 2^{-\alpha_{\delta,\gamma}}.
  \end{align*}
\end{lemma}
\begin{IEEEproof}
Note that~\eqref{eq:final_bound_s2} still holds. Upon using Lemma~\ref{lm:subadditivity_basics}, we obtain
  \begin{align}
    \E[\rvC_n]{\metric{2}(\rvC_n)} &\leq 2\E[\rvC_n]{\V{p_{\bar{\mathbf{Z}}|\bar{\mathbf{U}}={\mathbf{U}}_1, \rvM_1=1,\rvC_n},p_{{\mathbf{Z}}|{\mathbf{U}}={\mathbf{U}}_1}}}\nonumber{}\\
    &\leq 2\E[\rvC_n]{\V{p_{\bar{\mathbf{Z}}|\bar{\mathbf{U}}={\mathbf{U}}_1,
          \rvM_1=1,\rvC_n},p_{\tilde{\mathbf{Z}}|\tilde{\mathbf{U}}={\mathbf{U}}_1}}}\nonumber\\
    &\phantom{--}  +    2\E[\rvC_n]{\V{p_{\tilde{\mathbf{Z}}|\tilde{\mathbf{U}}={\mathbf{U}}_1},p_{{\mathbf{Z}}|{\mathbf{U}}={\mathbf{U}}_1}}}.
    \label{eq:bound_dmc_init}
  \end{align}
First, we bound the second term on the right-hand side of~\eqref{eq:bound_dmc_init}. For all ${\mathbf{u}}_1\in\calG_n$, we obtain the bound shown in Equation~\eqref{eq:bound_var_distance} on the next page, 
\begin{figure*}[!t]
\normalsize
\setcounter{equation}{42}
  \begin{align}
\V{{p}_{\tilde{\mathbf{Z}}|\tilde{\mathbf{U}}={\mathbf{u}}_1},{p}_{{\mathbf{Z}}|{\mathbf{U}}={\mathbf{u}}_1}}
  &\stackrel{(a)}{\leq} \V{{p}_{\tilde{\mathbf{X}}|\tilde{\mathbf{U}}={\mathbf{u}}_1},{p}_{{\mathbf{X}}|{\mathbf{U}}={\mathbf{u}}_1}}\nonumber\\
  &=2\sup_{\calA\subseteq\calX^n}\abs{\P[{\mathbf{X}}|{\mathbf{U}}={\mathbf{u}}_1]{\calA}-\P[\tilde{\mathbf{X}}|\tilde{\mathbf{U}}={\mathbf{u}}_1]{\calA}}\nonumber\\
  &\leq\sup_{\calA\subseteq\calX^n}\sum_{\calB\in\{\calA,\calA^c\}}\left(\abs{\P[{\mathbf{X}}|{\mathbf{U}}={\mathbf{u}}_1]{\calB}-\P[\tilde{\mathbf{X}}|\tilde{\mathbf{U}}={\mathbf{u}}_1]{\calB}}\right) \nonumber\\
  &\stackrel{(b)}{\leq} \sup_{\calA\subseteq\calX^n}\sum_{\calB\in\{\calA,\calA^c\}}\left(\abs{\P[{\mathbf{X}}|{\mathbf{U}}={\mathbf{u}}_1]{\calB\cap\calP_n}-\P[\tilde{\mathbf{X}}|\tilde{\mathbf{U}}={\mathbf{u}}_1]{\calB\cap\calP_n^c}-\P[\tilde{\mathbf{X}}|\tilde{\mathbf{U}}={\mathbf{u}}_1]{\calB\cap\calP_n}}\right) \nonumber\\
  &\stackrel{(c)}{\leq} \sup_{\calA\subseteq\calX^n}\sum_{\calB\in\{\calA,\calA^c\}}\left(\P[\tilde{\mathbf{X}}|\tilde{\mathbf{U}}={\mathbf{u}}_1]{\calB\cap\calP_n}\left(\frac{1}{\gamma_n}-1\right) + \P[\tilde{\mathbf{X}}|\tilde{\mathbf{U}}={\mathbf{u}}_1]{\calB\cap\calP_n^c}\right) \nonumber\\
  &\leq (\frac{1}{\gamma_n}-1) + (1-\gamma_n),\label{eq:bound_var_distance}
\end{align}
\setcounter{equation}{43}
\hrulefill
\vspace*{4pt}
\end{figure*}
where $(a)$ follows from Lemma~\ref{lm:dpi_variational}, $(b)$~follows because $\P[\mathbf{X|\mathbf{U}=\mathbf{u}_1}]{\calB\cap\calP_n^c}=0$ by Eq.~\eqref{eq:def_trunc_proba}, and $(c)$~follows from the bound in Eq.~\eqref{eq:bound_pxnun}; therefore, for $n$ large enough, there exists $\beta_\delta>0$, such that
  \begin{align}
   \E[\rvC_n]{\V{p_{\tilde{\mathbf{Z}}|\tilde{\mathbf{U}}={\mathbf{U}}_1},p_{{\mathbf{Z}}|{\mathbf{U}}={\mathbf{U}}_1}}} \leq     2^{-\beta_\delta n}\label{eq:bound_dmc_var_truncated}
  \end{align}
 
We now bound the first term on the right-hand side of~\eqref{eq:bound_dmc_init}.  Applying~\cite[Lemma 6.3.1]{InformationSpectrumMethods}, we obtain \vspace{-8pt}
  \begin{multline}
    2\E[\rvC_n]{ \V{p_{\bar{\mathbf{Z}}|\bar{\mathbf{U}}={\mathbf{U}}_1 \rvM_1=1},p_{\tilde{\mathbf{\rvZ}}|\tilde{\mathbf{U}}={\mathbf{U}}_1}}} 
    \leq 4\frac{\tau}{\log e} \\+ 4 \E[\rvC_n]{\P[\bar{\mathbf{Z}}|\bar{\mathbf{U}}={\mathbf{U}}_1
    \rvM_1=1]{\log\frac{\p[\bar{\mathbf{Z}}|\bar{\mathbf{U}}={\mathbf{U}}_1
        \rvM_1=1]{\bar{\mathbf{Z}}}}{p_{\tilde{\mathbf{\rvZ}}|\tilde{\mathbf{U}}={\mathbf{U}}_1}(\bar{\mathbf{Z}})}>\tau}}.\label{eq:bound_dmc_2}
  \end{multline}
  Note that~\eqref{eq:bound_dmc_2} is similar to~\eqref{eq:bound_delta_n}, and the only difference is the presence of $p_{\tilde{\mathbf{\rvZ}}|\tilde{\mathbf{\rvU}}}$ instead of $p_{\mathbf{\rvZ}|\mathbf{\rvU}}$ in the denominator; using the definition of $p_{{\mathbf{Z}}{\mathbf{X}}{\mathbf{U}}}$ in~\eqref{eq:def_trunc_proba}, the bounds in~\eqref{eq:bound_pun} and~\eqref{eq:bound_pxnun}, and repeating the steps leading from~\eqref{eq:bound_delta_n} to~\eqref{eq:bound_delta_total}, one obtains after some calculations
  \begin{multline}
    2\E[\rvC_n]{ \V{p_{\bar{\mathbf{Z}}|\bar{\mathbf{U}}={\mathbf{U}}_1
          \rvM_1=1},p_{\tilde{\rvZ}_1^n|\tilde{\mathbf{U}}={\mathbf{U}}_1}}}\\
    \leq 4\frac{\tau}{\log e} +
    \frac{4}{\gamma_n^2}\P[\tilde{\mathbf{U}}\tilde{\mathbf{X}}\tilde{\mathbf{Z}}]{\frac{1}{n}\I{\tilde{\mathbf{X}};\tilde{\mathbf{Z}}|\tilde{\mathbf{U}}}
      \geq\frac{\log
        M'}{n}+\frac{\log\rho}{n}}\\
    + \frac{4}{\gamma_n^3}\P[\tilde{\mathbf{U}}\tilde{\mathbf{X}}\tilde{\mathbf{Z}}]{\frac{1}{n}\I{\tilde{\mathbf{X}};\tilde{\mathbf{Z}}|\tilde{\mathbf{U}}}\geq\frac{\log M'}{n}}  + \frac{4 \cdot 2^{-n\gamma}}{\gamma_n(\gamma_n\rho+\gamma_n-1)^2}\\
    +\frac{4}{\gamma_n(\gamma_n\rho+\gamma_n-1)^2}\P[\tilde{\mathbf{U}}\tilde{\mathbf{X}}\tilde{\mathbf{Z}}]{\frac{1}{n}\I{\tilde{\mathbf{X}};\tilde{\mathbf{Z}}|\tilde{\mathbf{U}}}\geq\frac{\log
        M'}{n}-\gamma}\label{eq:bound_dmc_final}.
  \end{multline}
  If $\frac{1}{n}\log M'\geq\avgI{\tilde\rvX;\tilde\rvZ|\tilde\rvU}+2\gamma$, then
  Lemma~\ref{lm:chernov} guarantees there exists $\alpha_\gamma>0$
  such that
  \begin{align}
    \P[\tilde{\mathbf{U}}\tilde{\mathbf{X}}\tilde{\mathbf{Z}}]{\frac{1}{n}\I{\tilde{\mathbf{X}};\tilde{\mathbf{Z}}|\tilde{\mathbf{U}}}\geq\frac{\log
        M'}{n}-\gamma}\leq 2^{-\alpha_\gamma
      n}.\label{eq:bound_chernov_outage}
  \end{align}
  Set $\tau=2^{-\eta n}$ for some $\eta$ such that
  $0<2\eta<\min(\gamma,\alpha_\gamma)$; note that $\rho = \frac{\ln 2}{2}2^{-\eta n
  }+o(2^{-\eta n })$. Therefore, for $n$ large enough,
  \begin{align}
    \frac{1}{n}\log \rho &\geq -\gamma,\quad \frac{1}{\gamma_n(\gamma_n\rho
      +\gamma_n-1)}\leq 2\cdot 2^{\eta n},\quad
    \frac{1}{\gamma^3_{n}}\leq2.\label{eq:bound_technical}
  \end{align}
  Consequently,
  combining~\eqref{eq:bound_dmc_init},~\eqref{eq:bound_dmc_var_truncated},~\eqref{eq:bound_dmc_final},~\eqref{eq:bound_chernov_outage}
  and~\eqref{eq:bound_technical}, we obtain for $n$ large enough,
  \begin{multline*}
    \E[\rvC_n]{\metric{2}(\rvC_n)} \leq 4\cdot\frac{2^{-\eta n}}{\log e} +8\cdot
    2^{-\alpha_\gamma n} + 8\cdot 2^{-\alpha_\gamma n} \\+16\cdot
    2^{-(\gamma-2\eta)n}+ 16\cdot 2^{-(\alpha_\gamma-2\eta)n} +2\cdot
    2^{-\beta_\delta n}.
  \end{multline*}
  Therefore, for $n$ large enough, there exists $\alpha_{\gamma,\delta}>0$ such that $\E{\metric{2}(\rvC_n)}\leq 2^{-\alpha_{\gamma,\delta} n}$.
\end{IEEEproof}
Using Markov's inequality and for $n$ sufficiently large, we conclude that if
\begin{align*}
      \begin{array}{l}
          R_0 \leq \min\left(\avgI{\tilde\rvU;\tilde\rvY}-2\gamma,\avgI{\tilde\rvU;\tilde\rvZ}-2\gamma\right)\\
R_1\leq \avgI{\tilde\rvX;\tilde\rvY|\tilde\rvU}-\avgI{\tilde\rvX;\tilde\rvZ|\tilde\rvU}-4\gamma,
    \end{array}
\end{align*}
then there exists a specific code $\calC_n$ such that $\perr(\calC_n)\leq 2\epsilon$ and $\metric{2}(\calC_n)\leq 2^{-\frac{\alpha_{\gamma,\delta}}{2}n}$. Using~\cite[Lemma 1]{Csiszar1996} with $n$ large enough,
we obtain $\metric{1}(\calC_n)\leq 2^{-{\beta_{\gamma,\delta}}n}$ for some $\beta_{\gamma,\delta}>0$.


\begin{IEEEbiography}[{\includegraphics[width=1in,height =1.25in,clip,keepaspectratio]{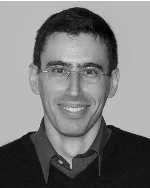}}]{Matthieu Bloch} received the Engineering degree from Sup\'elec, Gif-sur-Yvette, France, the M.S. degree in Electrical Engineering from the Georgia Institute of Technology, Atlanta, in 2003, the Ph.D. degree in Engineering Science from the Universit\'e de Franche-Comt\'e, Besan\c{c}on, France, in 2006, and the Ph.D. degree in Electrical Engineering from the Georgia Institute of Technology in 2008. In 2008-2009, he was a postdoctoral research associate at the University of Notre Dame, South Bend, IN, USA. Since July 2009, Dr. Bloch has been on the faculty of the School of Electrical and Computer Engineering at the Georgia Institute of Technology, where he is currently an Assistant Professor. His research interests are in the areas of information theory, error-control coding, wireless communications, and cryptography. Dr. Bloch is a member of the IEEE and has served on the organizing committee of several international conferences; he is the current chair of the Online Committee of the IEEE Information Theory Society. He is the co-recipient of the IEEE Communications Society and IEEE Information Theory Society 2011 Joint Paper Award and the co-author of the textbook \emph{Physical-Layer Security: From Information Theory to Security Engineering} published by Cambridge University Press.
\end{IEEEbiography}

\begin{IEEEbiography}[{\includegraphics[width=1in,height =1.25in,clip,keepaspectratio]{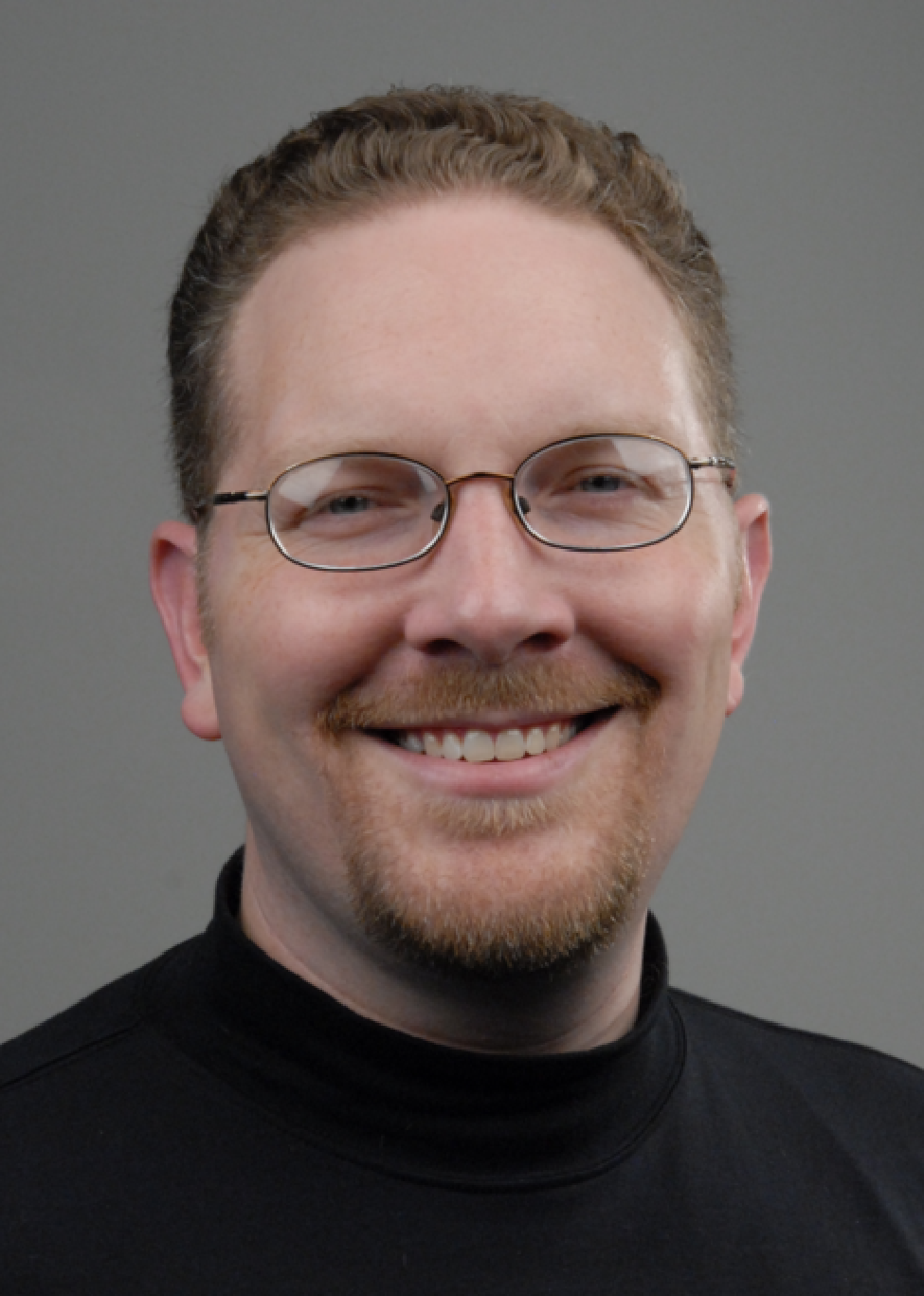}}]{J. Nicholas Laneman} is Founding Director of the Wireless Institute in the College of Engineering, an Associate Professor of Electrical Engineering, and a Fellow of the John J. Reilly Center for Science, Technology, and Values at the University of Notre Dame. He joined the faculty in August 2002 shortly after earning a Ph.D. in Electrical Engineering and Computer Science from the Massachusetts Institute of Technology (MIT). His research and teaching interests are in communications architecture -- a blend of information theory, error-control coding, signal processing for communications, network protocols, and hardware design -- with current emphasis on wireless systems.

Laneman has received a 2006 Presidential Early-Career Award for Scientists and Engineers (PECASE), a 2006 National Science Foundation (NSF) CAREER Award, a 2003 Oak Ridge Associated Universities (ORAU) Ralph E. Powe Junior Faculty Enhancement Award, and the 2001 MIT EECS Harold L. Hazen Graduate Teaching Award. He is an IEEE Senior Member and has served as an Associate Editor for IEEE Transactions on Communications, as a Guest Editor for Special Issues of IEEE Transactions on Information Theory and IEEE Journal on Selected Areas in Communications, and as the first Online Editor for the IEEE Information Theory Society.
\end{IEEEbiography}

\end{document}

%% file: rv_defs.tex
\DeclareMathAlphabet{\eurm}{U}{eur}{m}{n}
\DeclareMathAlphabet{\mathbsf}{OT1}{cmss}{bx}{n}
\DeclareMathAlphabet{\mathssf}{OT1}{cmss}{m}{sl}
\DeclareMathAlphabet{\mathcsf}{OT1}{cmss}{sbc}{n}

\newcommand{\samplevalue}[1]{{\lowercase{#1}}}
\newcommand{\randomvalue}[1]{{\uppercase{#1}}}

\DeclareSymbolFont{bsfletters}{OT1}{cmss}{bx}{n}  
\DeclareSymbolFont{ssfletters}{OT1}{cmss}{m}{n}
\DeclareMathSymbol{\bsfGamma}{0}{bsfletters}{'000}
\DeclareMathSymbol{\ssfGamma}{0}{ssfletters}{'000}
\DeclareMathSymbol{\bsfDelta}{0}{bsfletters}{'001}
\DeclareMathSymbol{\ssfDelta}{0}{ssfletters}{'001}
\DeclareMathSymbol{\bsfTheta}{0}{bsfletters}{'002}
\DeclareMathSymbol{\ssfTheta}{0}{ssfletters}{'002}
\DeclareMathSymbol{\bsfLambda}{0}{bsfletters}{'003}
\DeclareMathSymbol{\ssfLambda}{0}{ssfletters}{'003}
\DeclareMathSymbol{\bsfXi}{0}{bsfletters}{'004}
\DeclareMathSymbol{\ssfXi}{0}{ssfletters}{'004}
\DeclareMathSymbol{\bsfPi}{0}{bsfletters}{'005}
\DeclareMathSymbol{\ssfPi}{0}{ssfletters}{'005}
\DeclareMathSymbol{\bsfSigma}{0}{bsfletters}{'006}
\DeclareMathSymbol{\ssfSigma}{0}{ssfletters}{'006}
\DeclareMathSymbol{\bsfUpsilon}{0}{bsfletters}{'007}
\DeclareMathSymbol{\ssfUpsilon}{0}{ssfletters}{'007}
\DeclareMathSymbol{\bsfPhi}{0}{bsfletters}{'010}
\DeclareMathSymbol{\ssfPhi}{0}{ssfletters}{'010}
\DeclareMathSymbol{\bsfPsi}{0}{bsfletters}{'011}
\DeclareMathSymbol{\ssfPsi}{0}{ssfletters}{'011}
\DeclareMathSymbol{\bsfOmega}{0}{bsfletters}{'012}
\DeclareMathSymbol{\ssfOmega}{0}{ssfletters}{'012}

\newcommand{\rvA}{{\randomvalue{A}}}	

\newcommand{\rvC}{{\randomvalue{C}}}

\newcommand{\rvH}{{\randomvalue{H}}}
\newcommand{\rvI}{{\randomvalue{I}}}

\newcommand{\rvK}{{\randomvalue{K}}}
\newcommand{\rvM}{{\randomvalue{M}}}	
\newcommand{\rvN}{{\randomvalue{N}}}	
\newcommand{\rvR}{{\randomvalue{R}}}	
\newcommand{\rvU}{{\randomvalue{U}}}	
\newcommand{\rvV}{{\randomvalue{V}}}	
\newcommand{\rvW}{{\randomvalue{W}}}	
\newcommand{\rvX}{{\randomvalue{X}}}  	
\newcommand{\rvY}{{\randomvalue{Y}}}	
\newcommand{\rvZ}{{\randomvalue{Z}}}	

\newcommand{\svh}{{\samplevalue{h}}}	

\newcommand{\svu}{{\samplevalue{u}}}

\newcommand{\rvx}{{\randomvalue{x}}}	
\newcommand{\svx}{{\samplevalue{x}}}

\newcommand{\rvy}{{\randomvalue{y}}}	
\newcommand{\svy}{{\samplevalue{y}}}

\newcommand{\rvz}{{\randomvalue{z}}}	

\newcommand{\svz}{{\samplevalue{z}}}

\newcommand{\calA}{{\mathcal{A}}}
\newcommand{\calB}{{\mathcal{B}}}
\newcommand{\calC}{{\mathcal{C}}}
\newcommand{\calD}{{\mathcal{D}}}
\newcommand{\calE}{{\mathcal{E}}}

\newcommand{\calG}{{\mathcal{G}}}

\newcommand{\calK}{{\mathcal{K}}}

\newcommand{\calM}{{\mathcal{M}}}
\newcommand{\calN}{{\mathcal{N}}}
\newcommand{\calO}{{\mathcal{O}}}
\newcommand{\calP}{{\mathcal{P}}}
\newcommand{\calQ}{{\mathcal{Q}}}
\newcommand{\calR}{{\mathcal{R}}}
\newcommand{\calT}{{\mathcal{T}}}
\newcommand{\calS}{{\mathcal{S}}}
\newcommand{\calU}{{\mathcal{U}}}
\newcommand{\calV}{{\mathcal{V}}}
\newcommand{\calX}{{\mathcal{X}}}
\newcommand{\calY}{{\mathcal{Y}}}
\newcommand{\calW}{{\mathcal{W}}}
\newcommand{\calZ}{{\mathcal{Z}}}

%% file: fct_defs.tex
\newcommand{\E}[2][]{{\mathbb{E}_{#1}}{\left(#2\right)}}       
       
\renewcommand{\P}[2][]{{\mathbb{P}_{#1}}{\left(#2\right)}}
\newcommand{\p}[2][]{\ensuremath{p_{#1}\!\left({#2}\right)}}
\newcommand{\avgD}[2]{{{\mathbb{D}}\!\left({#1\Vert#2}\right)}}
\newcommand{\V}[1]{{{\mathbb{V}}\!\left(#1\right)}}
\newcommand{\avgI}[1]{{{\mathbb{I}}\!\left(#1\right)}}

\newcommand{\I}[1]{{\randomvalue{I}}{\left(#1\right)}}
\newcommand{\avgH}[1]{{\mathbb{H}}\!\left(#1\right)}
\renewcommand{\H}[1]{{\randomvalue{H}}\!\left(#1\right)}

\newcommand{\plimsup}[2][n\rightarrow\infty]{\underset{#1}{\mbox{\textnormal{p-limsup}}\;}{#2}}
\newcommand{\pliminf}[2][n\rightarrow\infty]{\underset{#1}{\mbox{\textnormal{p-liminf}}\;}{#2}}

\newcommand{\card}[1]{\ensuremath{\left|{#1}\right|}}           
\newcommand{\abs}[1]{\ensuremath{\left|#1\right|}}              
\newcommand{\eqdef}{\ensuremath{\triangleq}}                    
\newcommand{\intseq}[2]{\ensuremath{\llbracket{#1},{#2}\rrbracket}}  

\newcommand{\dd}{\text{\textnormal{d}}}


%% file: th_defs.tex
\newtheorem{lemma}{Lemma}
\newtheorem{example}{Example}
\newtheorem{theorem}{Theorem}
\newtheorem{proposition}{Proposition}
\newtheorem{definition}{Definition}
\newtheorem{remark}{Remark}
\newtheorem{corollary}{Corollary}